\documentclass[11pt]{article}
\usepackage[utf8]{inputenc}
\usepackage[tbtags]{amsmath}
\usepackage{amsthm}
\usepackage{mathtools}
\usepackage{amssymb}
\usepackage{amsfonts}
\usepackage[english]{babel}
\usepackage{hyperref}
\usepackage{bm}
\usepackage{graphicx}
\usepackage{subfigure}
\usepackage{multirow}
\usepackage{geometry}
\usepackage{multicol}
\usepackage{tabu}
\usepackage{longtable}
\usepackage{csquotes}
\usepackage[linesnumbered,ruled,vlined]{algorithm2e}
\usepackage[backend=biber,style=ieee,sorting=nty]{biblatex}
\allowdisplaybreaks

\usepackage[table]{xcolor}
\usepackage{multirow}
\usepackage{multicol}
\usepackage{tabu}
\usepackage{tikz}
\usetikzlibrary{shapes,arrows}
\usepackage{subfigure}
\usepackage{float}

\hypersetup{
    colorlinks=true,
    linkcolor=blue,
    filecolor=magenta,      
    urlcolor=cyan,
}

\newtheorem{theorem}{Theorem}

\newtheorem{remark}{Remark}

\SetKwInput{KwInput}{Input}
\SetKwInput{KwOutput}{Output}
\SetKwInput{KwInit}{Initialization}

\DeclareMathOperator*{\argmin}{arg\,min}

\title{\bf A Fast Detection Method of Break Points in Effective Connectivity Networks
\footnote{This article has been accepted by IEEE Transactions on Medical Imaging. (\url{https://ieeexplore.ieee.org/abstract/document/9627661)}}}
\author{Peiliang Bai\footnote{P. Bai is with the Department of Statistics, University of Florida, Gainesville, FL 32611 USA (email: baipl92@ufl.edu).} , Abolfazl Safikhani\footnote{A. Safikhani is with the Department of Statistics and the Informatics Institute, University of Florida, Gainesville, FL 32611 USA (email: a.safikhani@ufl.edu).}, George Michailidis\footnote{G. Michailidis is with the Departments of Statistics and Computer \& Information Sciences \& Engineering and the Informatics Institute, University of Florida, Gainesville, FL 32611 USA (email: gmichail@ufl.edu).}}

\begin{document}

\maketitle
\begin{abstract}
    There is increasing interest in identifying changes in the underlying states of brain networks. The availability of large scale neuroimaging data creates a strong need to develop fast, scalable methods for detecting and localizing in time such changes and also identify their drivers, thus enabling neuroscientists to hypothesize about potential mechanisms. This paper presents a fast method for detecting break points in exceedingly long time series neurogimaging data, based on vector autoregressive (Granger causal) models. It uses a multi-step strategy based on a regularized objective function that leads to fast identification of candidate break points, followed by clustering steps to select the final set of break points and subsequent estimation with false positives control of the underlying Granger causal networks. The latter provide insights into key changes in network connectivity that led to the presence of break points. The proposed methodology is illustrated on synthetic data varying in their length, dimensionality, number of break points, strength of signal and also applied to EEG data related to visual tasks. 
\end{abstract}


\section{Introduction.}\label{sec:intro}
There is strong interest in the study of whole-brain \textit{connectivity patterns} and in identifying \textit{changes} at different time scales (seconds or minutes as measured by fMRI technology or milliseconds as measured by EEG/MEG technologies). Such connectivity patterns reflect large
scale neuronal synchrony across distributed regions of the brain and play a significant role in memory and high-order associative processing \cite{buzsaki2012brain}, and also serve in part to enhance the efficiency of neuronal communication \cite{deco2011dynamical,sporns2013human}. 
It is further thought that disruption of this synchrony underlies various neurological and psychiatric diseases \cite{buzsaki2012brain,sporns2013structure}. This is supported by a growing number of neuroimaging studies which show that neural communication between distant regions is impaired in conditions such as a mental health disorders or neurodegenerative diseases \cite{cisler2013altered, ma2011abnormal}, or due to aging processes \cite{nakagawa2013bottom,betzel2014changes}. 

Recent work has shown that interesting information is contained within dynamically evolving connectivity patterns, even in the resting state \cite{hutchison2013dynamic,jeong2016connectivity}. 
Given the importance of identifying time evolving connectivity patterns, a number of approaches have been proposed in the literature - for a recent review see 
\cite{preti2017dynamic}. Further, an important distinction between the concepts of functional versus effective connectivity that can be derived from neuroimaging data is presented in \cite{friston1994functional,friston2011functional}. Proposed approaches capture connectivity patterns through covariance/correlation \cite{lindquist2014evaluating,xu2015dynamic} or (sparse) inverse covariance matrices \cite{cribben2012dynamic, zhu2018sparse}, hidden Markov models \cite{zhang2019estimating}, vector autoregressive (Granger causal) models \cite{seth2015granger,teng2019bayesian,zhang2017linear, antonacci2020information} and reduced rank vector autoregressive models \cite{ting2017estimating}. Further, strategies for detection of break points involve search procedures based on information criteria \cite{cribben2013detecting}, setting up a hypothesis test, \cite{hindriks2016can}, or using penalized estimation procedures \cite{monti2014estimating,kundu2018estimating}, clustering procedures \cite{allen2014tracking,esposito2003real}, or regime switching models \cite{ting2017estimating}. Note that (sparse) VAR (Granger causal) models have been extensively used for estimating {effective connectivity} networks (without considering break points) \cite{friston2011functional,valdes2005estimating,haufe2010sparse,monti2014estimating,siggiridou2015granger}. Further, another stream of literature aims to capture nonlinearities in the times series data obtained from neurimaging data \cite{deco2017dynamics,garrett2003comparison,acharya2015nonlinear,natarajan2004nonlinear,chua2011application}.

This brief literature review shows that different statistical models have been used for representing {effective connectivity (for discussion on this concept and connection to functional connectivity see \cite{friston1994functional,friston2011functional})} coupled with strategies to detect break points. However, most of these algorithmic approaches do not scale well to data sets comprising of very large numbers of time points and also lack strong theoretical foundations in the form of statistical guarantees for consistently detecting break points. Thus, the goal of this paper is to introduce a fast algorithmic strategy for detection of multiple break points in exceedingly long neuroimaging data based on vector autoregressive (VAR) models, followed by estimation of the Granger causal networks in the stationary segments identified. Note that it reformulates the break point detection problem as a \textit{variable selection} problem for a regularized  high-dimensional linear regression model exhibiting temporal dependence amongst the observations, as described in Section \ref{sec:method}. It adds significant flexibility that leads to substantial computational gains to the detection procedure developed as presented in Section \ref{comp_complexity} and coupled with highly accurate estimates of the break points and the underlying Granger causal networks, as numerically illustrated in Section \ref{section:3}. 

Note that (sparse) VAR (Granger causal) models have been extensively used for estimating {effective connectivity} networks (without considering break points) \cite{friston2011functional,valdes2005estimating,haufe2010sparse,monti2014estimating,siggiridou2015granger}. Further, there is another stream of literature that aims to capture nonlinearities in the times series data obtained from neurimaging data \cite{deco2017dynamics,garrett2003comparison,acharya2015nonlinear,natarajan2004nonlinear,chua2011application}.


\section{Methods}\label{sec:method}

\subsection{Sparse VAR Models with Structural Breaks}\label{sec:model}

A sparse VAR model with structural breaks is defined as: suppose there exist $ m_0 $ break points $ 0 = t_0 < t_1 < \cdots < t_{m_0} < t_{m_0 + 1} = T + 1 $, so that for $t_{j-1} \leq t < t_j, j = 1, \cdots, m_0 + 1$,
\begin{equation}\label{eq:model}
y_t = \Phi^{(1,j)} y_{t-1} + \cdots + \Phi^{(q,j)} y_{t-q} +  {\Sigma_j}^{1/2} \epsilon_t,
\end{equation}
where $ y_t $ is a $p$-vector of observed time series at time $ t $; $ \Phi^{(l,j)} \in \mathbb{R}^{p \times p}$ is the coefficient matrix corresponding to the $l$-th lag ($l\in\{1,\cdots,q\}$) for the $j$-th stationary segment, where $ j = 1, \cdots, m_0+1 $. Note that in the sequel, we assume that the coefficient matrices $\Phi^{(l,j)}$ are {(weakly)} \textit{sparse}; {in the strict sparse case, not too many elements of the transition matrix are non-zero, whereas in the weakly sparse one, there are a few elements in $\Phi^{(l,j)}$ with ``large" magnitudes and the remaining ones have either ``small" magnitudes or are zero. } The elements of $\Phi^{(l,j)}$ correspond to Granger causal effects \cite{lin2017regularized} and their collection is referred to as a Granger causal network \cite{friston2013analysing}. Hence, a weakly sparse VAR model can accommodate small magnitude Granger causal effects, thus leading to a more connected network that may better reflect effective connectivity patterns. Note that the distinction between estimating a strict versus a weak sparse VAR (or more generally a penalized regression) model is manifested both theoretically and in applications through the selection of the tuning parameter \cite{negahban2012unified}, but \textit{does not impact} the nature of the estimation procedure (penalized least squares, as discussed in the sequel). Finally, $\epsilon_t $ is a Gaussian noise process  with $ \Sigma_j$ denoting the covariance matrix for the $ j$-th segment. In each segment $ [t_{j-1}, t_j] $, all model parameters are assumed to be fixed. However, elements of both the autoregressive  matrices and the covariance matrix of the noise process can potentially change structure and magnitudes of its entries between segments. To simplify notation, we occasionally denote the noise process as $ \epsilon_t $ without specifying the covariance $\Sigma_j$. For ease of presentation of the break point detection procedure, we ignore the contemporaneous covariance matrices $\Sigma_j$ for the moment and return to them in Section \ref{section:4} (see also discussion in \cite{safikhani2017joint}). 

Note that the VAR model focuses on modeling the temporal dependencies in the data. Hence, it can be seen that in the proposed change point model for a VAR process, changes in the transition matrices $\Phi^{(l,j)}, j=1,2,\dots, m_0+1$ induce changes in the temporal dynamics of the process. 

The key objective is then to detect the break points $t_j$, in a computationally efficient manner that is also scalable for very large values of $T$. Of interest is also to estimate accurately the VAR parameters $ \Phi^{(l,j)}$, under a high dimensional scaling ($p^2 \gg T$).

\medskip
\noindent
{\bf Notation:} Throughout this paper, the transpose of a matrix $A$ is denoted by $A^\prime$, and the cardinality of a set $\mathcal{A}$ is denoted as $|\mathcal{A}|$. We use $\|A\|_1$ and $\|A\|_F$ to denote the $\ell_1$ norm and Frobenius norm of matrix A, and use $\|v\|_1$ and $\|v\|_2$ to represent the $\ell_1$ and $\ell_2$ norm of vector $v$, respectively. Denote by $\Phi^{(\cdot, j)} = \left(\Phi^{(1,j)}, \dots, \Phi^{(q,j)}\right) \in \mathbb{R}^{p\times pq}$, and define the number of nonzero elements in the $k$-th row of $\Phi^{(.,j)}$ by $d_{kj}$, $ k=1,2,\dots,p, \ j = 1,2,\dots,m_0$. Note that our analysis deals with the high-dimensional case with many break points, where $p$, $m_0$ and the sparsity level $d_{kj}$ increase with the number of observations, $T$. 
\vspace{-0.22cm}

\subsection{A Thresholded Block Segmentation Scheme (TBSS) Algorithm}\label{sec:methods}

The main idea of the proposed algorithm is to partition the time axis into blocks of size $b_T$ and fix the VAR parameters within each block. Obviously, the maximum block size $b_T$ that can be afforded is {dictated by the minimum distance $D_T$ allowed between break points; namely $D_T=\min_{j\in\{0,\cdots,m_0\}} |t_j-t_{j+1}|$ (see the following Assumption A3)}. Further, as discussed in the sequel, there is a trade-off between the computational cost of TBSS and the block size $b_T$.

To formally summarize the key steps of the TBSS algorithm, {we define a sequence of time points $q = r_0 < r_1 < \cdots < r_{k_T} = T$} which play the role of end points for the blocks; i.e. $ r_{i+1} - r_i = b_T $ for $ i = 0, ..., T-1 $, and $ k_T = \lceil \frac{T}{b_T} \rceil $ is the total number of blocks. {Further, without loss of generality, we assume that $T=k_Tb_T$.}
We define $\theta_1 = \Phi^{(\cdot, 1 )}$ and for $i=2,3,\dots,k_T$ set
\begin{equation}
    \label{eq:theta}
    \theta_i = 
    \begin{cases}
        \Phi^{(\cdot,j+1)} - \Phi^{(\cdot,j)}, &\  \text{if}\  {t_j \in [r_{i-1}, r_i)}\  \text{for some }j, \\
        0, &\  \text{otherwise.}
    \end{cases}
\end{equation}
Next, we form the following multivariate linear regression with the newly defined $\theta_i$ in \eqref{eq:theta} as the regression parameters. We define the design matrix as:
\begin{equation*}
    \mathcal{Z} = 
    \begin{pmatrix} 
    \mathbf{Y}_0^\prime & \mathbf{0} & \cdots & \mathbf{0} \\
    \mathbf{Y}_1^\prime & \mathbf{Y}_1^\prime & \cdots & \mathbf{0} \\
    \vdots & \vdots & \ddots & \vdots \\
    \mathbf{Y}_{k_T-1}^\prime & \mathbf{Y}_{k_T-1}^\prime & \cdots & \mathbf{Y}_{k_T-1}^\prime
    \end{pmatrix},
\end{equation*}
{where the block matrices $\mathbf{Y}_j \in \mathbb{R}^{b_T \times pq}$ are defined as: $\mathbf{Y}_{j} = (Y_{r_j}, Y_{r_j+1}, \cdots, Y_{r_{j+1}-1})$, for $j=0, 1, \cdots, k_T-1$. Further, the elements in each block matrix $\mathbf{Y}_j$ correspond to the $pq$-dimensional vectors {$Y_l = (y_l^\prime, \cdots, y_{l-q+1}^\prime)^\prime$}, for time points $l=q-1, q, \cdots, T$. Thus, the dimensionality of the design matrix is $\mathcal{Z} \in \mathbb{R}^{T\times k_Tpq}$.}
Similarly, the response matrix is: $\mathcal{Y} = (y_q, \cdots, y_{r_1}, \cdots, y_{r_{k_T-1}}, \cdots, y_T)^\prime$, {the model parameters can be  defined compactly as: $\Theta = (\theta_1, \theta_2, \dots, \theta_{k_T})^\prime \in \mathbb{R}^{k_Tpq \times p}$}, and the corresponding error term can be similarly given by $E = (\epsilon_q, \cdots, \epsilon_{r_1}, \epsilon_{r_1+1}, \cdots, \epsilon_{r_2}, \cdots, \epsilon_{r_{k_T-1}}, \cdots, \epsilon_T)^\prime$. Note that in this parameterization, $ \theta_i \neq 0$ for $i \geq 2$ implies a change in the VAR coefficients $\Phi^{(.,j)}$ in the block $[r_{i-1}, r_i)$. Therefore, for $j = 1, \cdots, m_0$, the structural break points $ t_j$ would correspond to time points $r_i \geq 1$, for which $\theta_i \neq 0$. 

{
The multivariate regression model can be succinctly written as $\mathcal{Y}=\mathcal{Z}\Theta+E$ and expressed as a regular regression one, with a vector response, as:
\begin{equation}
    \label{eqn:vectorform_block}
    \textbf{Y} = \textbf{Z} \mathbf{\Theta} + \textbf{E}, 
\end{equation}
where $\textbf{Y} = \mbox{vec}(\mathcal{Y})$, $\textbf{Z} = I_p \otimes \mathcal{Z}$, $\bm{\Theta} = \mbox{vec}(\Theta)$, and the noise term $\textbf{E} = \mbox{vec}(E)$, with $ \otimes $ denoting the Kronecker product of two matrices, and $ \pi_b = k_T p^2 q $. Then, the dimensions of the dependent variable, design matix and parameter vector are $\textbf{Y} \in \mathbb{R}^{Tp \times 1}$, $\textbf{Z} \in \mathbb{R}^{Tp \times \pi_b}$, $\mathbf{\Theta} \in \mathbb{R}^{\pi_b \times 1}$, and $\textbf{E} \in \mathbb{R}^{Tp \times 1}$.}

\noindent
{\bf Step 1. Identification of candidate break points:} Based on the linear regression representation in \eqref{eqn:vectorform_block}, the model parameters $\mathbf{\Theta}$ will be estimated via a regularized least squares objective function. Regularization is necessary to handle the growing number of parameters due to the presence of break points, as well as the number of time series $p$. Therefore, an initial estimate of the $\mathbf{\Theta}$ parameters is obtained by solving the following $\ell_1$-penalized least squares regression:
\begin{equation}
    \label{eqn:objectivefunc}
    \widehat{\mathbf{\Theta}} 
    = \argmin_{\mathbf{\Theta}}\frac{1}{T-q+1} \| \textbf{Y} - \textbf{Z} \mathbf{\Theta} \|_2^2 + \lambda_{1,T}  \| \mathbf{\Theta} \|_1
    + \lambda_{2,T} \sum_{i=1}^{k_T} \left \| \sum_{j=1}^{i} \theta_j \right \|_1.
\end{equation}
{
There are two penalty terms, the first term controlling the number of break points through a fused lasso penalty, and the second one ensuring the sparsity of the parameters in accordance with the sparse nature of the posited model.} The corresponding optimization problem is convex and hence can be solved efficiently through standard algorithms \cite{tibshirani2005sparsity}. Denote the set of estimated break points obtained from solving \eqref{eqn:objectivefunc} by 
$
    \widehat{\mathcal{A}}_T = \left\lbrace i \geq 2 :  \widehat{\theta}_i \neq 0 \right\rbrace.
$
The cardinality of this set corresponds to the estimated number of break points; i.e., $\widehat{m} = | \widehat{\mathcal{A}}_T |$. Further, let $\widehat{t}_j$, $j= 1, \ldots, \widehat{m}$ denote their estimated locations. Then, the relationship between $ \widehat{\theta}_j $ and $ \widehat{\Phi}^{(.,j)}$ in each of the estimated segments is given by:
\begin{equation}
    \label{eqn:psi}
    \widehat{\Phi}^{(.,1)} = \widehat{\theta}_1\quad \mbox{and}\quad \widehat{\Phi}^{(.,j)} = \sum_{i=1}^{\widehat{t}_j} \widehat{\theta}_i,\quad j = 1, 2, \ldots, \widehat{m}. 
\end{equation}
Note that the block size $b_T$ acts as a tuning parameter that \textit{regulates} the number of model parameters to be estimated, given by 
$\pi_b = \lceil \frac{T}{b_T}  \rceil p^2 q $. In the extreme case with $b_T=1$, TBSS reverts to an exhaustive search of all time points to locate the structural breaks. Nevertheless, $b_T$ can not also be too large as the theoretical results in the sequel show. Specifically, $b_T$ can range in $\{1,\cdots,\lfloor \sqrt{T}\rfloor\}$. To see this, note that large values of the block size $b_T$ result in reduced computational cost in Step 1, while increasing the computation time in Step 2, as explained later on. To keep the balance between the computation times of both steps and to minimize their total computation time, it is appropriate to select the block size $b_T$ within the range $\{1,\cdots,\lfloor \sqrt{T}\rfloor\}$. On the other hand, clearly $b_T$ can not be larger than the minimum spacing $D_T$ between consecutive true break points since it that case, there may be more than one true break point in a certain block, while the strategy in \eqref{eqn:objectivefunc} can at most detect a single break point within each block. Therefore, any selection of block size will implicitly impose restrictions on the minimum spacing $D_T$ between consecutive break points or equivalently, it puts an upper bound on the total number of break points allowed in the model. Specifically, $D_T$ should be much large than the block size $b_T$ (asymptotically, we should have ${b_T}/{D_T} \rightarrow 0$ as $T \rightarrow +\infty$). For example, if $b_T = \lfloor \sqrt{T}\rfloor$, then $D_T$ should be of order at least $T^{0.5+\nu}$ for some small positive $\nu$ which means the total number of true break points $m_0$ must be at most of order $T^{0.5-\nu}$. Any block size larger than $\lfloor \sqrt{T}\rfloor$ restricts even further the total number of true break points allowed in the model, while not helping in reducing computation time. Thus, the range $\{1,\cdots,\lfloor \sqrt{T}\rfloor\}$ can be seen as the feasible set of possible block sizes. Note that restrictions on the minimum spacing $D_T$ between consecutive break points are natural/typical in break point detection literature \cite{chan2014group}. Even in the case of $b_T=1$, as mentioned in \cite{safikhani2017joint}, $D_T$ has a lower bound of order $ (\log p)^{1+\nu} $ for some positive $\nu$. A practical procedure to determine a good value for $b_T$, in the absence of good prior information on the spacing $D_T$ between consecutive break points, is outlined in Section \ref{section:4}. 

\noindent
{\bf Step 2. Hard-thresholding for screening out redundant candidate change points:} the magnitude of the required threshold $\eta$ is selected by a completely data-driven method. The main idea is to combine the $K$-means clustering algorithm with the BIC criterion to cluster the changes in the parameter matrix into two subgroups. The main steps are given by:
\begin{itemize}
\item[A.] (Initialization): Define the jumps for each partitioned block by $v_k = \|\widehat{\theta}_k\|_2^2$, for $k=2,3,\dots, k_T$ and let $v_1 = 0$. Define the set $V$ as $V = (v_1, v_2, \dots, v_{k_T})$. Denote the set of selected blocks with large jumps as $J$ and set it to be the empty set. Also, set the value of the BIC function by $\text{BIC}^{\text{old}} = \infty$.
\item[B.] (Recursion state): Apply $K$-means clustering to the obtained jumps $V$ with $K=2$ clusters. Cluster $V_S$ contains small jump values, while cluster $V_L$ includes those with large jump sizes. Define $\eta = \left( \min V_L + \max V_S  \right)/2$ as the threshold. Then we add the corresponding partitioned blocks in the large jump size into $J$, and compute the BIC by using the estimated parameters $\widehat{\Theta}$ with $\widehat{\theta}_i=\mathbf{0}$ for blocks $i \notin J$ and denoted as $\text{BIC}^{\text{new}}$. Compute the difference $\text{BIC}^{\text{diff}} = \text{BIC}^{\text{new}} - \text{BIC}^{\text{old}}$. Stop if $\text{BIC}^{\text{diff}} \geq 0$; otherwise, set $\text{BIC}^{\text{old}} = \text{BIC}^{\text{new}}$ and $V=V \backslash J$.
\end{itemize}
{\bf Step 3. Block clustering step:} In this step, we use the Gap statistic \cite{friedman2001elements} to determine the number of clusters of the candidate break points selected in Step 2. The output of this step corresponds to clusters $C_i, i=1,\cdots,\widetilde{m}$ of candidate break points. \\
{\bf Step 4. Exhaustive search for identifying a single break point for each cluster:} For each selected cluster of break points $C_i, i=1,2,\cdots,\widetilde{m}$ from Step 3, {we define the \textit{search interval} $(l_i, u_i)$ whose lower and upper bounds are given by:
\begin{equation*}
    l_i =
    \begin{cases}
        c_i - b_T,&\ \text{if}\ |C_i|=1, \\
        \min\{C_i\},&\ \text{otherwise},
    \end{cases}\quad \text{and}\quad 
    u_i =
    \begin{cases}
        c_i + b_T,&\ \text{if}\ |C_i|=1, \\
        \max\{C_i\},&\ \text{otherwise}.
    \end{cases}
\end{equation*}
where $c_i$ is the unique element in $C_i$, whenever $|C_i|=1$.} Denote the subset of corresponding block indices in the interval $(l_i,u_i)$ by $J_i$ with $J_0 = \{1\}$ and $J_{\widetilde{m}+1} = \{k_T\}$. Further denote the closest block end to $\left( \max J_{i-1} + \min J_i \right)/2$ as $w_i$. Now, the \textit{local} parameter estimators are given by:
\begin{equation}
    \label{eqn:localparam}
    \widetilde{\Phi}^{(.,i)} = \sum_{k=1}^{w_i}\widehat{\theta}_k, 
\end{equation}
for $i=1,2,\cdots, \widetilde{m}+1$, where $\widehat{\theta}_k$ is derived by block fused lasso step from \eqref{eqn:objectivefunc},  for $k=1,2,\cdots, k_T$. Finally, for each $i=1,2,\cdots, \widetilde{m}$, the final estimated break points is defined as:
\begin{equation}
    \label{eqn:final}
    \widetilde{t}_i 
    = \argmin_{s\in (l_i, u_i)}\left\{ \sum_{t=l_i}^{s-1}\|y_{t+1} - \widetilde{\Phi}^{(.,i)}Y_t \|_2^2 + \sum_{t=s}^{u_i-1}\|y_{t+1} - \widetilde{\Phi}^{(.,i+1)}Y_t \|_2^2 \right\},
\end{equation}
where $ \widetilde{\Phi}^{(.,i)}$ are the local parameter estimates obtained by \eqref{eqn:localparam}. The final set of estimated break points obtained by solving \eqref{eqn:final} are denoted by $\widetilde{\mathcal{A}}_T = \left\{ \widetilde{t}_1, \dots, \widetilde{t}_{\widetilde{m}}\right\}$. 

\noindent
{\bf Step 5. Model parameter estimation:} Once the final set of break points have been identified from Step 4, we can {estimate} the transition matrices (and thus the Granger causal networks) by using the algorithms developed in \cite{lin2017regularized} for stationary data. To ensure that the data in the time segments between break points are strictly stationary, we remove all time points in a $R_T$-radius neighborhood of the break points obtained in Step 4. The length of $R_T$ needs to be at least $b_T$. Specifically, denote by $s_{j1} = \widetilde{t}_j - R_T - 1 $, $ s_{j2} = \widetilde{t}_j + R_T + 1 $ for $ j = 1, \cdots, \widetilde{m} $, and set $ s_{02} = q $ and $ s_{(\widetilde{m}+1)1} = T $. Next, define the intervals $ I_{j+1} = [ s_{j2}, s_{(j+1)1}  ] $ for $ j = 0, \cdots, \widetilde{m}$. The idea is to form a linear regression on $ \cup_{j=0}^{\widetilde{m}} I_{j+1} $ and estimate the auto-regressive parameters by minimizing an $\ell_1$-regularized least squares criterion. Specifically, we form the following linear regression similar to \eqref{eqn:vectorform_block}:
$
     \mathcal{Y}_s = \mathcal{X}_s B + E_s,
$
where $\mathcal{Y}_s = (y_q, \cdots, y_{s_{11}}, \cdots, y_{s_{\widetilde{m}2}}, \cdots, y_T)^\prime$, $B = (\beta_1, \beta_2, \cdots, \beta_{\widetilde{m}+1})$, the corresponding error term $E_s = (\zeta_q, \cdots, \zeta_{s_{11}}, \dots, \zeta_{s_{\widetilde{m}2}}, \cdots, \zeta_{T})^\prime$, and the design matrix is given by:
\begin{equation*}
 \mathcal{X}_s = 
 \begin{pmatrix} 
     \widetilde{\mathbf{Y}}_1 & \mathbf{0} & \cdots & \mathbf{0} \\
     \mathbf{0} & \widetilde{\mathbf{Y}}_2 & \cdots & \mathbf{0} \\
     \vdots & \vdots & \ddots & \vdots \\
     \mathbf{0} & \mathbf{0} & \cdots & \widetilde{\mathbf{Y}}_{\widetilde{m}},
 \end{pmatrix}.
\end{equation*}
where the diagonal elements are given by: $\widetilde{\mathbf{Y}}_j^\prime = (Y_{s_{j2}-1}, \cdots, Y_{s_{(j+1)1}-1})$, and $j=0,1,\dots, \widetilde{m}-1$. Then, we can rewrite it in compact form as: 
\begin{equation*}
 {\textbf{Y}}_{\textbf{s}} =  \textbf{Z}_{\textbf{s}} \textbf{B}  + \textbf{E}_{\textbf{s}},
\end{equation*}
where $ \textbf{Y}_{\textbf{s}} = \mbox{vec}(\mathcal{Y}_{\textbf{s}}) $, $ {\textbf{Z}_{\textbf{s}}} = I_p \otimes \mathcal{X}_{\textbf{s}} $, $ \textbf{B} = \mbox{vec}(B) $, $ {\textbf{E}}_{\textbf{s}} = \mbox{vec}(E_{\textbf{s}}) $, and $ {\textbf{s}} $ is the collection of all $ s_{j1}$ and $ s_{j2} $ for $ j = 0, \ldots, m_0+1 $. Let $ \widetilde{\pi} = ( \widetilde{m} +1) p^2 q $, $ N_j = s_{(j+1)1}-s_{j2} $ be the length of the interval $I_{j+1}$ for $ j = 0, \cdots, \widetilde{m} $ and $ N = \sum_{j=1}^{\tilde{m}} N_j $. {Then, $ \textbf{Y}_{\textbf{s}} \in \mathbb{R}^{N p \times 1} $, $ {\textbf{Z}_{\textbf{s}}} \in \mathbb{R}^{N p \times \widetilde{\pi}} $, $ \textbf{B} \in \mathbb{R}^{\tilde{\pi} \times 1} $, and $ {\textbf{E}}_{\textbf{s}} \in \mathbb{R}^{N p \times 1} $.} Therefore, we estimate the VAR parameters by solving the following $\ell_1$ regularized optimization problem:
\begin{equation}
 \label{eq:estimation_third}
 \widehat{\textbf{B}} = \argmin_{\textbf{B}}\left\{ \frac{1}{N}\left\| \textbf{Y}_{\textbf{s}} - {\textbf{Z}_{\textbf{s}}} \textbf{B} \right\|_2^2 + \rho_T  \left\| \textbf{B} \right\|_1\right\}.
\end{equation}

\noindent
{
{\bf Step 6 (optional): Stability selection:} It aims to remove false positive entries in the estimated Granger causal networks $\widehat{B}_j$ from Step 5.
}

{
The main steps are listed next (see also \cite{meinshausen2010stability}):
\begin{itemize}
    \item[(a)] Define a sequence of regularization parameters $R = (\rho_1, \cdots, \rho_L)$ for optimization problem \eqref{eq:estimation_third}, and a number $N_s$ of subsamples.
    \item[(b)] For each $\rho \in R$, do:
    \begin{itemize}
        \item[1.] Start with the full data set $X_{(full)} = (X_1, X_2, \dots, X_T)$;
        \item[2.] For each $n$ in $1,2,\cdots, N_s$, do:
        \begin{itemize}
            \item[i.] Subsample from $X_{(full)}$ without replacement to generate a smaller dataset of size $\lfloor T/2 \rfloor$, given by $X_{(n)}$;
            \item[ii.] Solve optimization problem \eqref{eq:estimation_third} using data $X_{(n)}$ with tuning parameter $\rho$ to obtain a selection set $\widehat{S}^{\rho}_{(n)}$ of non-zero elements in $\{\widehat{B}_j\}_{j=1}^{\hat{m}+1}$.
        \end{itemize}
        \item[3.] Given the selection sets from each subsample, calculate the empirical selection probability for each element $e$ in $\widehat{B}_j$:
        \begin{equation*}
            \widehat{\Pi}_e^\rho = \mathbb{P}(e \in \widehat{S}^{\rho}) = \frac{1}{N_s}\sum_{n=1}^{N_s} \mathbb{I}(e \in \widehat{S}_{(n)}^\rho).
        \end{equation*}
        The selection probability for element $e$ (an edge in the corresponding Granger causal network) is the corresponding probability of being selected by the solution of optimization problem \eqref{eq:estimation_third}.
    \end{itemize}
    \item[(c)] Given the selection probabilities for each edge $e$ in the network, construct the stable set according to the following criterion:
    \begin{equation*}
        \widehat{S}^{stable} := \left\{e: \max_{\rho \in R}\widehat{\Pi}^\rho_e \geq \tau \right\},
    \end{equation*}
    where $\tau \in (0,1)$ is some pre-determined threshold.
\end{itemize}
\begin{remark}
\label{remark:stabs}
Step 6 provides control of the false positive rate and provides better estimates of the underlying Granger causal network, as our numerical investigation in Section \ref{sec:application-results} demonstrates. 
\end{remark}
}

{
The main steps of TBSS for detection of break points (1-4) are summarized in the Supplement and illustrated in the flowchart shown in Figure~\ref{fig:steps}. The illustrative results in the sub-Figures are based on synthetic data,  with $p=20$ and $T=600$ and the presence of two true change points at $t=200, 400$, respectively. The default value for the block size is used: $b_T = \lfloor \sqrt{T} \rfloor \approx 25$. Note that TBSS accurately estimates the change points: $\widehat{t}_1 = 200$ and $\widehat{t}_2 = 400$. }
\tikzstyle{block} = [rectangle, draw, fill=blue!20,
    text width=12em, text centered, rounded corners, minimum height=4em]
\tikzstyle{smallblock} = [rectangle, draw, fill=purple!20,
    text width=8em, text centered, minimum height=3em, node distance=5cm]
\tikzstyle{thin-cloud} = [rectangle, draw, fill=red!20, node distance=7cm, text width=14em, 
    minimum height=5em]
\tikzstyle{line} = [draw, -latex']
\tikzstyle{plots} = [rectangle, draw, fill=yellow!20, node distance=5cm, text width=8em,
    minimum height=5.5em]
\tikzstyle{cloud} = [draw, rectangle, fill=red!20, node distance=7cm, text width=14em,
    minimum height=10em]

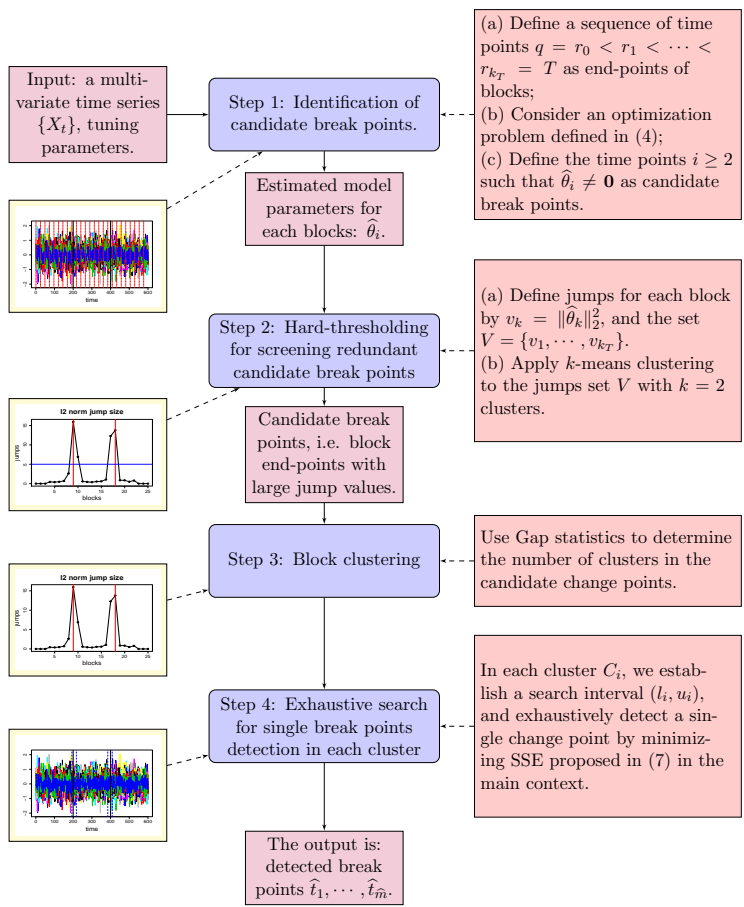
\begin{figure}[!ht]
    \centering
    \resizebox{.65\textwidth}{!}{
    \begin{tikzpicture}[node distance = 2cm]
        \node [block] (step1) {Step 1: Identification of candidate break points.};
        
        \node [cloud, right of=step1, node distance=6cm] (step1-steps) {
        (a) Define a sequence of time points $q=r_0<r_1<\cdots < r_{k_T}=T$ as end-points of blocks; \\
        (b) Consider an optimization problem defined in (4); \\
        (c) Define the time points $i \geq 2$ such that $\widehat{\theta}_i \neq \mathbf{0}$ as candidate break points.
        };
        \node [smallblock, left of=step1](input){
        Input: a multivariate time series $\{X_t\}$, tuning parameters.
        };
        
        \node [smallblock, below of=step1, node distance=2cm] (step1-result){
        Estimated model parameters for each blocks: $\widehat{\theta}_i$.
        };
        
        \node [plots, below of=input, node distance=3cm](step1-plot){
        \includegraphics[height=5.5em, width=8em]{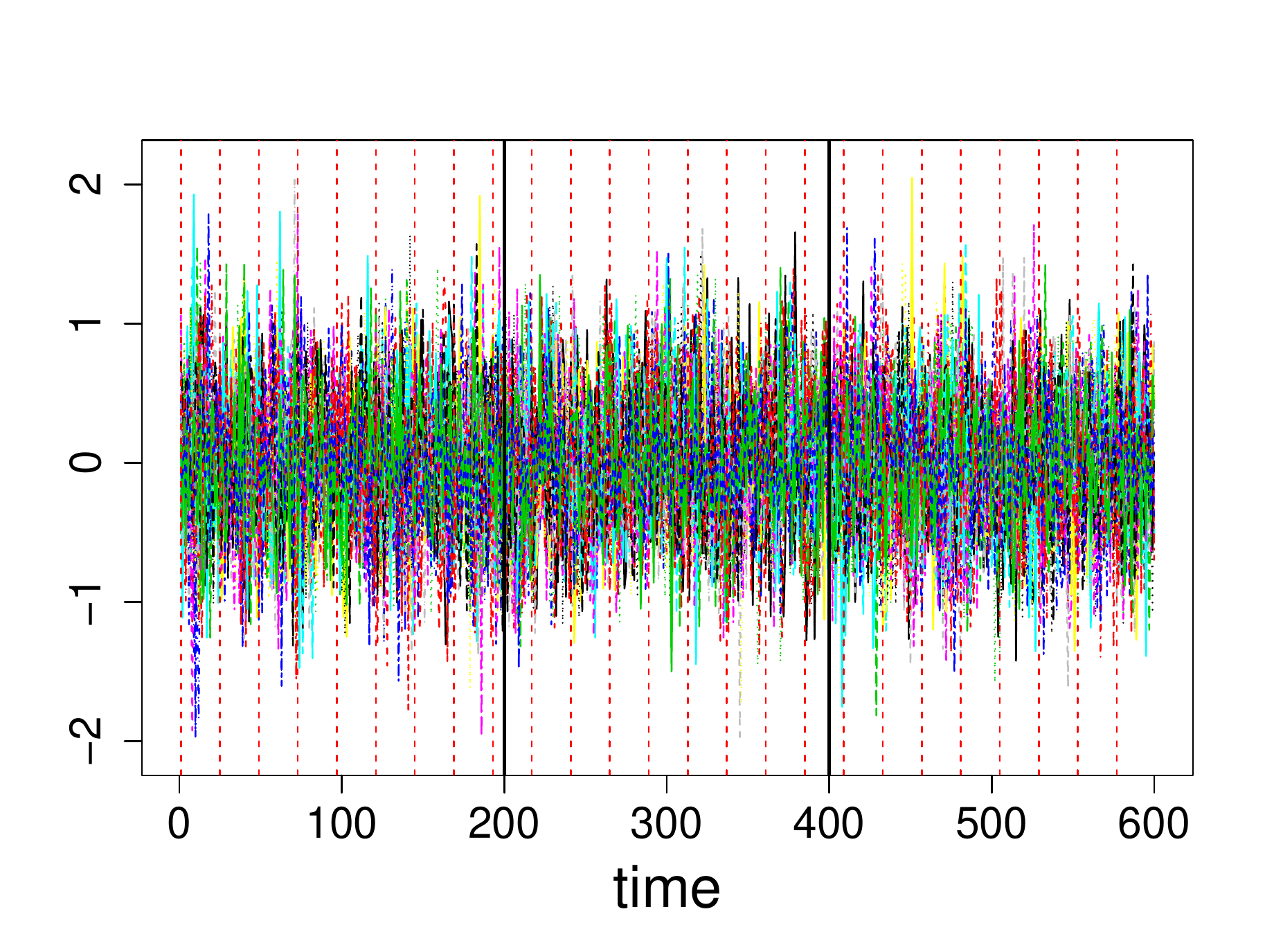}
        };
        
        \node [block, below of=step1-result, node distance=3cm] (step2) {Step 2: Hard-thresholding for screening redundant candidate break points};
        
        \node [cloud, right of=step2, node distance=6cm] (hard-threshold) {
        (a) Define jumps for each block by $v_k = \|\widehat{\theta}_k\|_2^2$, and the set $V=\{v_1, \cdots, v_{k_T}\}$. \\
        (b) Apply $k$-means clustering to the jumps set $V$ with $k=2$ clusters.
        };
        
        \node[smallblock, below of=step2, node distance=2.2cm] (step2-result) {
        Candidate break points, i.e. block end-points with large jump values.
        };
        
        \node[plots, left of=step2-result, node distance=5cm] (step2-plot) {
        \includegraphics[height=5.5em, width=8em]{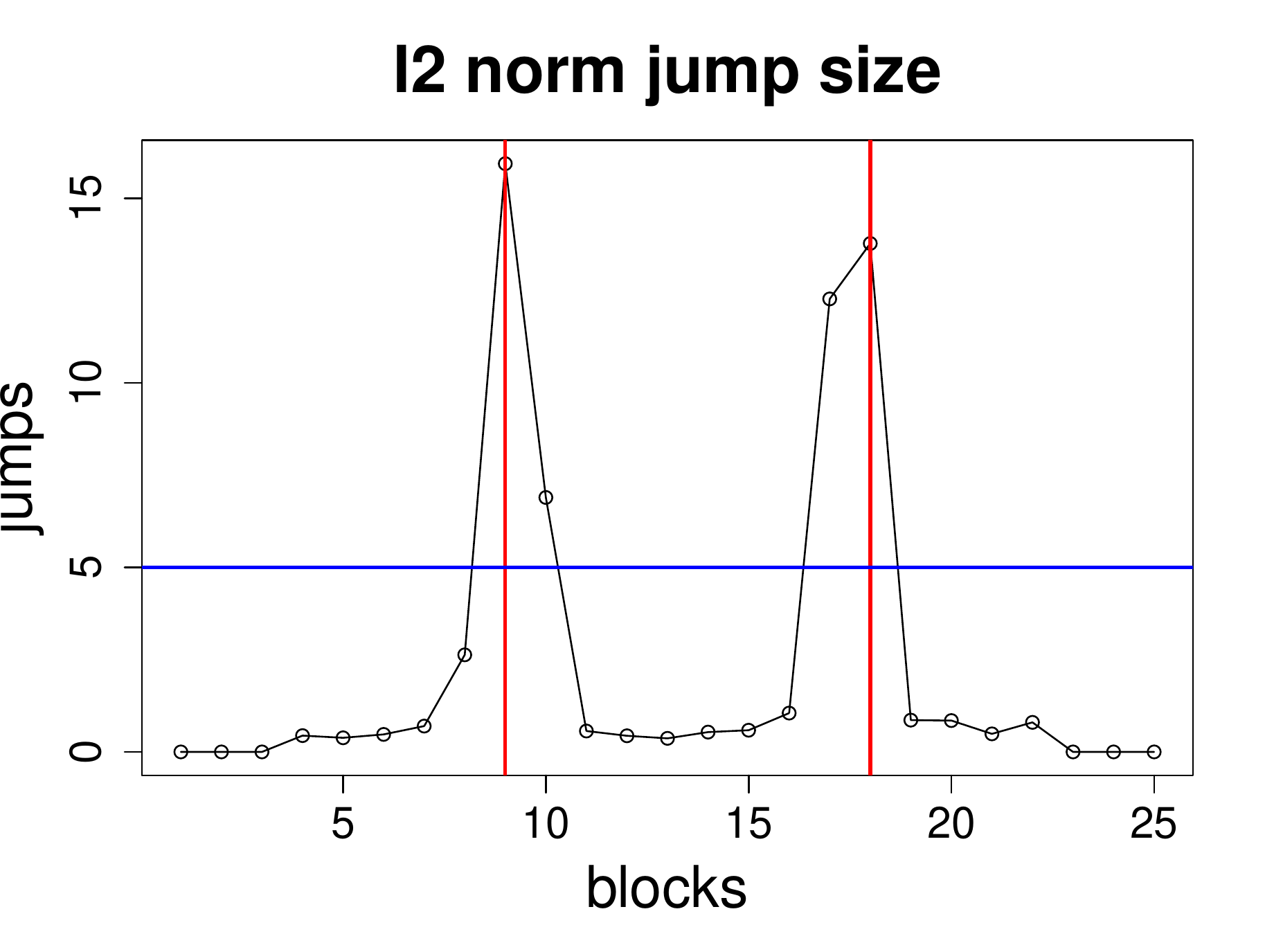}
        };
        
        \node[block, below of=step2-result, node distance=2.25cm](step3) {
        Step 3: Block clustering
        };
        
        \node[thin-cloud, right of=step3, node distance=6cm](block-cluster) {
        Use Gap statistics to determine the number of clusters in the candidate change points.
        };
        
        \node[block, below of=step3, node distance=3.5cm](step4) {
        Step 4: Exhaustive search for single break points detection in each cluster
        };
        
        \node[cloud, right of=step4, node distance=6cm](exhaustive) {
        In each cluster $C_i$, we establish a search interval $(l_i, u_i)$, and exhaustively detect a single change point by minimizing SSE proposed in (7) in the main context.
        };
        
        \node[smallblock, below of=step4, node distance=3cm](output) {
        The output is: detected break points $\widehat{t}_1, \cdots, \widehat{t}_{\widehat{m}}$.
        };
        
        \node[plots, below of=step2-plot, node distance=3.5cm](step3-plot) {
        \includegraphics[height=5.5em, width=8em]{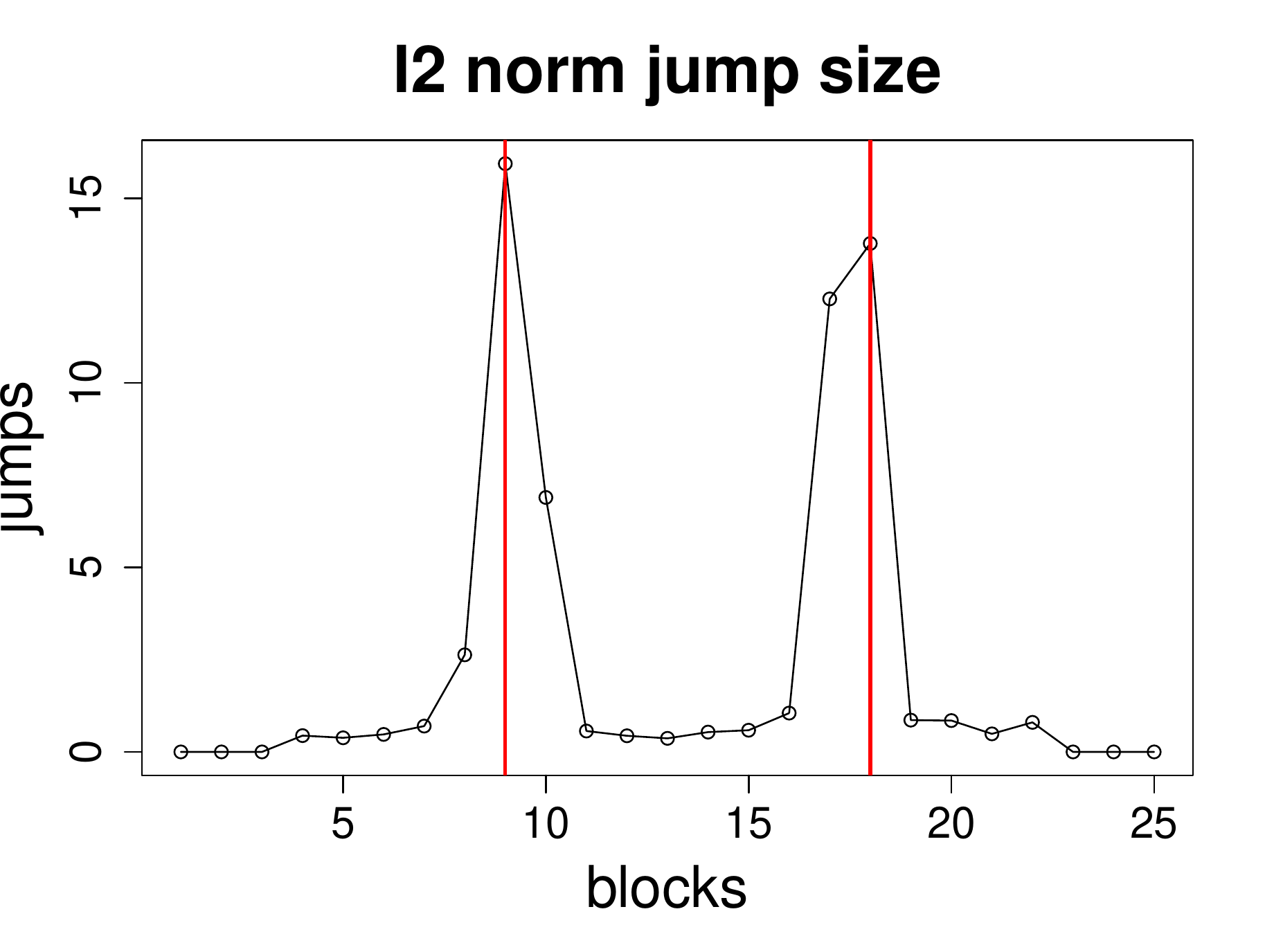}
        };
        
        \node[plots, below of=step3-plot, node distance=3.5cm](step4-plot) {
        \includegraphics[height=5.5em, width=8em]{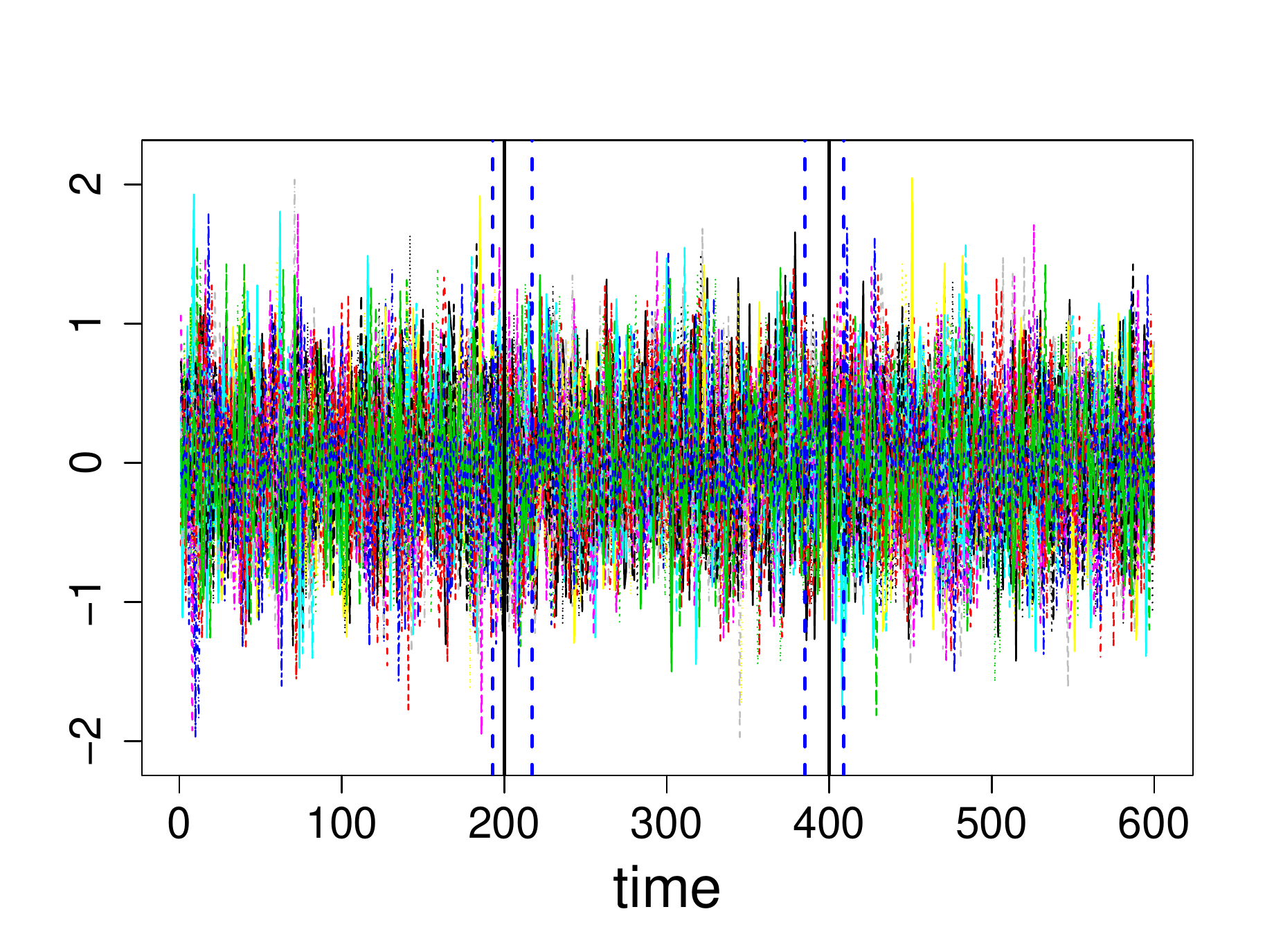}
        };
        
        \path [line] (input) -- (step1);
        \path [line] (step1) -- (step1-result);
        \path [line] (step1-result) -- (step2);
        \path [line] (step2) -- (step2-result);
        \path [line] (step2-result) -- (step3);
        \path [line] (step3) -- (step4);
        \path [line] (step4) -- (output);
        \path [line, dashed] (step1-steps) -- (step1);
        \path [line, dashed] (hard-threshold) -- (step2);
        \path [line, dashed] (block-cluster) -- (step3);
        \path [line, dashed] (exhaustive) -- (step4);
        \path [line, dashed] (step1-plot) -- (step1);
        \path [line, dashed] (step2-plot) -- (step2);
        \path [line, dashed] (step3-plot) -- (step3);
        \path [line, dashed] (step4-plot) -- (step4);
    \end{tikzpicture}}
    \caption{Algorithm flowchart for TBSS algorithm Step 1--4.}
    \label{fig:steps}
\end{figure}
\vspace{-.22cm}

\subsection{Computational Complexity of TBSS:} 
\label{comp_complexity}
The first four steps of TBSS correspond to estimating the number of true break points and their locations. To calculate the combined computational complexity of these steps, we assume that $m_0$ is finite. The effective sample size in Step 1 is $k_T$, which implies that the cost of solving the penalized regression in \eqref{eqn:objectivefunc} is $O \left( k_T p^2 q \right) $. 
In Step 2, the initialization step in which model parameters in all blocks are estimated, requires the computational cost $O\left( k_T p^2q \right)$, while the recursively $K$-means clustering step needs the computational cost of order $O \left(k_T \right)$ \cite{friedman2001elements}. Thus, the computational complexity of Step 2 is $O\left( k_T p^2 q + k_T \right)$. The computational cost of Step 3 is similar to Step 2, since we cluster all candidates into $\widetilde{m}$ subsets, and the fact that $\widetilde{m} \leq k_T$ indicates that the order is upper bounded by $O\left( k_T \right)$. The complexity of Step 4 is also similar to Step 2. According to the definition of intervals $(l_i, u_i)$ in \eqref{eqn:final}, the order for Step 4 is $O\left(b_T p^2 q\right)$. Therefore, the total computational complexity of TBSS is $O\left( (k_T+b_T)p^2q + k_T \right)$. Note that the number of blocks $k_T$ is in fact $T / b_T$ (rounded to the closest integer). Thus, the total computational complexity of TBSS can be written as a function of $b_T$ only, i.e., $O\left( \left(\frac{T}{b_T} + b_T\right)p^2q + \frac{T}{b_T} \right)$. Selecting the block size $b_T = O(\sqrt{T})$ yields the optimal computational cost $O\left(\sqrt{T}(p^2q + 1)\right)$.

\begin{figure}[!ht]
    \centering
    \includegraphics[scale=.42, trim={0 0.5cm 0 2cm}, clip]{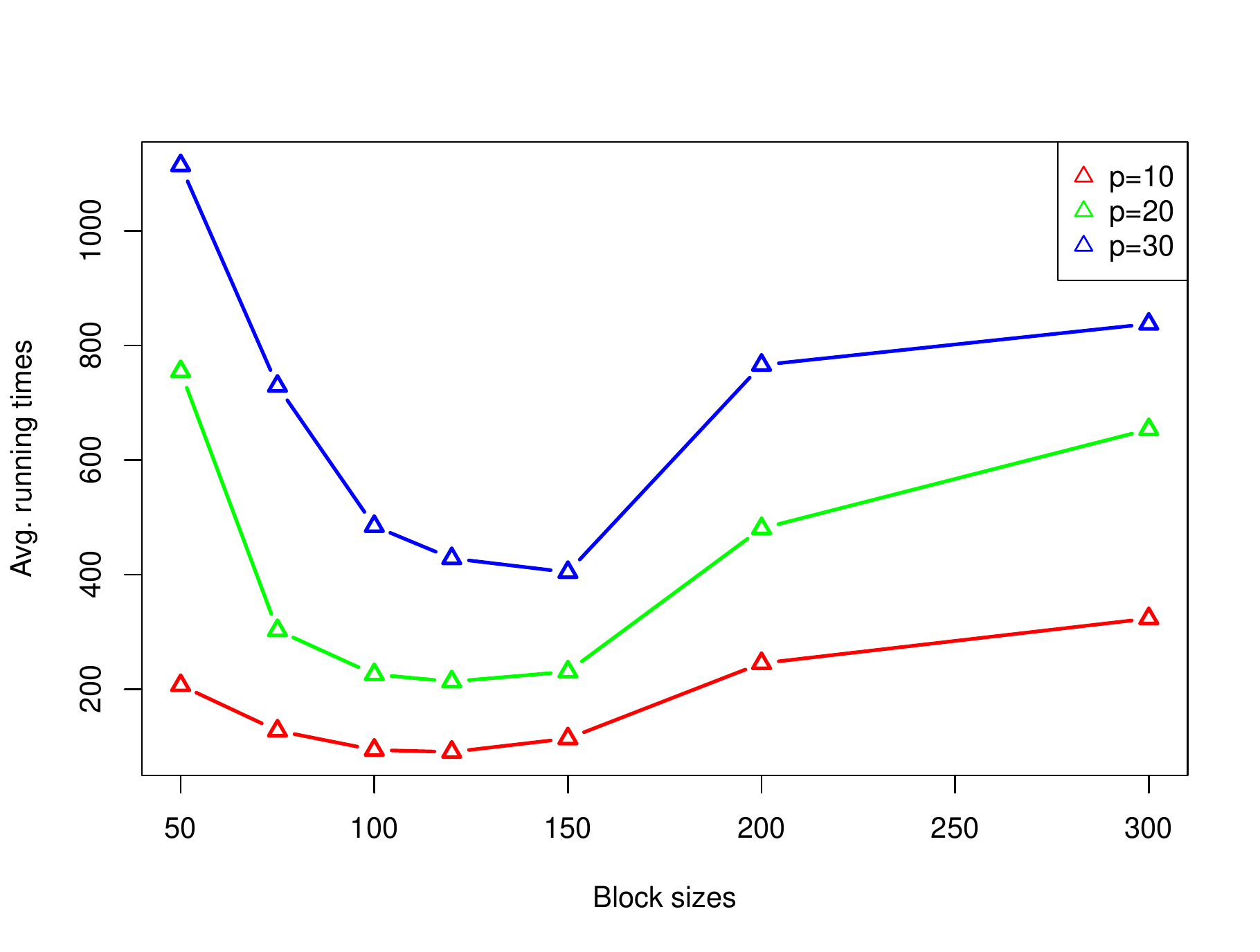}
    \caption{Averaged running time (in secs) for different block sizes $b_T$, for a lag 1 VAR model with $p=10, 20, 30$ time series, respectively (plotted in red, green, and blue), $T=20000$ time points and $m_0=4$ equally spaced break points.}
    \label{fig:running_times}
\end{figure}

\noindent
{\bf Illustration of Computational Complexity:} We consider the following setting for a sparse lag-1 VAR model, whose transition matrix has non-zero elements in the first off diagonal, analogous to the pattern explored in Scenario A in the simulation studies (see Section \ref{sec:sim_studies}, Figure \ref{fig:1}). The number of time series under consideration is $p=20$ with the number of time points $T=20000$. Further, there are four equally-spaced break points located at $t_1=4000$, $t_2=8000$, $t_3=12000$, and $t_4=16000$. We investigate the running time of the first four steps of TBSS for the following seven block sizes $b_T:$ 50, 75, 100, 120, 150, 200, and 300, respectively. The corresponding mean running times averaged over 50 replicates are depicted in Figure~\ref{fig:running_times}. It can be seen that initially the running time rapidly decreases as the block size increases from 50 to 120, but once it becomes larger than 150 it keeps increasing. Note that for a very small block size $b_T=5$, the average running time for $p=20$ exceeds 2.5 hours ($\approx$10,000 secs), while a method employing mean shift models and binary segmentation \cite{cho2015multiple} requires over 5 hours. Hence, the TBSS is suitable for exceedingly long neuroimage data.

These numerical results are consistent with the discussion above, wherein it was shown that the computational complexity of the detection phase (first four steps) of TBSS is $O \left( \left( \frac{T}{b_T} + b_T \right) p^2 q + \frac{T}{b_T} \right)$, ignoring any (unknown) constants involved in the calculations for Steps 1-4 that may also differ across Steps. For this setting, the empirical optimal step size is 120 (for $p=10,20$) or 150 (for $p=30$), which are close to $\lfloor \sqrt{T} \rfloor = 141$. Hence, it is recommended to set $b_T=\sqrt{T}$ in applications, provided that the underlying break points are not located too close to each other -see previous discussion on the interplay between the minimum spacing between consecutive break points $D_T$ and the block size $b_T$.


%
\vspace{-0.22cm}

\subsection{Theoretical Results for TBSS}
{
In this section, we establish theoretical guarantees for the proposed TBSS algorithm. The proofs of the following three Theorems are given in the Supplement Section I.
}

{
Theorem \ref{thm:1} states that Step 1 of TBSS detects break points that are close to the true ones under certain regularity conditions specified next. Following \cite{chan2014group} and \cite{safikhani2017joint}, we define the Hausdorff distance between two countable sets on the real line as:
\begin{equation*}
    d_H(A, B) = \max_{b \in B}\min_{a \in A}|b-a|.
\end{equation*}
\begin{itemize}
    \item[A1] For any fixed $j$, denote the covariance matrices $\Gamma_j(h) = \text{cov}\left(y_t^{(j)}, y_{t+h}^{(j)}\right)$ for $t,h \in \mathbb{Z}$. Assume that for $\kappa \in [-\pi, \pi]$, the spectral density matrices $f_j(\kappa) = (2\pi)^{-1}\sum_{l\in \mathbb{Z}}\Gamma_j(l)\exp(-i\kappa l)$ exist. Further, assume that the largest and smallest eigenvalues of the spectral density matrices are uniformly bounded (over all segments) from above and below, respectively. 
    \item[A2] The matrices $\Phi^{(\cdot, j)}$ are (weakly) sparse and there exists a positive constant $M>0$ such that
    \begin{equation*}
        \max_{1\leq j \leq m_0+1}\|\Phi^{(\cdot, j)}\|_\infty \leq M.
    \end{equation*}
    \item[A3] There exists a positive constant $v$ such that 
    \begin{equation*}
        \min_{1\leq j \leq m_0}\|\Phi^{(\cdot, j+1)} - \Phi^{(\cdot, j)}\|_2 \geq v > 0.
    \end{equation*}
    Moreover, there exists a vanishing positive sequence $\gamma_T$ such that, as $T \to +\infty$, 
    \begin{equation*}
        \begin{aligned}
            &\frac{D_T}{T\gamma_T} \to +\infty,\  d^\star_{\max}\sqrt{\frac{\log p}{T\gamma_T}} \to 0, \ b_T = o(D_T), \\
            &\limsup \frac{b_T}{T\gamma_T} \leq \frac{1}{2},\ \text{and}\ \frac{d_{\max}^\star\log p}{b_T} \to 0,
        \end{aligned}
    \end{equation*}
    where $d^\star_{\max} = \max_{1\leq j \leq m_0+1}d_j^\star$.
\end{itemize}
\begin{remark}\label{remark:1}
Assumption A1 ensures that the piecewise VAR models between break points are \textit{stable and hence stationary}. Assumption A2 is standard in the high-dimensional linear regression literature, and it specifies that the maximum entry in each (weakly) sparse transition matrix is bounded, which is needed for technical reasons. The first part of Assumption A3 requires that there is adequate cumulative ``signal", as captured by the norm differences between transition matrices across change points. This combined with the spacing $D_T$ between break points and the choice of the tuning parameter in the optimization problem \eqref{eqn:objectivefunc} ensures detectability of the break points. Assumption A3 can be regarded as an extension of Assumptions H2 and H3 in \cite{chan2014group}, and Assumptions A2 and A3 in \cite{harchaoui2010multiple}. Also, the proposed vanishing sequence $\{\gamma_T\}$ is directly related to the detection rate of the break points. 
\end{remark}
}

{
Denote the set of the relative locations of candidate break points ($\widehat{t}_j / T$) identified by Step 1 of TBBS, by $\widehat{\mathcal{A}}$, and the set of the relative locations of the true break points by $\mathcal{A}^\star$. Then, we can establish the following result.}

{
\begin{theorem}
\label{thm:1}
Suppose Assumptions A1-A3 hold, and choose tuning parameters $\lambda_{1,T} = \lambda_{2,T} = \mathcal{O}\left( \frac{\sqrt{b_T(2\log p + \log q)}}{T} \right)$. Then, as $T \to +\infty$, the following holds:
\begin{equation*}
    \mathbb{P}(d_H(\widehat{\mathcal{A}}_T, \mathcal{A}_T) \leq \frac{b_T}{T}) \to 1,
\end{equation*}
with $\gamma_T$ implicitly specified in Assumption A3. 
\end{theorem}
\begin{remark}\label{remark:2} 
Theorem \ref{thm:1} provides the rate of consistency for break point detection after Step 1 of TBSS. It shows that for each true break point $t_j$, there exists a candidate break point $\widehat{t}_j$ obtained by Step 1, whose relative location satisfies $|\frac{\widehat{t}_j - t_j}{T}| \leq \frac{b_T}{T}$. Further, based on the Assumption A3, we have $b_T/T \leq \gamma_T$, and the sequence $\gamma_T$ also depends on the minimum spacing $D_T$ as well as the model dimension $p$ (as seen from Assumption A3) and it can be chosen as $\gamma_T = (\log p\log T)/T$ or $\gamma_T = (\log\log T\log p)/T$. This implies that the convergence rate for estimating the relative locations of the break points, could be as small as $(\log\log T\log p)/T$. 
\end{remark}
}
{
Theorem \ref{thm:2} establishes that the locations of the final detected break points $\widetilde{t}_j$'s obtained by Step 4 of TBSS are consistently estimated. Let $\widetilde{\alpha}_j = \widetilde{t}_j / T$ and $\alpha_j = t_j / T$, and obtain:
\begin{theorem}
    \label{thm:2}
    Suppose Assumptions A1-A3 hold. Then, as $T \to +\infty$, there exists a large enough constant $K>0$, we have
    \begin{equation*}
        \mathbb{P}\left(\max_{1\leq j \leq m_0}|\widetilde{\alpha}_j - \alpha_j| \leq \frac{K d^\star_{\max}\log p}{T\min_j v_j^2}\right) \to 1,
    \end{equation*}
    where $v_j \overset{\text{def}}{=} \|\Phi^{(\cdot, j+1)} - \Phi^{(\cdot, j)}\|_2$.
\end{theorem}
\begin{remark}
Theorem \ref{thm:2} shows that the relative locations of the final set of detected break points converge to the true ones as the sample size increases. Also, it is possible to relax Assumption A3 by allowing the jump size to vanish as a function of the number of observations $T$ and the consistency rate depends on how fast the minimum jump size vanishes; in words, as long as there is adequate sample size $T$, TBSS can detect break points driven by small changes in the temporal dynamics of the VAR process.
\end{remark}
}

{
After solving optimization problem \eqref{eq:estimation_third} in Step 5, we can establish the following result on the quality of the estimated model parameters (Granger causal networks). 
Before that, we introduce some required assumptions to establish the theorem. For the given design matrix $\mathbf{Z}_s$, and error term $\mathbf{E}_s$, we have:
\begin{itemize}
    \item[A4] {\it Restricted strong convexity (RSC)}: There exist constants $\xi > 0$ and $\alpha_{RSC} > 0$ such that
    \begin{equation*}
        \frac{1}{2}\|\mathbf{Z}_s\Delta\|_F^2 \geq \frac{\xi}{2}\|\Delta\|_F^2 - \alpha_{RSC}\|\Delta\|_1^2,\quad \text{for all }\Delta \in \mathbb{R}^{\tilde{\pi}}.
    \end{equation*}
    \item[A5] {\it Deviation conditions}: There exists a deterministic function $\mathcal{G}({\Phi}, \Sigma)$ of the model parameters $\Phi$ and $\Sigma$ such that:
    \begin{equation*}
        \left\| \frac{\mathbf{Z}_s^\prime \mathbf{E}_s}{T} \right\|_\infty \leq \mathcal{G}(\Phi, \Sigma)\sqrt{\frac{2\log p + \log q}{T}}.
    \end{equation*}
\end{itemize}
\begin{theorem}
\label{thm:3}
Suppose Assumptions A4-A5 hold, and further choose tuning parameter $\rho_T = \mathcal{O}\left(\sqrt{\frac{\log p}{T}}\right)$; then, minimizer $\widehat{\mathbf{B}}$ of \eqref{eq:estimation_third} satisfies
\begin{equation*}
\begin{aligned}
    \|\widehat{\mathbf{B}} - \text{vec}(\Phi)\|_2 &= \mathcal{O}_p\left(\sqrt{\frac{d_{\max}^\star\log p}{T}}\right),
\end{aligned}
\end{equation*}
where $d^\star_{\max} = \max_{1 \leq j \leq m_0+1} d_j^\star$.
\end{theorem}
\begin{remark}\label{remark:3}
Theorem \ref{thm:3} provides an error bound for the estimation accuracy of the model parameters in each stationary segment obtained after detection of the break points $\widetilde{t}_j$. It can be seen that the bound vanishes as $T$ grows and therefore it guarantees accurate estimates of the Granger causal networks. The $\ell_2$ rate $\sqrt{d_{\max}^\star\log p/T}$ is of similar order as that obtained in penalized multivariate regression problems with independent and identically distributed samples - the mild impact of temporal dependence is captured in a constant term; see \cite{loh2012high, basu2015regularized}. 
\end{remark}
}

\section{Results}
\label{section:3}
We assess the performance of the proposed TBSS algorithm based on an extensive set of experiments on synthetic data and also illustrate it on neuroimaging data.
\footnote{All \texttt{R} scripts used in the simulation experiments and the real application example are available at \url{https://github.com/peiliangbai92/TMI_code}} 
We start with an extensive discussion of how to select the tuning parameters in the algorithm, followed by a presentation of the simulation settings considered and the results obtained. Finally, an application to EEG data is presented.

\vspace{-0.22cm}

\subsection{Tuning parameter selection} \label{sec:tuning_parameters}
The TBSS algorithm is governed by a few tuning parameters. In Section \ref{sec:method}, we provided a range for some of them based on some theoretical calculations. Next, we discuss in detail how these were selected in the simulation studies below and in the neuroimage application. \\
{\bf (1)} $\lambda_{1,T}$: This parameter is selected through cross-validation. In the simulation study, we randomly select $20\%$ of the blocks equally spaced with a random initial point. Denote the last time point (block ends) in these selected blocks by $\mathcal{T}$. Data without observations in $\mathcal{T}$ can then be used in the first step of our procedure to estimate $\Theta$ for a range of values for $\lambda_{1,T}$. The parameters estimated in Step 1 are used to predict the time series data at time points in $\mathcal{T}$. The value of $\lambda_{1,T}$ which minimizes the mean squared prediction error over $\mathcal{T}$ is the cross-validated choice of $\lambda_{1,T}$. The candidate sequence for $\lambda_{1,T}$ is selected as follows (see also \cite{friedman2010regularization}): pick a sequence of $K_1$ values for decreasing from $\lambda_{1,\max}$ to $\lambda_{1,\min}$ on the log-scale, where the maximum value $\lambda_{1,\max}$ is the smallest value that yields $\widehat{\theta}_i = 0$, for all $i=1,2,\cdots, k_T$, and the minimum value is set to be $\lambda_{1,\min}=\epsilon\lambda_{1,\max}$. We selected $\epsilon = 10^{-3}$ if the blocks size $b_T \leq 2p$ and $\epsilon = 10^{-4}$ otherwise. Finally, we selected $K_1 = 10$. \\
{\bf (2)} $\lambda_{2,T}$: This tuning parameter is also selected through cross-validation. Specifically, we set $\lambda_{2,T} = c\sqrt{{\log p}/{T}}$, where $c$ is selected by similar procedure as $\lambda_{1,T}$: pick a decreasing sequence of $K_2$ values from $c_{\max}$ to $c_{\min}$, we set $c_{\max}=0.1$ and $c_{\min} = \epsilon c_{\max}$, where $K_2 = 5$ and $\epsilon = 10^{-3}$. {We construct a 2-dimension grid search with size $K_1 \times K_2$ to simultaneously select tuning parameters $\lambda_{1,T}$ and $\lambda_{2,T}$ by cross-validation.}\\
{\bf (3)} $\rho_T$: This parameter controls the regularization term in optimization problem \eqref{eq:estimation_third}, and is selected by using cross-validation or BIC. In Algorithm \ref{algo:1} in the supplement, the $\ell_1$ regularized optimization problems in Steps 1 and 5 are solved by using the \texttt{glmnet} package in \texttt{R}. \\
{\bf (4)} ${b_T}$:  The default selection of block size is given by $\lfloor \sqrt{T} \rfloor$. In practice, $b_T$ can be selected by using BIC.  \\
{\bf (5)} ${R_T}$: As we mentioned in \ref{sec:methods}, we set $R_T$ equal to $b_T$ in the last step of TBSS. 
\vspace{-0.22cm}

\subsection{Simulation Studies} \label{sec:sim_studies}
Next, we evaluate the performance of TBSS with respect to its accuracy of detecting both the number and location of break points, as well as the quality of the Granger causal network estimates ($\Phi^{(\cdot,j)}$s). 

To evaluate the performance of the first 4 steps of TBSS, we consider the mean and the standard deviation of the estimated break point locations relative to the sample size, i.e. $\widetilde{t}_j/T$, and the percentage of simulation runs where the correct number of break points are correctly detected. A detected break point is counted as a \textit{success} for the $j$-th true break point, if it falls into the \textit{selection interval}: $[t_{j-1} + \frac{t_j - t_{j-1}}{5}, t_j + \frac{t_{j+1}-t_j}{5}]$. 

To measure the accuracy of the estimation of the sparse transition matrices ($\Phi^{(\cdot,j)}$s), we use sensitivity (SEN), specificity (SPC), {Matthews Correlation Coefficient (MCC)}, and relative error in Frobenius norm (RE) as the evaluation criteria, defined next: 
\begin{equation*}
\begin{aligned}
    &\text{SEN} = \frac{\text{TP}}{\text{TP} + \text{FN}},\  \text{SPC} = \frac{\text{TN}}{\text{FP} + \text{TN}},\ 
    \text{RE} = \frac{\|\text{Est.} - \text{Truth}\|_F}{\|\text{Truth}\|_F}, \\
    &\text{MCC} = \frac{\text{TP} \times \text{TN} - \text{FP}\times \text{FN}}{\sqrt{(\text{TP}+\text{FP})(\text{TP}+\text{FN})(\text{TN}+\text{FP})(\text{TN}+\text{FN})}},
\end{aligned}
\end{equation*}
{
wherein for any sparse matrix $A$ and its estimator $\widehat{A}$, denote by $A_{ij}$ its $(i,j)$-th element; then,
\begin{equation*}
    \begin{aligned}
         \text{TP} &= \#\left\{ \{\widehat{A}_{ij} \neq 0\} \cap \{A_{ij} \neq 0\}\right\}, \\
         \text{FP} &= \# \left\{ \{\widehat{A}_{ij} \neq 0\} \cap \{A_{ij} = 0\} \right\}, \\
         \text{FN} &= \#\left\{ \{\widehat{A}_{ij} = 0\}\cap \{A_{ij} \neq 0\} \right\}, \\
         \text{TN} &= \#\left\{ \{\widehat{A}_{ij} = 0\}\cap \{A_{ij} = 0\} \right\},
    \end{aligned}
\end{equation*}
and ``Est." stands for the \emph{estimated} transition matrix, while ``Truth" is the \emph{true} transition matrix. 
}

\subsubsection{Numerical Scenarios and Model Parameter Settings}
There are a number of factors influencing the performance of proposed algorithm, including the dimension of the model $p$, the number of observations $T$, the number of break points $m_0$, the spacing between them $D_T$, the lag of the auto-regressive model time series $q$, the sparsity level $d_{kj}$ of transition matrices, and the block size $b_T$ for selecting the candidate break points. 

\begin{figure}[!ht]
    \centering
    \includegraphics[scale=.28]{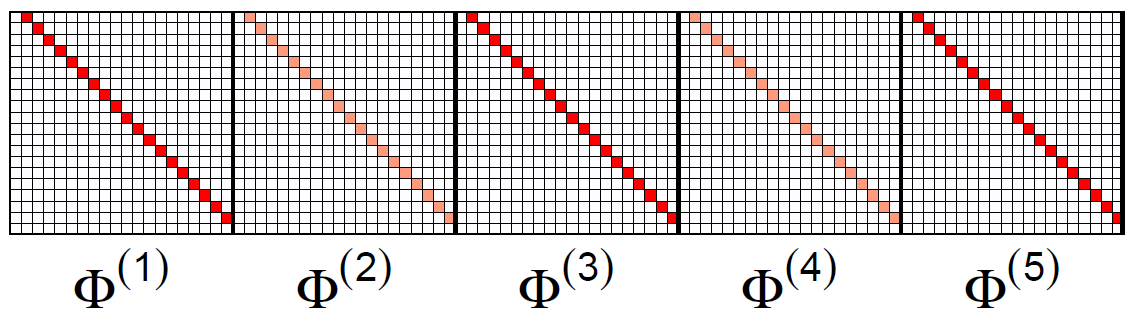}\hfill
    \includegraphics[scale=0.14]{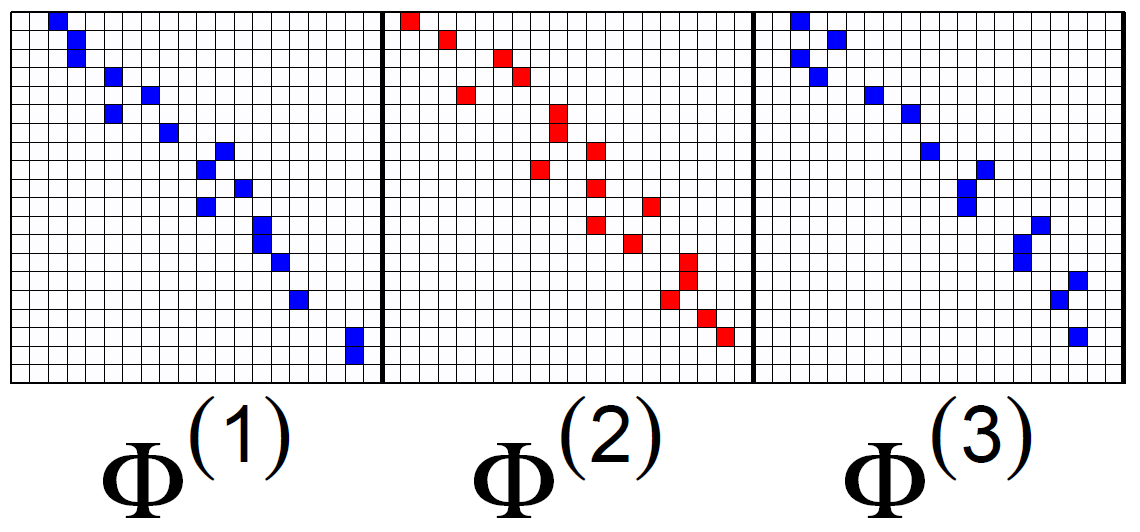}%
    \includegraphics[scale=0.14]{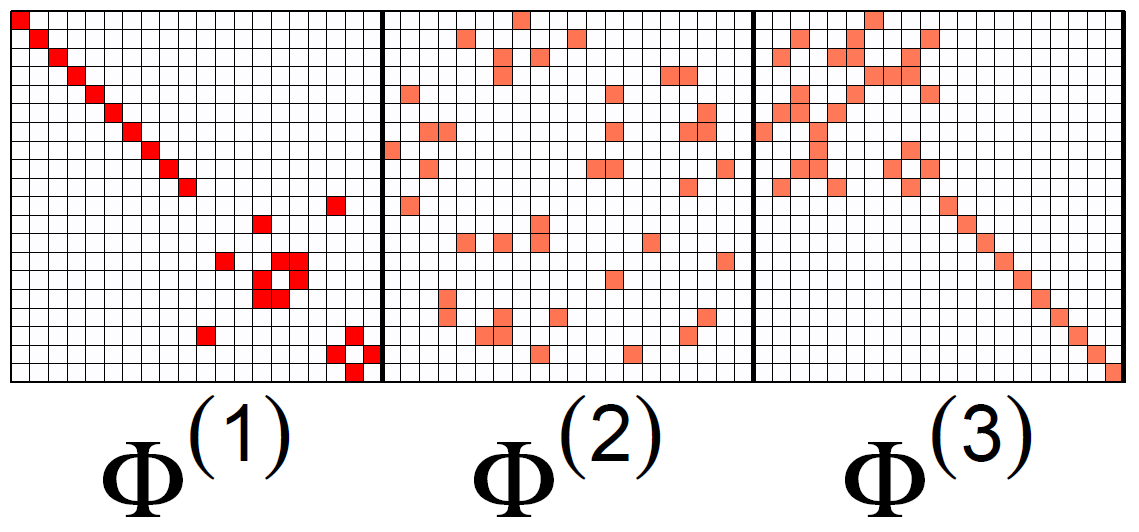}\hfill
    \includegraphics[scale=0.295]{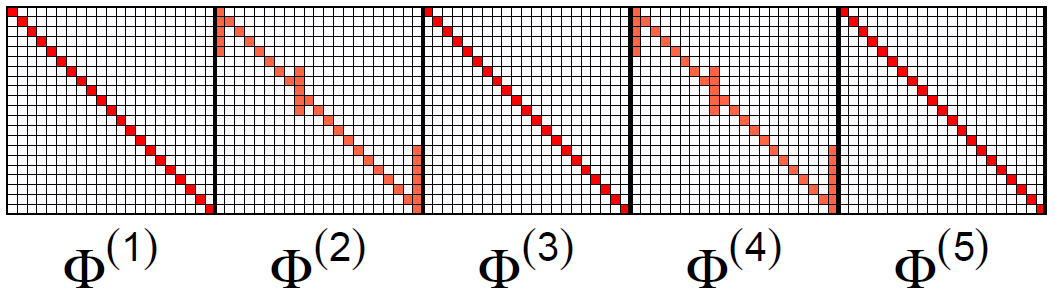}
    \vspace{-5pt}
    \caption{True transition matrices for different simulation scenarios. Top: 1-off diagonal matrices for five segments in scenarios~A; Middle Left: Random designed sparsity pattern for scenario E.1; Middle Right: Random designed sparsity pattern for scenario E.2; Bottom: Simulated brain connectivity structure for scenario F.}
    \label{fig:1}
\end{figure}

In this section, we first examine the impact of the block size of $b_T$ in detection accuracy and computational speed based on scenario A discussed below. For all subsequent scenarios B-F, the block size is fixed to $b_T=\lfloor\sqrt{T}\rfloor$. The experimental scenarios considered are described next:
\begin{itemize}
    \item[A.] In this scenario, we investigate the impact of different block sizes $b_T$. For other parameters, we set $p=20$, $T=6000$ with four break points: $t_1=1200, t_2 = 2400, t_3 = 3600$, and $t_4=4800$, the sparse transition matrices have all non-zero entries at 1-off diagonal, which is illustrated in the top panel of Figure~\ref{fig:1}.
    \item[B.] In this scenario, we examine the influence of the number of break points $m_0$, for the following case: $p=20$, $T=21000$, with 6 equally spacing break points: $t_i = 3000i$ for $i=1,2,\cdots,6$. The sparsity pattern is as in scenario A.
    \item[C.] {This scenario considers high-dimensional settings. Specifically, we set $p=40, 60, 200$ with $T=1000$ and a single break point at $t_1=500$, also, we set $p=100$, $T=1000$ with two change points at $t_1=333$, $t_2=666$. The sparsity pattern is as in scenario A.}
    \item[D.] In this scenario, we consider the effects of the time lag $q$ in the presence of few break points; specifically, we examine a VAR(2) model in case D.1 with $p=20$, $T=3000$ and two break points at $t_1=1000$ and $t_2=2000$, and we assume that only the lag 1 autoregressive coefficients exhibit changes while the lag 2 ones remain the same. In D.2, we consider the setting $p=20$ and $T=2000$ with only one break point $t_1=1000$ with the assumption that both the lag 1 and lag 2 coefficients exhibit a change. The sparsity pattern is as in scenario A.
    \item[E.] This scenario considers a random sparsity pattern in the transition matrices $\Phi^{(.,j)}$, illustrated in the middle panel of Figure~\ref{fig:1}.  The other model parameters are set to $p=20$ and $T=1000$ with two break points at $t_1=333$ and $t_2=667$.
    \item[F.] This scenario investigates the case that the transition matrices have similar patterns as real brain connectivity networks ({after pre-processing and de-trending}), which is presented in Figure~\ref{fig:1} (bottom panel). In this scenario, we set $p=21$, $T=6000$ with four break points: $t_1=1200$, $t_2=2400$, $t_3=3600$, and $t_4=4800$. 
    \item[G.] {The last scenario considers a \textit{nonlinear} VAR model with $p=10$, and $T=400$ with a single change point located at $t_1=200$, described in the sequel. Hence, the linear VAR model used for detection of the change point is \textit{misspecified}. }
\end{itemize}
Table \ref{tab:1} provides a summary of the specific settings and the exact location of the break points. We also provide some additional numerical experiments in the supplementary material. The tuning parameters are selected according to the guidelines in Section \ref{sec:tuning_parameters}. 
All numerical experiments are run in \texttt{R} version 3.6.3, on a PC equipped with 4 CPU cores and 16GB memory.
\begin{table}[!ht]
    \centering
    \caption{Parameters settings aiming to large-scale time series data break point detection.}
    \label{tab:1}
    \begin{tabular}{c|c|c|c|c|c}
        \hline\hline
         & $b_T$ & $p$ & $T$ & $t_j/T$ & $q$  \\
        \hline
        A.1 & 60 & \multirow{3}*{20} & \multirow{3}*{6000} & \multirow{3}*{(0.200, 0.400, 0.600, 0.800)} & \multirow{3}*{1} \\
        A.2 & 80 &  &  & & \\
        A.3 & 100 &  &  & & \\
        \hline
        B.1 & 144 & 20 & 21000 & $i/7$ for $i=1,2,\dots,6$ & 1 \\
        \hline
        C.1 & \multirow{3}*{31} & 40 & \multirow{3}*{1000} & \multirow{3}*{$0.500$} & \multirow{3}*{1} \\
        C.2 & & 60 & & & \\
        C.3 & & 200 & & & \\
        \hline
        C.4 & 31 & 100 & 1000 & $(0.333, 0.667)$ & 1 \\
        \hline
        D.1 & {55} & {20} & {3000} & {$(0.333, 0.667)$} & {2} \\
        \hline
        D.2 & 45 & 20 & 2000 & 0.500 & 2 \\
        \hline
        E.1 & \multirow{2}*{31} & \multirow{2}*{20} & \multirow{2}*{1000} & \multirow{2}*{$(0.333, 0.667)$} & \multirow{2}*{1} \\
        E.2 & & & & & \\
        \hline
        F.1 & 80 & 21 & 6000 & $(0.200, 0.400, 0.600, 0.800)$ & 1 \\
        \hline
        G.1 & 20 & 10 & 400 & 0.500 & 1 \\
        \hline\hline
    \end{tabular}
\end{table}

Next, we describe the data generation mechanism. In all scenarios, {the transition matrices $\Phi^{(\cdot, j)}$ is of size $p\times pq$}, and the error terms are independent and identically distributed from a Gaussian distribution with zero mean and variance 0.1, i.e. $\epsilon_t \overset{iid}{\sim} \mathcal{N}(0, 0.1\mathbf{I}_p)$. In order to ensure the stationarity of the VAR processes in each segment, the spectral radius of the transition matrices $\Phi^{(.,j)}$ is equal to 0.8. In scenarios A, B, and C, the values of the non-zero elements of the transition matrices are set to $-0.8$ and $0.8$ in an alternating manner. In scenario D, we consider a higher time lag $q=2$, hence, the non-zero elements are set as follows: (a) in scenario D.1, we only allow the changes in the first lag, specifically, the values of lag 1 and lag 2 for all three segments equal to $(0.6, 0.3)$, $(-0.6, 0.3)$ and $(0.6, 0.3)$, respectively; (b) in scenario D.2, we set the magnitudes of lag 1 and lag 2 for all segments changes: $(0.5, 0.35)$ and $(-0.6, -0.3)$, respectively. For scenario E, we consider two different random sparsity patterns shown in the bottom panel of Figure~\ref{fig:1} with $p=20$, $T=1000$ and two change points. In the scenario F, we construct the transition matrices similar to the real patterns in the {effective connectivity} networks of brain (here we estimate the connectivity networks of brain based on the real EEG data provided in Section~\ref{sec:application}), and generate VAR(1) time series data of $p=21$, $T=6000$ with four equally-spaced change points, which is similar to the real EEG data set in the next section. {
Finally in scenario G, we examine the performance of TBSS for data generated by the following nonlinear VAR model. Analogously to the simulation settings in \cite{terasvirta2014specification}, we consider a $p$-dimensional vector time series $\{X_t\}$ for $t=1,2,\cdots, T$ with $m_0$ change points located at $1=t_0 < t_1 < \cdots < t_{m_0} < t_{m_0+1} = T$, the error term $\epsilon_t$ is independent from $X_t$, where $\epsilon_t \sim \mathcal{N}(0, 0.01\mathbf{I})$:
\begin{equation*}
    X_t = f_j(X_{t-1}) + \epsilon_t,\quad t_{j-1} \leq t < t_j,
\end{equation*}
where $f_j(X_{t-1}) = \left[ f_{j,1}(X_{t-1}), \cdots, f_{j,p}(X_{t-1}) \right]^\prime \in \mathbb{R}^p$, and $f_{j,l}(\cdot)$ is the $l$-th coordinate. We set $p=10$, $T=400$ with a single change point $t^\star=200$, and the nonlinear function $f(\cdot)$ is given by
\begin{equation*}
    f_{j,l}(X_{t-1}) = (\alpha_j + \beta_j\exp(-\gamma_j X_{t-1, l+1}^2))X_{t-1, l+1},\  (l<p)\  \text{and}\ 
    f_{j,p}(X_{t-1}) = 0,
\end{equation*}
where $(\alpha_j, \beta_j, \gamma_j)$ for $j=1,2$ are the main model parameters. We set $(\alpha_1, \beta_1, \gamma_1) = (-1, -1.4, 1)$ and $(\alpha_2, \beta_2, \gamma_2) = (-1, -1.8, 0.01)$, respectively, to control the signal-to-noise ratio which is around 3. Note that in this setting the change in the temporal dynamics is driven by the nonlinear component in the model.
}
\subsubsection{Simulation Results}
Table~\ref{tab:simulation_result} presents the performance of the TBSS algorithm for Scenarios A-F. Overall, the detection performance of TBSS is satisfactory with having more than $80\%$ selection rate in all simulation scenarios while in around half of them the selection rate is $100\%$. In scenarios A and B, the detection and estimation performance seem to be robust with respect to changes in the block size. Most of the selection rates are above $90\%$ in this scenario while the selection rate for the first change point in scenario~A.3 drops to around $80\%$. This may be due to the fact that the block size in simulation A.3 is large compared to scenarios~A.1 and A.2. Further, the estimated location of break points seem to align with the location of true break points in all scenarios as can be seen in the ``mean" and ``sd" columns of Table~\ref{tab:simulation_result}. {The \textit{mean} corresponds to the average of the relative locations of the estimated break points (i.e. $\widetilde{t}_j/T$) over all replicates while the \textit{sd} is simply the standard deviation of the relative location of the estimated change points. Note that TBSS yields perfect (100\%) SEN, almost perfect ($\sim 100\%$) SPC and good ($\sim 94\%$) MCC and fairly small relative error for the estimated transition matrices $\Phi^{(.,j)}$ across all scenarios considered. Of interest, is the performance in simulation scenario D that considers a lag-2 VAR model. We obtain $90\%$ SEN, almost $100\%$ SPC, and almost $90\%$ MCC}, while the detection rates ($100\%$) are still very satisfactory. {It is worth noting that in scenario~C, the computation time is high compared to other scenarios  with similar sample size $T$. Note that the total computational complexity is $\mathcal{O}(\sqrt{T}(p^2q))$ and thus primarily driven by the larger value of $p$ (40, 60, 100 and 200).} Next, TBSS performs very well in simulation scenarios E and F, wherein the sparsity pattern of transition matrices are selected randomly and mimics the brain connectivity patterns. This shows the robustness of the proposed algorithm with respect to changes in the sparsity pattern. {Finally, the performance of scenario G.1 (a nonlinear VAR) is very satisfactory, despite the model used for detection purposes being misspecified. Specifically, the detection rate (over 100 replicates) is 88\% and the mean estimate for the relative location of the single change is 0.465 (true change point located at 0.5)}
\begin{table}[!ht]
    \centering
    \caption{Simulation results for scenarios A-F; break point selection rate and model parameter estimation.}
    \label{tab:simulation_result}
    \resizebox{.65\textwidth}{!}{%
    \begin{tabular}{c|c|c|c|c|c|c|c|c|c}
        \hline\hline
           & CP & mean & sd & rate & time & SEN & SPC & MCC & RE \\
        \hline
       \multirow{4}*{A.1} & 1 & 0.200 & 0.000 & 0.92 & \multirow{4}*{21.61} & \multirow{4}*{1.00} & \multirow{4}*{0.99} & \multirow{4}{*}{0.94} & \multirow{4}*{0.048} \\
                          & 2 & 0.400 & 0.000 & 0.92 & & & & \\
                          & 3 & 0.601 & 0.002 & 1.00 & & & & \\
                          & 4 & 0.800 & 0.001 & 1.00 & & & & \\
       \hline
       \multirow{4}*{A.2} & 1 & 0.200 & 0.000 & 1.00 & \multirow{4}*{15.08} & \multirow{4}*{1.00} & \multirow{4}*{0.99} & \multirow{4}{*}{0.93} & \multirow{4}*{0.048} \\
                          & 2 & 0.400 & 0.000 & 1.00 & & & & \\
                          & 3 & 0.600 & 0.000 & 1.00 & & & & \\
                          & 4 & 0.800 & 0.000 & 1.00 & & & & \\
       \hline
       \multirow{4}*{A.3} & 1 & 0.200 & 0.000 & 0.80 & \multirow{4}*{9.16} & \multirow{4}*{1.00} & \multirow{4}*{0.99} & \multirow{4}{*}{0.94} & \multirow{4}*{0.047} \\
                          & 2 & 0.400 & 0.000 & 1.00 & & & & \\
                          & 3 & 0.600 & 0.003 & 1.00 & & & & \\
                          & 4 & 0.800 & 0.000 & 1.00 & & & & \\
       \hline
       \multirow{6}*{B.1} & 1 & 0.143 & 0.003 & 0.92 & \multirow{6}*{72.71} & \multirow{6}*{1.00} & \multirow{6}*{1.00} & \multirow{6}{*}{0.99} & \multirow{6}*{0.029} \\
                          & 2 & 0.286 & 0.000 & 0.84 & & & & \\
                          & 3 & 0.429 & 0.000 & 0.84 & & & & \\
                          & 4 & 0.571 & 0.000 & 0.84 & & & & \\
                          & 5 & 0.714 & 0.000 & 0.84 & & & & \\
                          & 6 & 0.857 & 0.001 & 0.92 & & & & \\
       \hline
       \multirow{1}*{C.1} & 1 & 0.500 & 0.000 & 1.00 & 12.75 & \multirow{1}*{1.00} & \multirow{1}*{0.99} & 0.99 & \multirow{1}*{0.075} \\                                        
       \hline
       \multirow{1}*{C.2} & 1 & 0.500 & 0.001 & 1.00 & 34.48 & \multirow{1}*{1.00} & \multirow{1}*{1.00 } & 0.99 & \multirow{1}*{0.079} \\
       \hline
       C.3 & 1 & 0.527 & 0.012 & 1.00 & 630.18 & 1.00 & 0.99 & 0.97 & 0.11 \\
       \hline
       \multirow{2}*{C.4} & 1 & 0.326 & 0.026 & 1.00 & \multirow{2}*{145.30} & \multirow{2}*{1.00} & \multirow{2}*{1.00} & \multirow{2}*{0.98} & \multirow{2}*{0.13} \\
       & 2 & 0.667 & 0.022 & 1.00 & & & & & \\
       \hline
       \multirow{2}*{D.1} & 1 & 0.331 & 0.005 & 1.00 & \multirow{2}*{13.67} & \multirow{2}*{0.87} & \multirow{2}*{0.99} & \multirow{2}*{0.86} & \multirow{2}*{0.42} \\
                          & 2 & 0.666 & 0.001 & 1.00 & & & & \\
       \hline
       {D.2} & 1 & 0.500 & 0.000 & 1.00 & 22.14 & {0.94} & {0.99} & 0.96 & {0.21} \\
       \hline
       \multirow{2}*{E.1} & 1 & 0.333 & 0.000 & 1.00 & \multirow{2}*{4.23} & \multirow{2}*{1.00} & \multirow{2}*{0.98} & \multirow{2}{*}{0.98} & \multirow{2}*{0.10} \\
                          & 2 & 0.666 & 0.001 & 1.00 & & & & \\
       \hline
       \multirow{2}*{E.2} & 1 & 0.333 & 0.001 & 0.92 & \multirow{2}*{4.59} & \multirow{2}*{0.95} & \multirow{2}*{0.98} & \multirow{2}{*}{0.94} & \multirow{2}*{0.28} \\
                          & 2 & 0.664 & 0.006 & 0.96 & & & & \\
       \hline
       \multirow{4}*{F.1} & 1 & 0.200 & 0.000 & 0.82 & \multirow{4}*{14.19} & \multirow{4}*{1.00} & \multirow{4}*{0.97} & \multirow{4}{*}{0.82} & \multirow{4}*{0.16} \\
                          & 2 & 0.400 & 0.000 & 1.00 & & & & \\
                          & 3 & 0.600 & 0.000 & 1.00 & & & & \\
                          & 4 & 0.800 & 0.000 & 1.00 & & & & \\
       \hline
       G.1 & 1 & 0.465 & 0.042 & 0.88 & 2.36 & - & - & - & - \\
       \hline\hline
    \end{tabular}}
\end{table}

\subsection{Application to EEG Data for a Visual Task}\label{sec:application}

We use the EEG data\footnote{The raw data can be downloaded from: \url{https://dataverse.tdl.org/dataverse/rsed2017}.} analyzed in \cite{trujillo2017effect}. In this database, EEG signals from active electrodes for 72 channels are recorded at a sampling frequency of 256Hz, for a total of 480 seconds ($T=122880$). The stimulus procedure tested on 22 subjects comprised of eight 1-min duration interleaved sessions with eyes open and closed. The preceding resting state data were removed and not considered in the analysis. The subjects were undergraduates at Texas State University (11 female, 11 male, mean age = 21.1, age range = 18—26) and participated in this study for course credit or monetary payment. {Note that due to a technical recording error, one subject only had 240 seconds of recording time \cite{trujillo2017effect}, and hence was excluded from the ensuing analysis.}

Note that there has been work in the literature on related experimental setups, wherein the subjects were asked to keep their eyes open or closed \cite{agcaoglu2019resting, weng2020open, wang2015investigating, liu2020dynamic}). Some studies recruited adolescents (e.g. \cite{agcaoglu2019resting,weng2020open}), while others college students (e.g. \cite{wang2015investigating}). Further, the design of the experiment differed, since the subjects did a session with eyes open and a session with eyes closed, as opposed to switching between them in a single session. Nevertheless, the focus of these studies was to identify both differences in {effective connectivity} brain networks in these two states, and also track fluctuations in them during the recorded sessions. Further, preprocessing issues were addressed, including discarding initial time points, motion correction and data smoothing.

On the other hand, the objective of our analysis is listed as follows:
\subsubsection{Data Pre-processing}
{
The analysis is based on $T=65,000$ observations located in the middle of the total recorded period of 254 seconds. According to the experimental setup described in \cite{trujillo2017effect}, there are three applications of the stimulus (break points) during the selected time period, and the switches between the stimulus states (closed vs open eyes) occur at: $t_1 = 16280$, $t_2 = 32600$, and $t_3 = 48920$, respectively. These are considered as the locations of the \textit{true} break points.
}

\begin{figure}[!ht]
    \centering
    \includegraphics[scale=.32, trim={0 0 0 1cm}, clip]{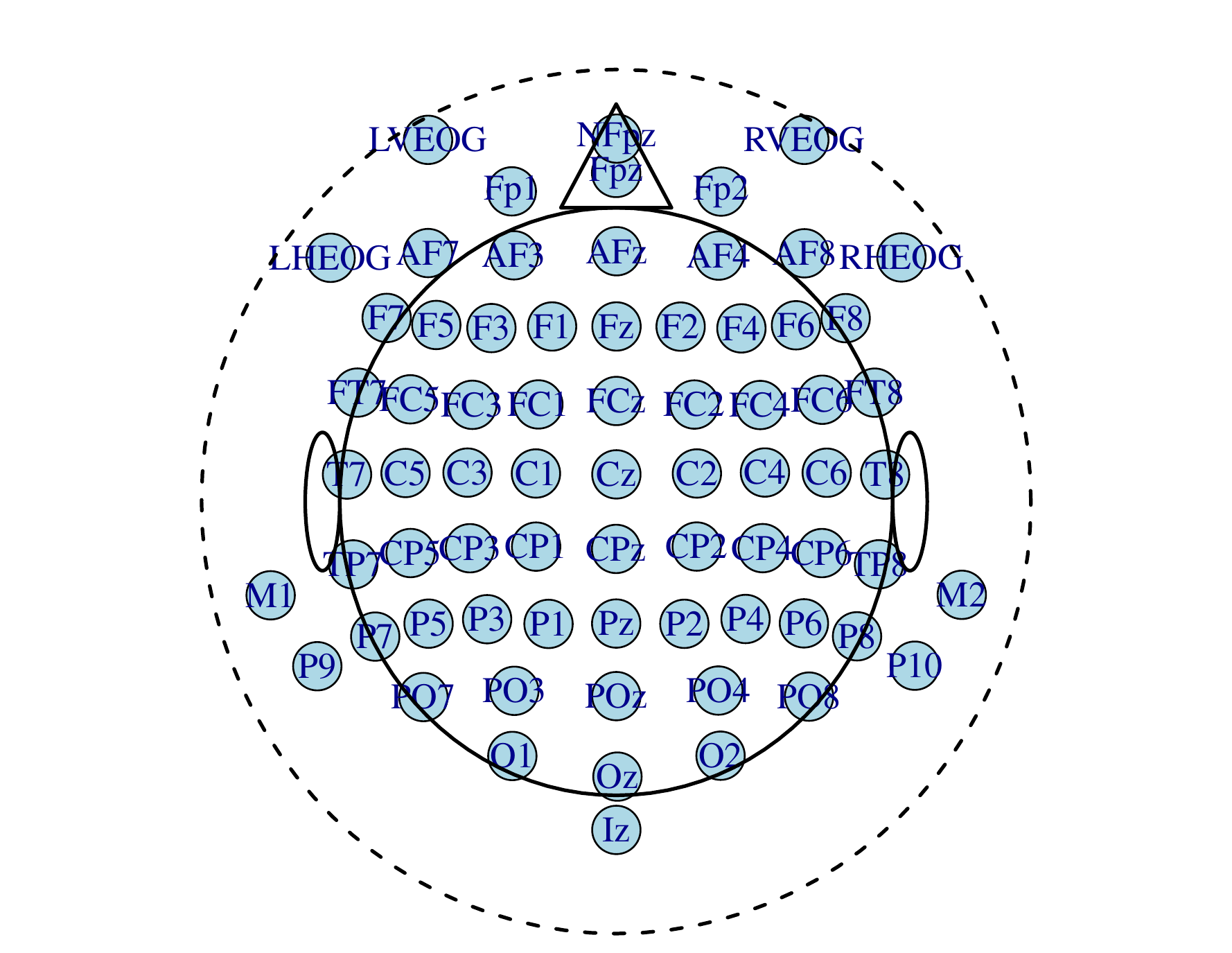}
    \caption{The selected 71 EEG channels used in the detection of break points, together with their names and locations.}
    \label{fig:eegcap}
\end{figure}

{
The raw EEG required detrending and filtering as shown and explained in the Supplement Section III A. We used the robust detrending method \cite{de2018robust} to remove trend patterns, and obtain the data used in the analysis.
We also filtered the raw detrended EEG data to extract the alpha-band (8-13Hz) signals (using the \texttt{eegkit} \texttt{R} package), also used in subsequent analysis. The alpha band is associated with visual tasks (\cite{chen2008eeg,klimesch1999eeg,tan2013difference}). The analysis is based on 71 channels (the NAS channel  was excluded due to an erroneous recording of its signal), whose locations on the scalp are depicted in Figure~\ref{fig:eegcap}. 
}

\subsubsection{Examining the Suitability of a linear VAR Model}
\begin{figure}[!ht]
    \centering
    \includegraphics[scale=.32, trim={0 0 0 1.2cm}, clip]{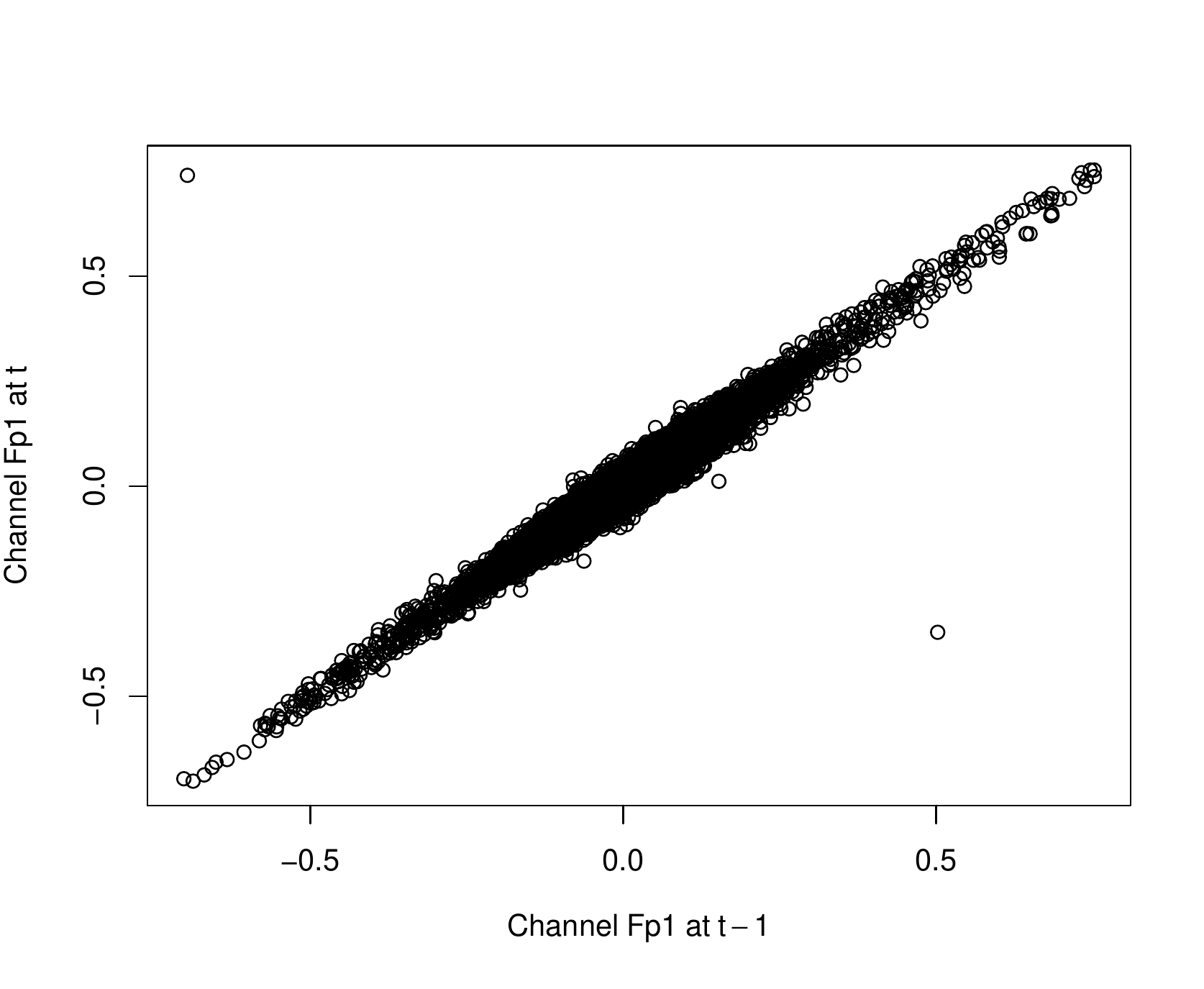}%
    \includegraphics[scale=.32, trim={0 0 0 1.2cm}, clip]{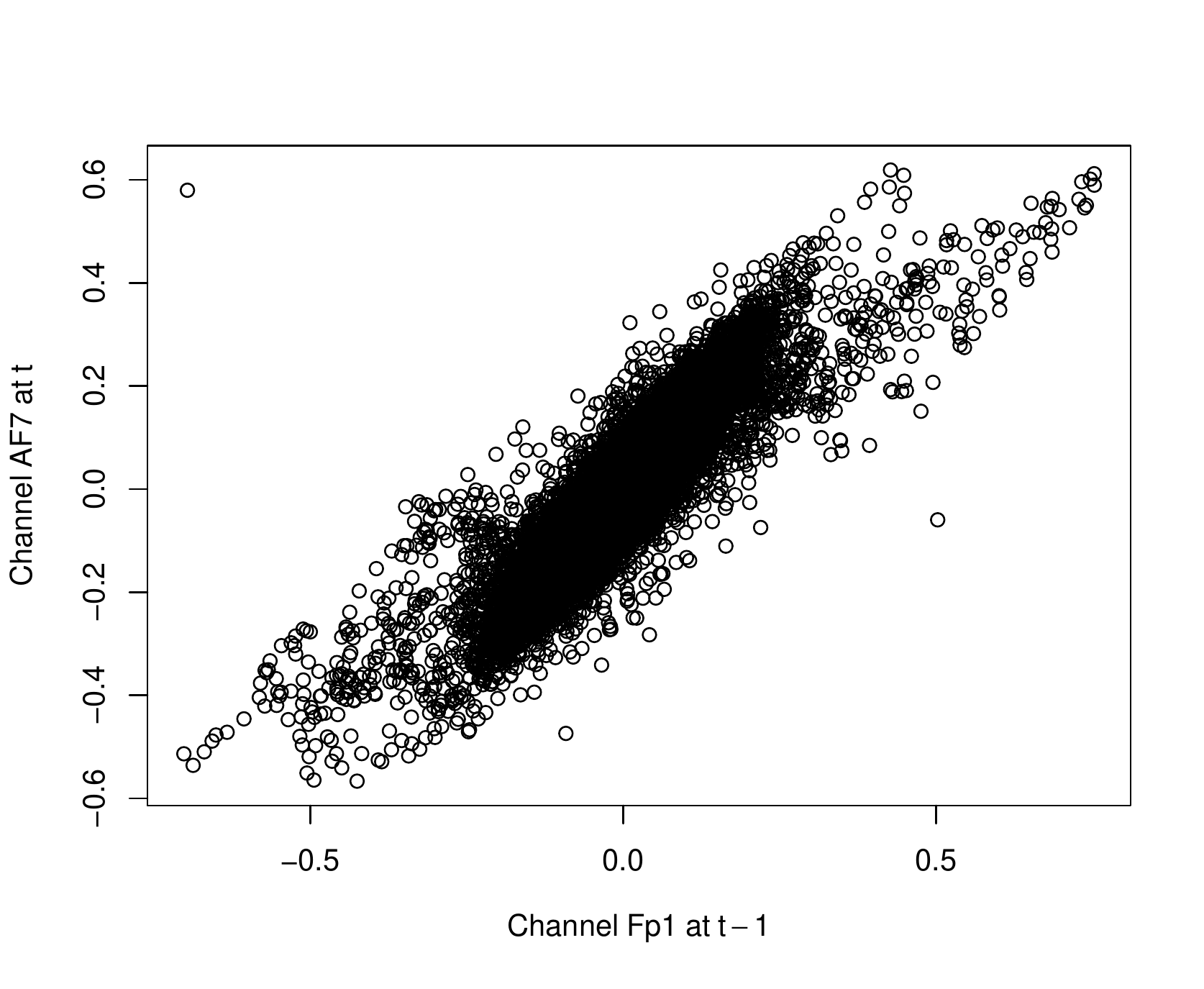}
    \caption{Scatter plots for verifying the sufficiency of VAR model. Left panel: Channel Fp1 vs Fp1; Right panel: Channel Fp1 vs AF7.}
    \label{fig:verify_var}
\end{figure}

In order to verify that a VAR(1) model is adequate for the EEG data under consideration, we examined the following scatter plots. (i) For the $i$-th channel at time point $t$: $y_i(t)$ versus the $i$-th channel at time point $t-1$: $y_i(t-1)$; (ii) for the $i$-th channel at time $t$: $y_i(t)$  versus the $j$-th channel ($j \neq i$) at time $t-1$: $y_j(t-1)$. In Figure~\ref{fig:verify_var}, we provide the scatter plots for channel Fp1 versus its 1-lagged data, and AF7 versus 1-lagged Fp1 data for illustration purposes. Scatter plots for other channels and other lags also exhibit similar linear patterns that support the use of a linear VAR model for identifying the break points, when switching from the open to the closed eyes state.

{
To select the optimal time lag for the VAR model, we employ a Bayesian Information Criterion (BIC) and the details of the procedure are provided in the supplement, Section III B(2). For this EEG data set, the time lag selected by BIC is 1 and hence we proceed with VAR(1) model. 
}

\subsubsection{Results}\label{sec:application-results}
To estimate break points for the EEG data of each of the 21 subjects, the block size used in the TBSS algorithm is set to $b_T \in \left[ \lfloor \sqrt{T} \rfloor, \lfloor 2.5\sqrt{T} \rfloor \right]$ depending on the subject under consideration, while the other tuning parameters were set according to the guidelines presented in Section \ref{sec:tuning_parameters}. Further, a VAR model with a single lag exhibited better fit in terms of mean squared error than a model with  more time lags for the majority of the subjects, and thus it was fitted to all subjects. 

\begin{figure*}[!ht]
    \centering
    \includegraphics[scale=.32]{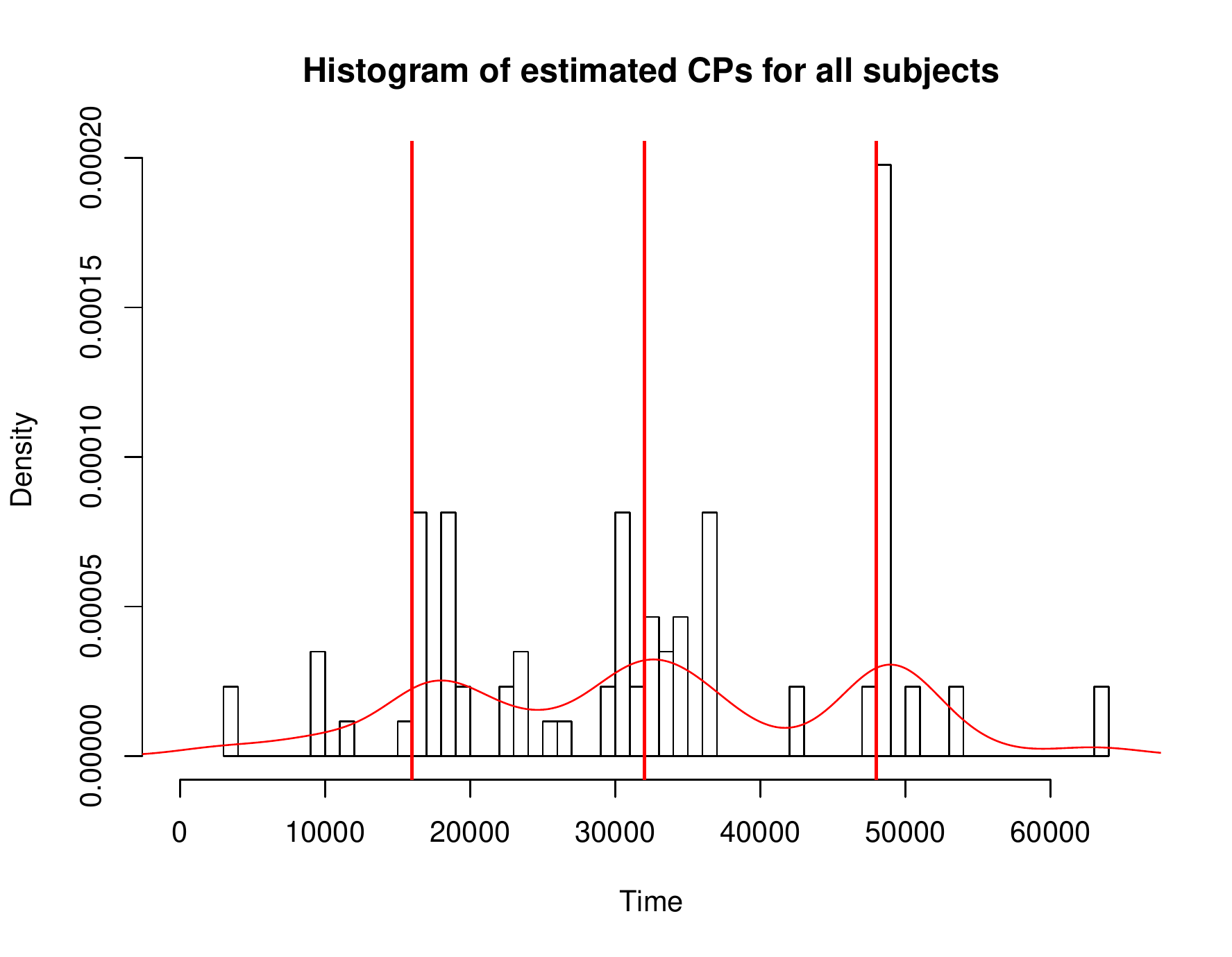}%
    \includegraphics[scale=.32]{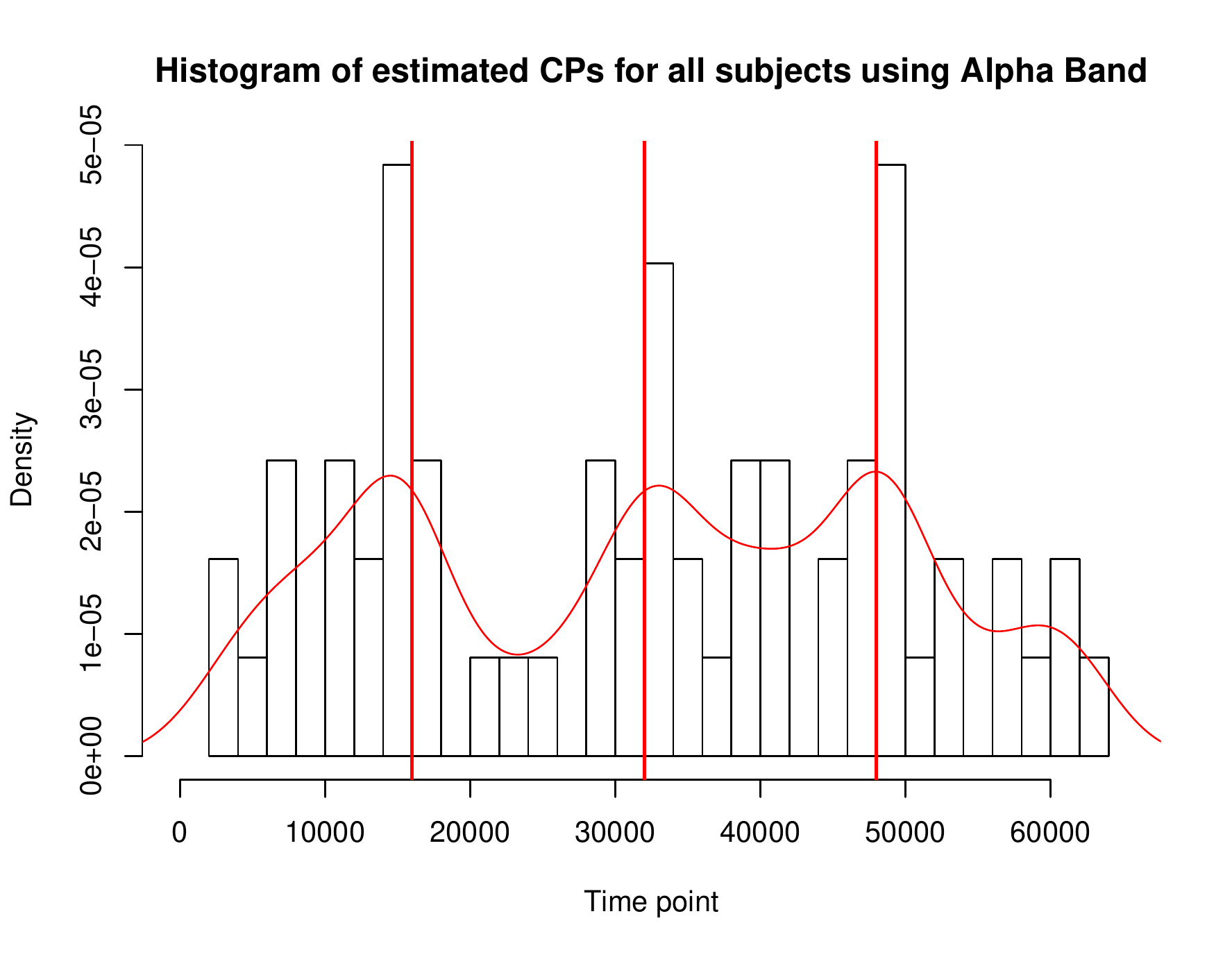}
    \caption{Histogram of estimated break points for all 21 subjects; the red curve depicts their estimated density function and the red vertical lines represent the location of the time points when the stimulus switched. Left panel: robust detrended data; Right panel: alpha-band data.
    }
    \label{fig:hist}
\end{figure*}

Figure~\ref{fig:hist} presents the histogram of \textit{all} estimated break points across \textit{all} 21 subjects using the detrended data (left panel) and alpha-band data (right panel), respectively. The red curve depicts the estimated density of the histogram. It can be clearly seen that the majority of the break points are located at time points close to 16000, 32000, and 48000, in accordance with changes in the stimulus. Specifically, the estimated change points by using the robust detrended data are located at: $\widehat{t}_1 = 15709$, $\widehat{t}_2 = 31555$, and $\widehat{t}_3 = 48499$, while the estimated change points based on the alpha-band data are located at: $\widehat{t}_1^{\alpha} = 17257$, $\widehat{t}_2^\alpha = 32343$, and $\widehat{t}_3^\alpha = 48673$. For both sets of data, TBSS provides very satisfactory break point estimates across subjects.
Hence, four stationary segments are obtained for the majority of the subjects and the corresponding Granger causal networks are estimated using Steps 5 and 6 (stability selection). For the latter step, $N_s=50$ subsamples are used together with a threshold $\tau$ of 0.8 to control the false positive rate in the estimated Granger causal networks.

A summary of the network characteristics (number of edges in the open and closed eyes states and the five most connected channels) is provided in Table \ref{tab:network_results}. The networks are depicted in Figure \ref{fig:openclose_network}. Analogous results for the beta-band data are shown in Section III C of the Supplement. 
\begin{table}[!ht]
    \centering
    \caption{Results for the estimated networks for eye-open/eye-closed states for all 21 subjects. }
    \label{tab:network_results}
    \resizebox{.85\textwidth}{!}{
    \begin{tabular}{c|c|c|c}
    \hline\hline
      Data & State & No. edges & Five most connected channels (degrees) \\
    \hline
    \multirow{2}{*}{Robust} & Open & 425 & PO3(70), P1(62), POz(62), P3(29), P2(29) \\
     & Closed & 439 & PO3(71), P1(67), P2(54), POz(43), P3(32) \\
    \hline
    \multirow{2}*{Alpha-band} & Open & 224 & PO3(19), PO8(13), Iz(12), Pz(12), P3(11) \\
     & Closed & 257 & Oz(32), PO3(20), P3(15), PO7(14), Iz(12) \\
    \hline\hline           
    \end{tabular}}
\end{table}

The results in Table~\ref{tab:network_results} show that the networks for the closed eyes segments (1 and 3) are denser than their open eyes counterparts (2 and 4), also shown in Figure~\ref{fig:openclose_network}. Further, the most connected channels are different across these two states. Note that channels identified as strongly connected by the analysis have been mentioned in vision-related tasks in the literature \cite{nezamfar2011decoding,das2016eeg}. For example, in the alpha-band estimated network, channel Oz exhibits a higher connectivity pattern in the closed eyes segments; it is located within the primary visual cortex which is the most studied visual area in the brain \cite{arden2003electrical,sharon2007advantage}.
In addition, channels PO3, PO8, and Iz exhibit also a high degree of connectivity in the closed eyes segments.

We also provide a summary measure of the differences between the estimated networks for the two {states}, by calculating their Hamming distance (HD) for both the robust detrended data ($\text{HD}_{robust}(open, close)=0.058$), and the alpha-band data ($\text{HD}_{alpha}(open, close)=0.066)$.  Further, Figure~\ref{fig:channel_boxplots} shows boxplots of the connectivity of each channel for the close and open eye segments across all 21 subjects (i.e., each is based on 21 measurements). The EEG channels are grouped in different ways to facilitate comparisons: eye-related, central, central left, central right, front, and posterior, respectively. {The channels located at the posterior of the scalp exhibit significantly bigger changes in their connectivity patterns, as verified by a permutation test whose $p$-value equals 0.0345. Specifically, we examine the hypothesis that $H_0: C_{open} = C_{closed}$ versus $H_1: C_{open} \neq C_{closed}$ for each EEG channel group, where $C_{open}$ and $C_{closed}$ represent the averaged connectivity over the specific EEG channel group for the eye open/closed status, respectively.}
Hence, the obtained Granger causal networks are in agreement with analogous findings in the literature.
\begin{figure}[H]
    \centering
    \includegraphics[scale=.3]{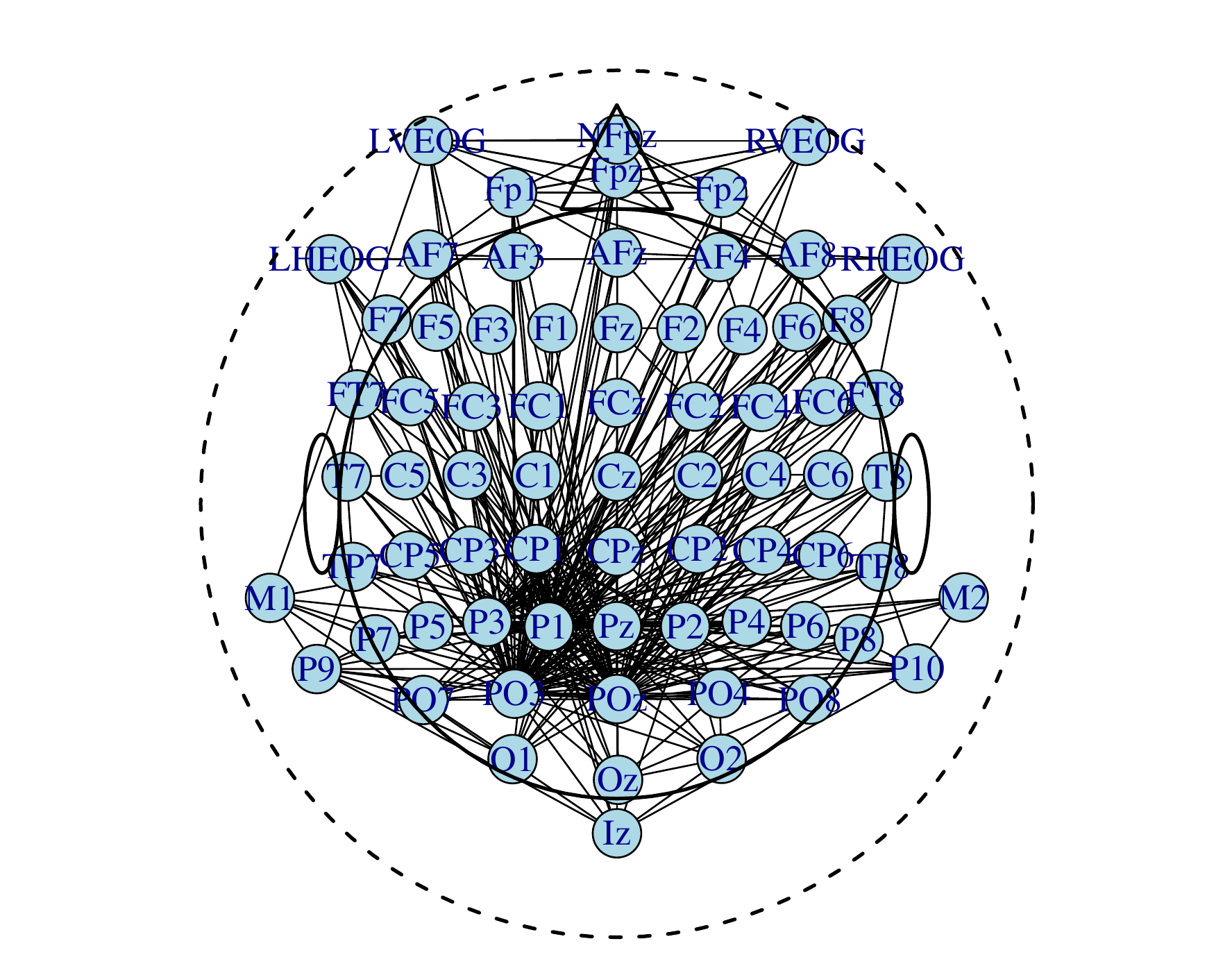}%
    \includegraphics[scale=.3]{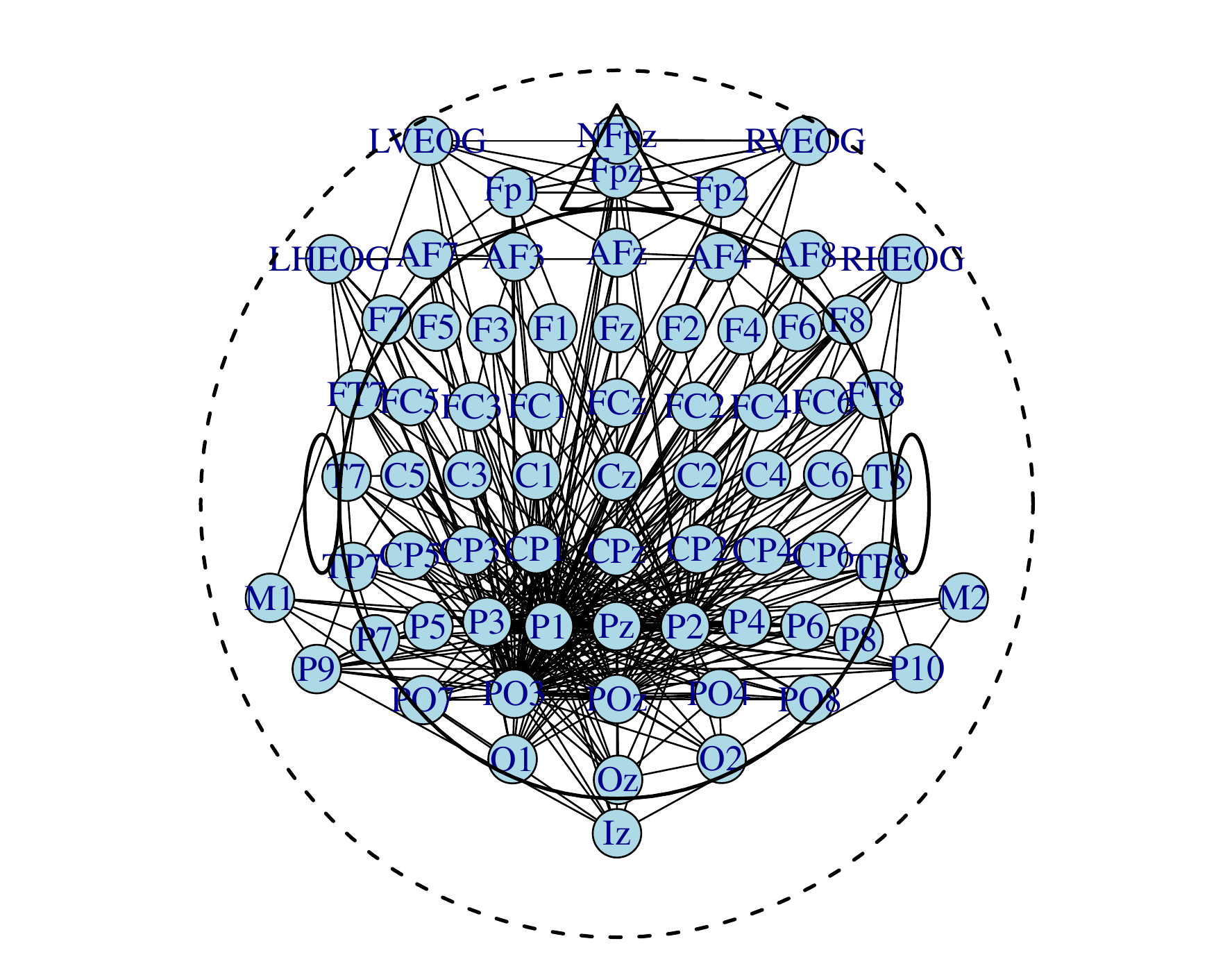}
    \vfill
    \includegraphics[scale=.3]{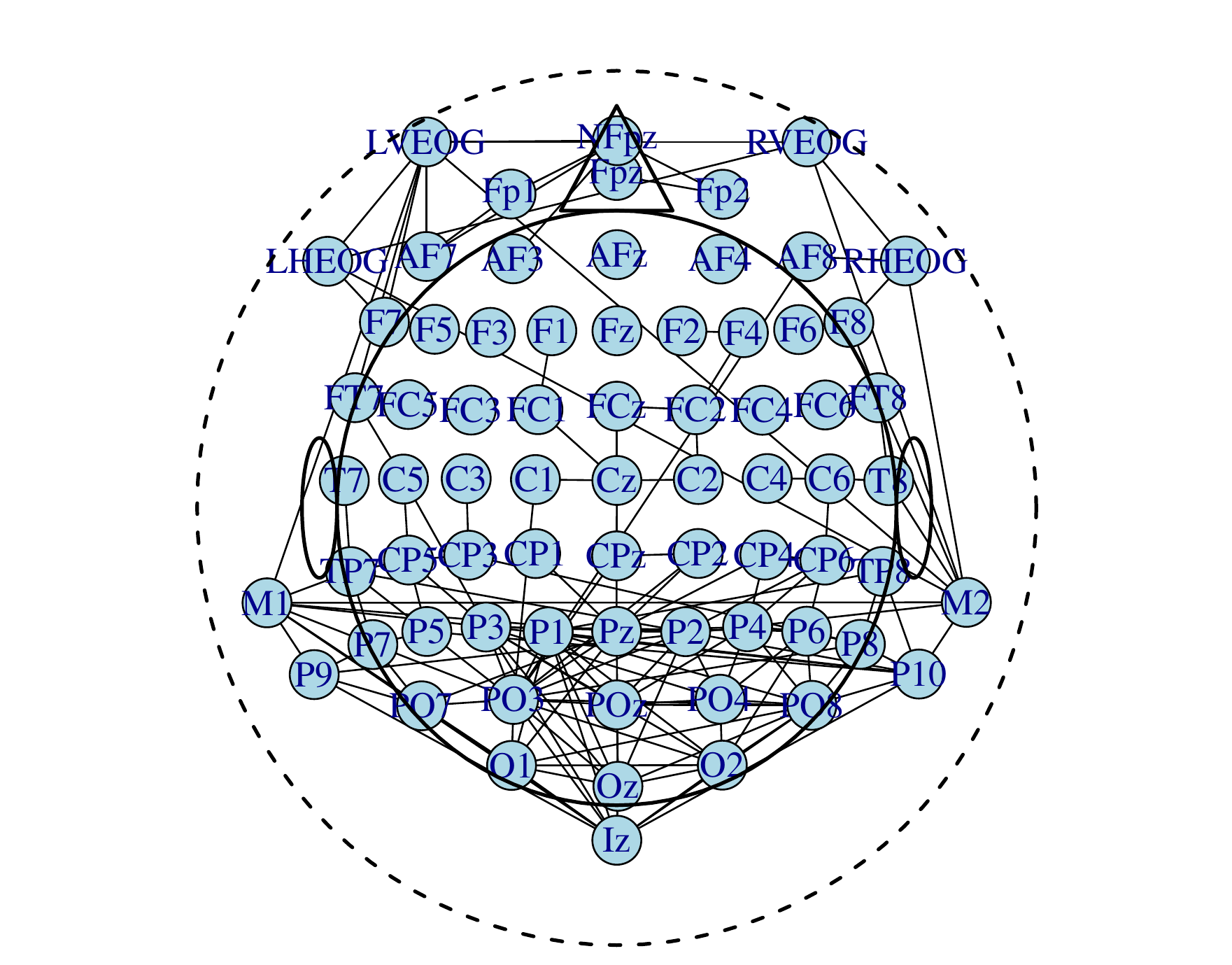}%
    \includegraphics[scale=.3]{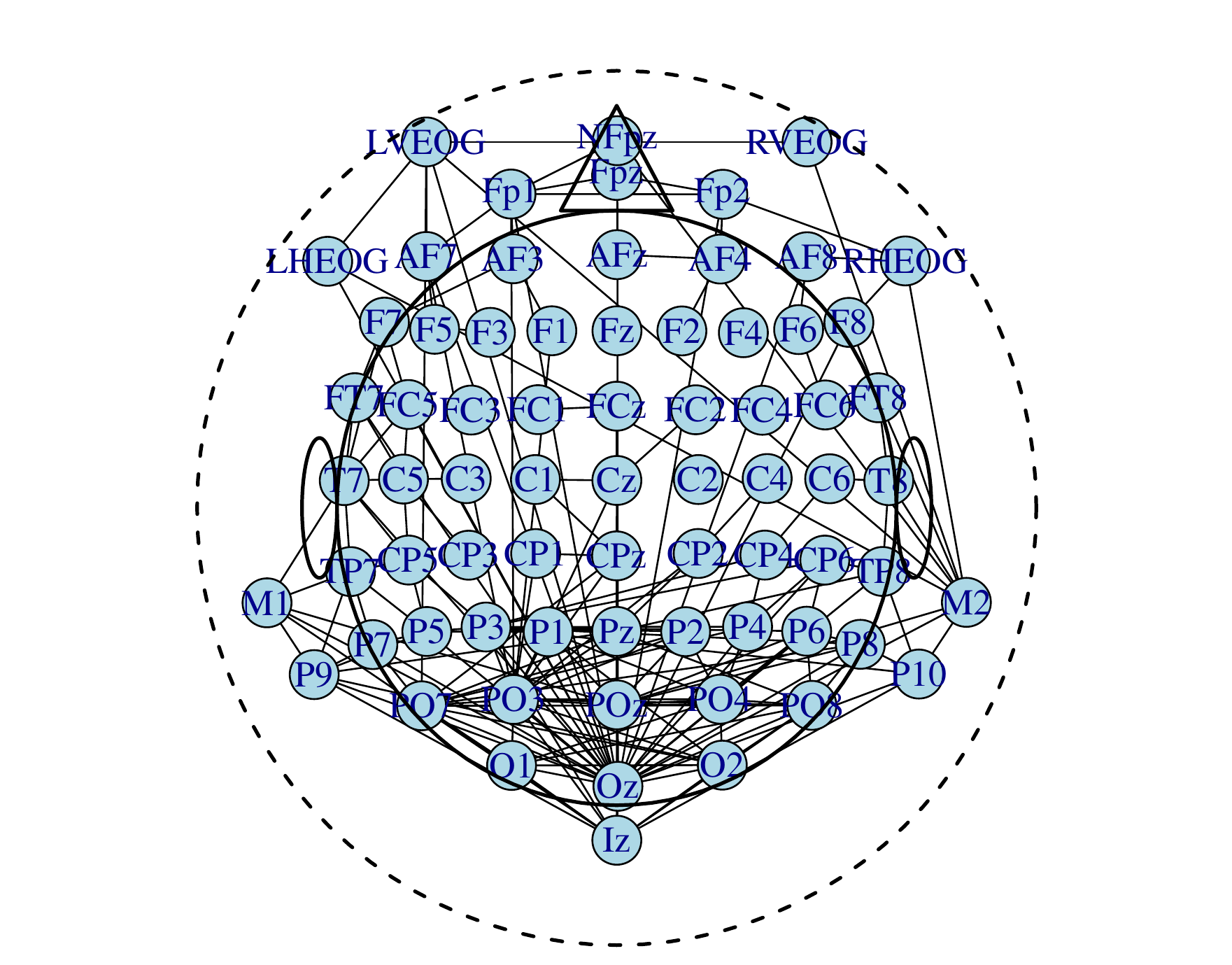}
    \caption{Top left: averaged estimated network for open-eye intervals over all 21 subjects using robust detrended data; Top right: averaged estimated network for close-eye intervals over all 21 subjects using original data; Bottom left: averaged estimated network for open-eye intervals using Alpha-band data; Bottom right: averaged estimated network for close-eye intervals using alpha-band data.}
    \label{fig:openclose_network}
\end{figure}

{
\begin{figure}[!ht]
    \centering
    \includegraphics[scale=.25]{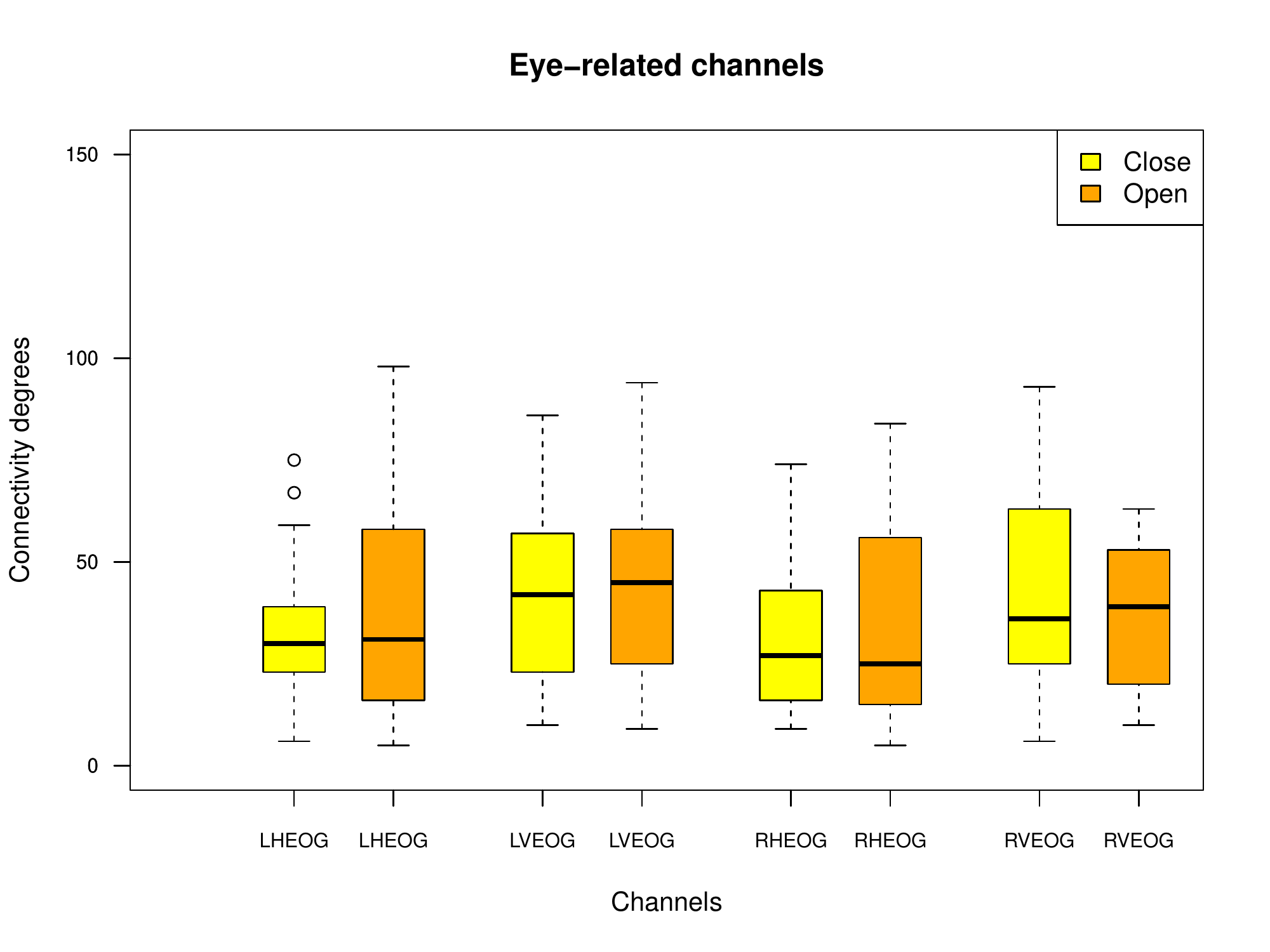}%
    \includegraphics[scale=.25]{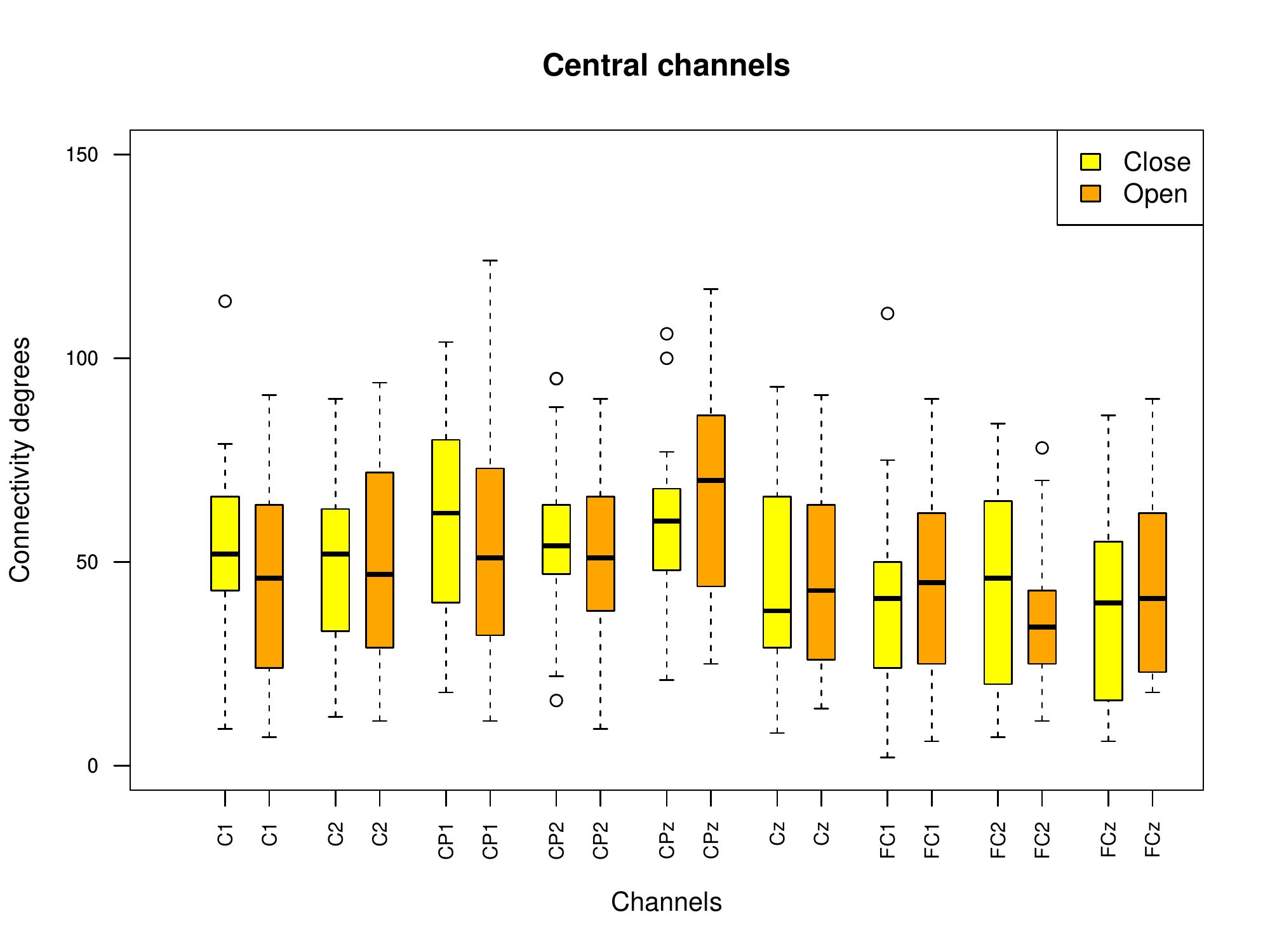}%
    \vfill
    \includegraphics[scale=.25]{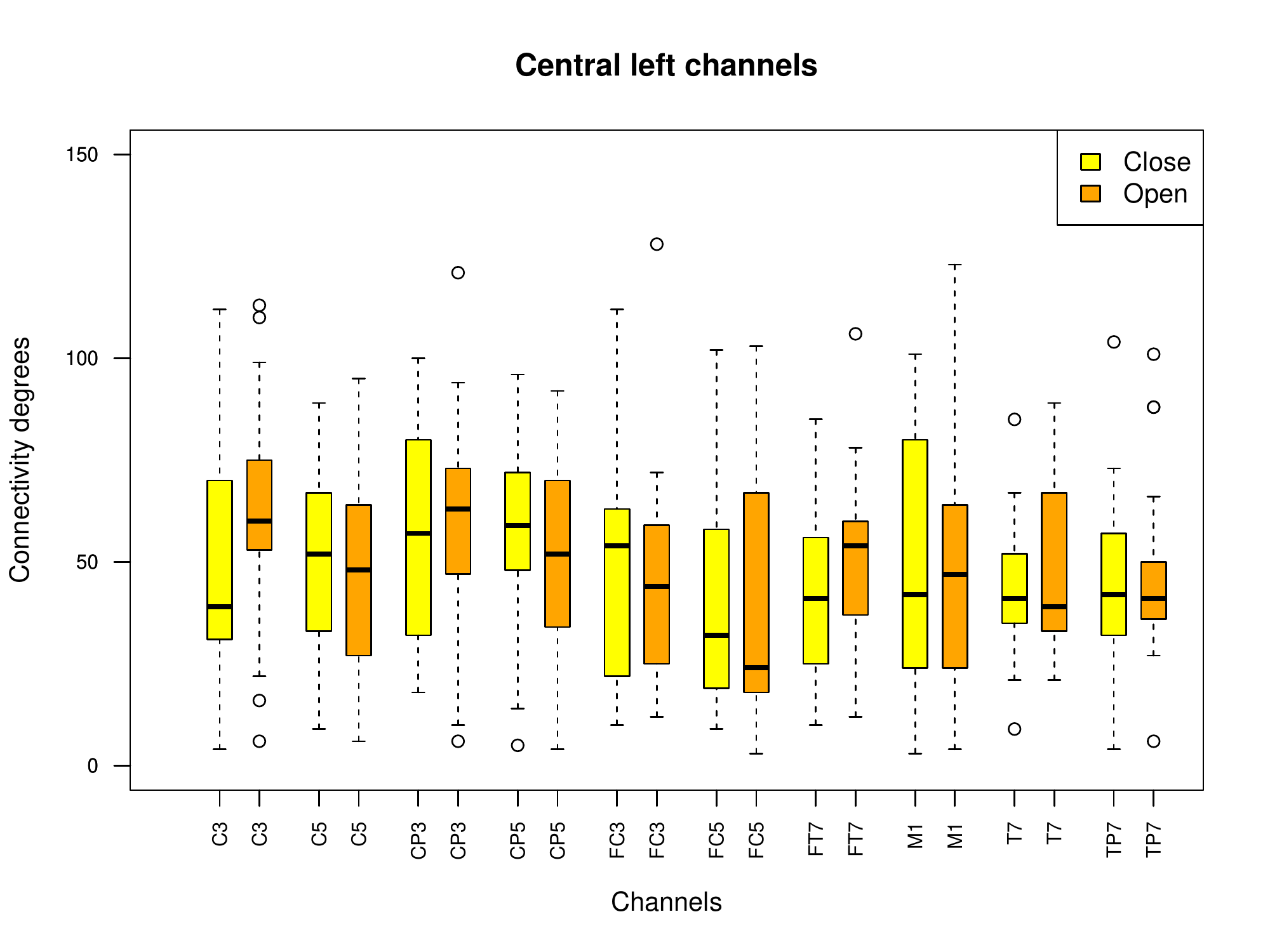}%
    \includegraphics[scale=.25]{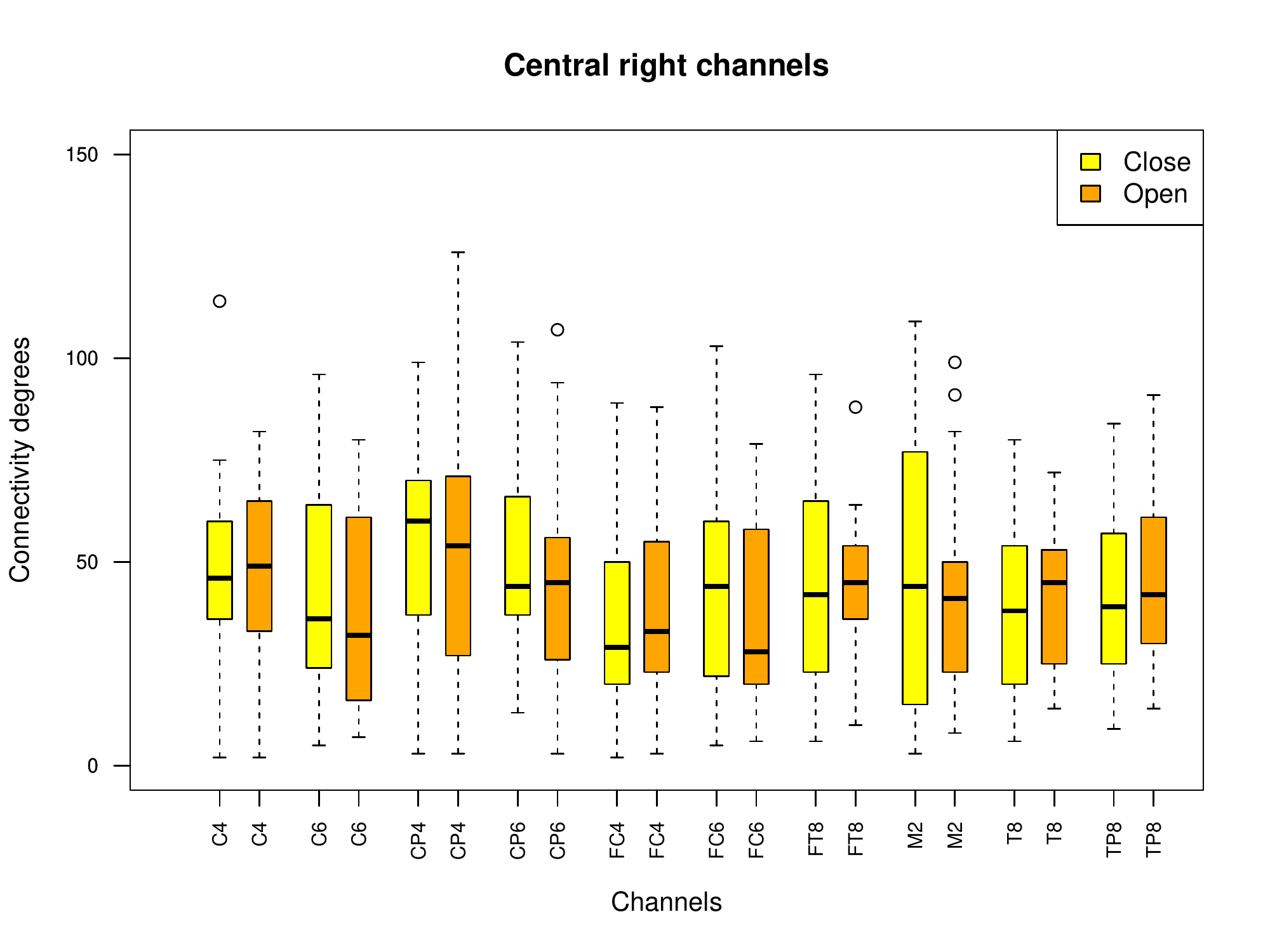}%
    \vfill
    \includegraphics[scale=.25]{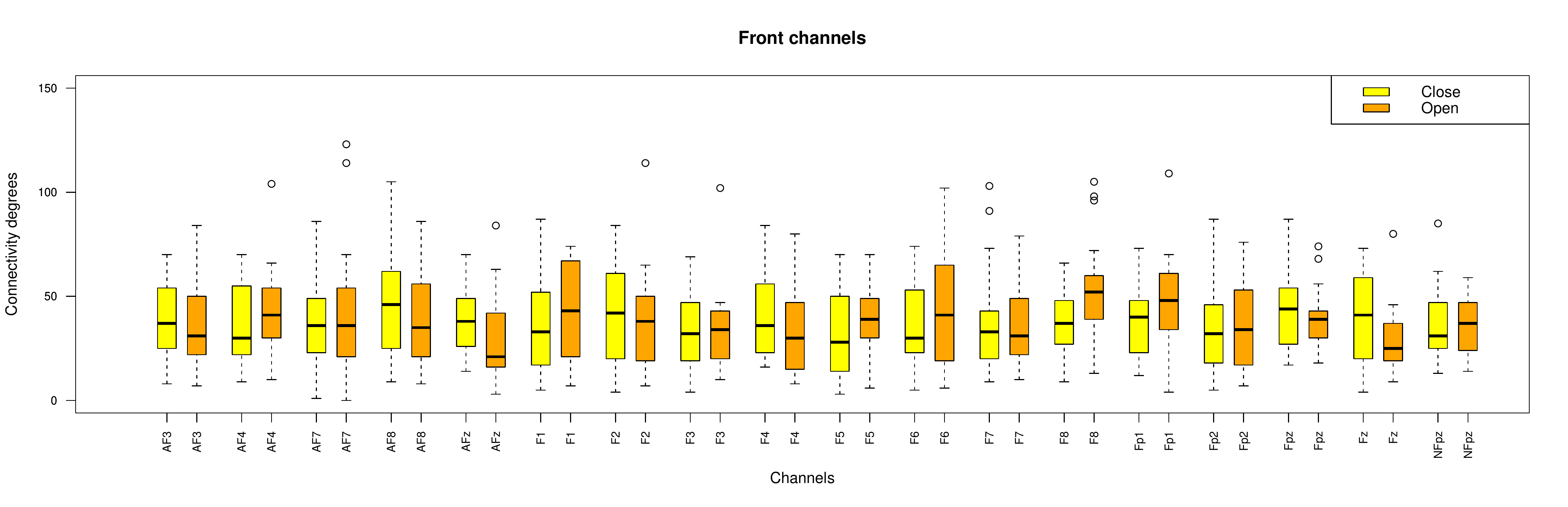}%
    \vfill
    \includegraphics[scale=.25]{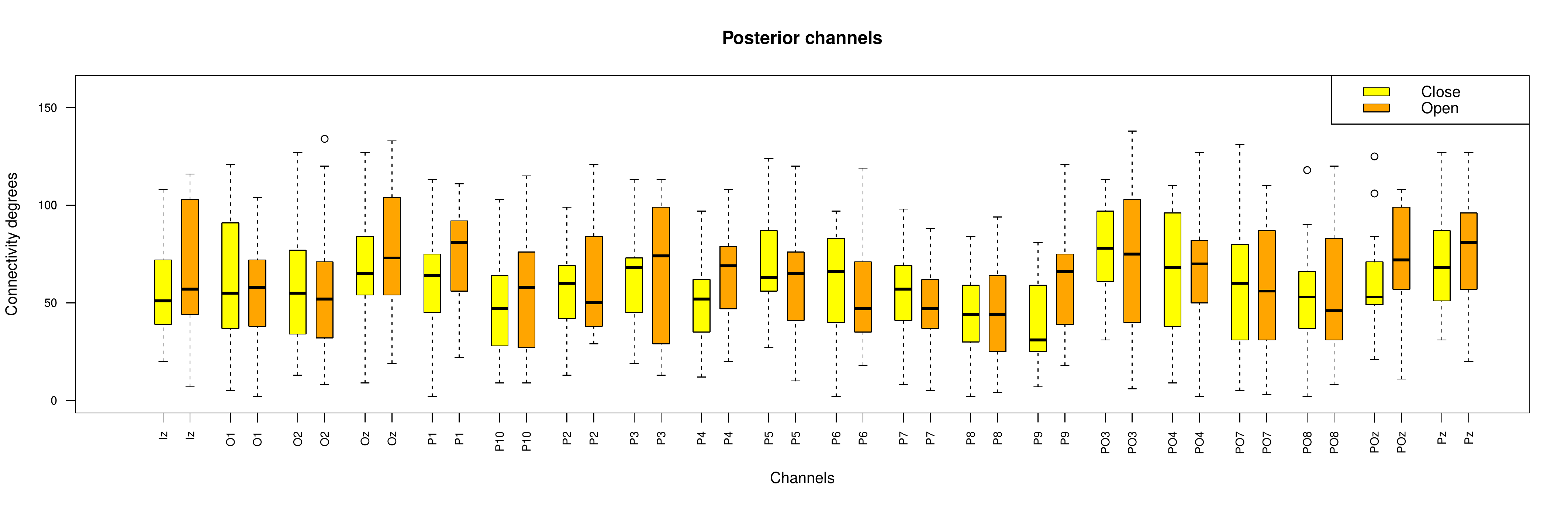}
    \caption{Boxplots for all channels in close/open eyes segments over all 21 subjects. Top panel: close-eyes segment; Bottom panel: open-eyes segment.}
    \label{fig:channel_boxplots}
\end{figure}
}

\subsubsection{Comparison with Other Methods}
{
Next, we provide comparisons of the results obtained by TBSS to those obtained by the fused lasso method \cite{tibshirani2005sparsity} (which corresponds to Step 1 of TBSS), the FreSpeD method \cite{schroder2019fresped}, and the HMM-VAR model \cite{vidaurre2016spectrally}, respectively. Figure \ref{fig:fusedlasso_hist} clearly shows that by only using fused lasso, it becomes difficult to clearly identify the true break points across the 21 subjects. This is not surprising, given the need for the subsequent Steps in TBSS, and the theoretical results provided in Section \ref{sec:methods}.  
\begin{figure}[H]
    \centering
    \includegraphics[scale=.32]{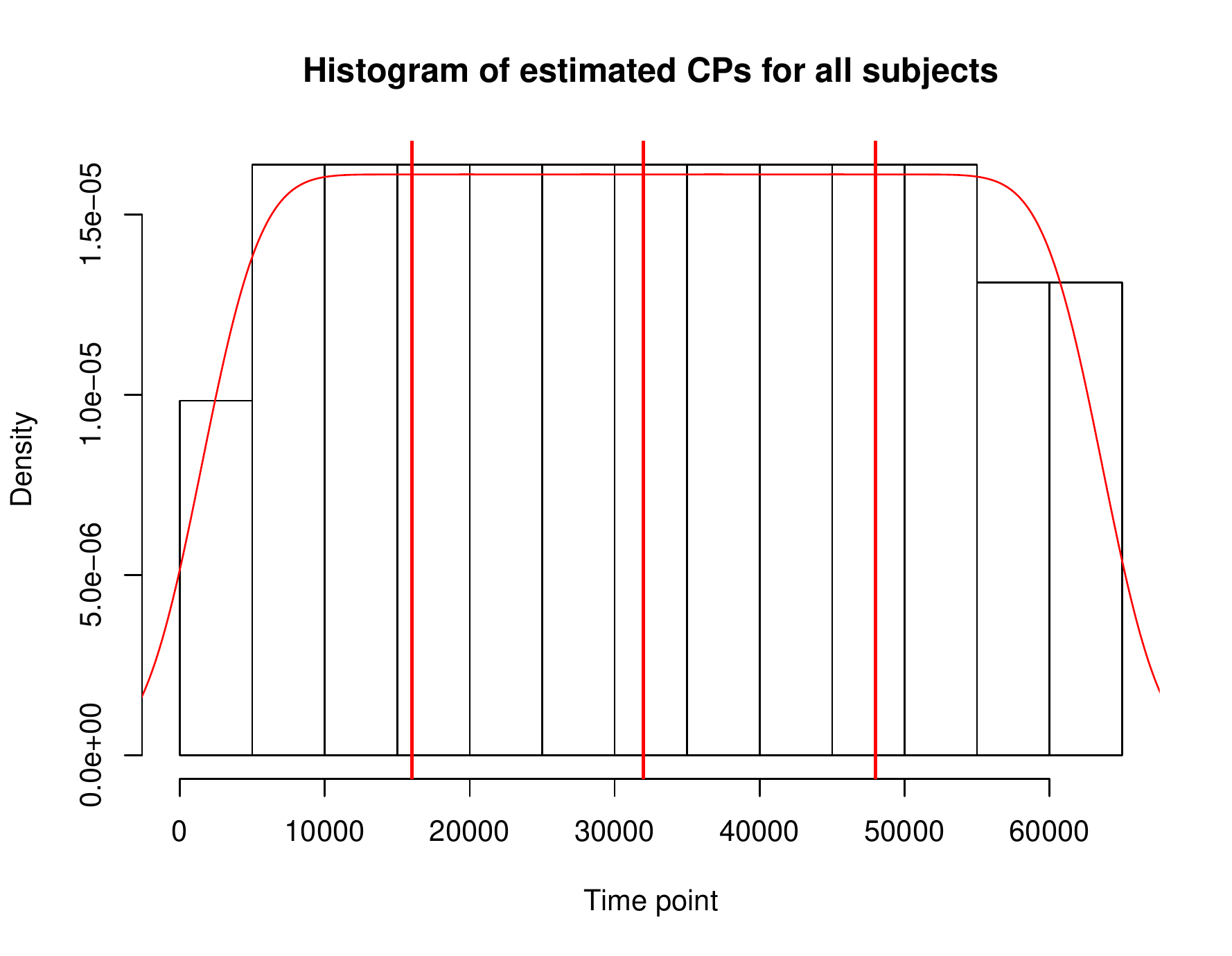}
    \caption{Histogram of selected break points by using the fused lasso method}
    \label{fig:fusedlasso_hist}
\end{figure}
}

{
Next, we provide the results based on the FreSpeD method \cite{schroder2019fresped}, which leverages a cumulative sum-type test statistic combined with a binary segmentation algorithm. 
\begin{figure}[H]
    \centering
    \includegraphics[scale=.175]{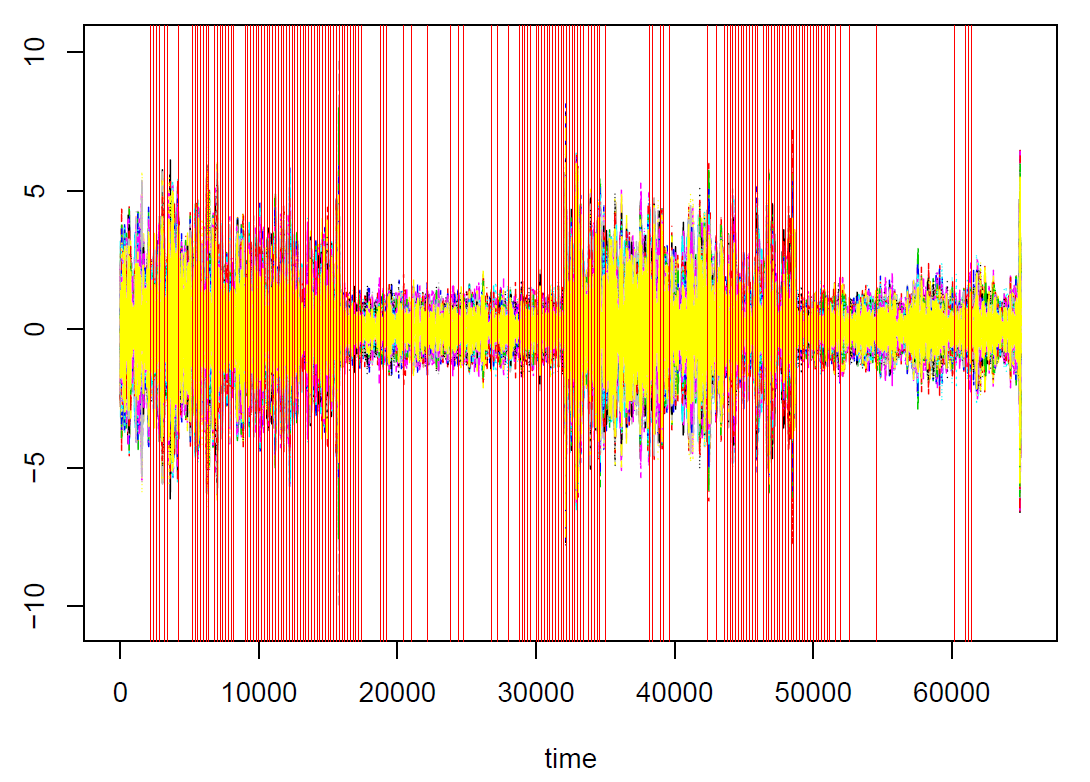}%
    \caption{Estimated change points by using FreSpeD algorithm.}
    \label{fig:fresped}
\end{figure}
Figure \ref{fig:fresped} shows that the method estimates a large number of spurious break points (red vertical lines) in the current selected subject; specifically, we obtain over 4000 break points in total for this subject.}

{
Finally, we present the results from the HMM-VAR model \cite{vidaurre2016spectrally}. We set the number of hidden states to two, that incorporates our prior knowledge about the nature of the experiment that gave rise to the data. Figure \ref{fig:hmm_application} shows the estimated posterior probabilities for these two states (purple and yellow bars indicate different states). 
\begin{figure}[H]
    \centering
    \includegraphics[scale=.15]{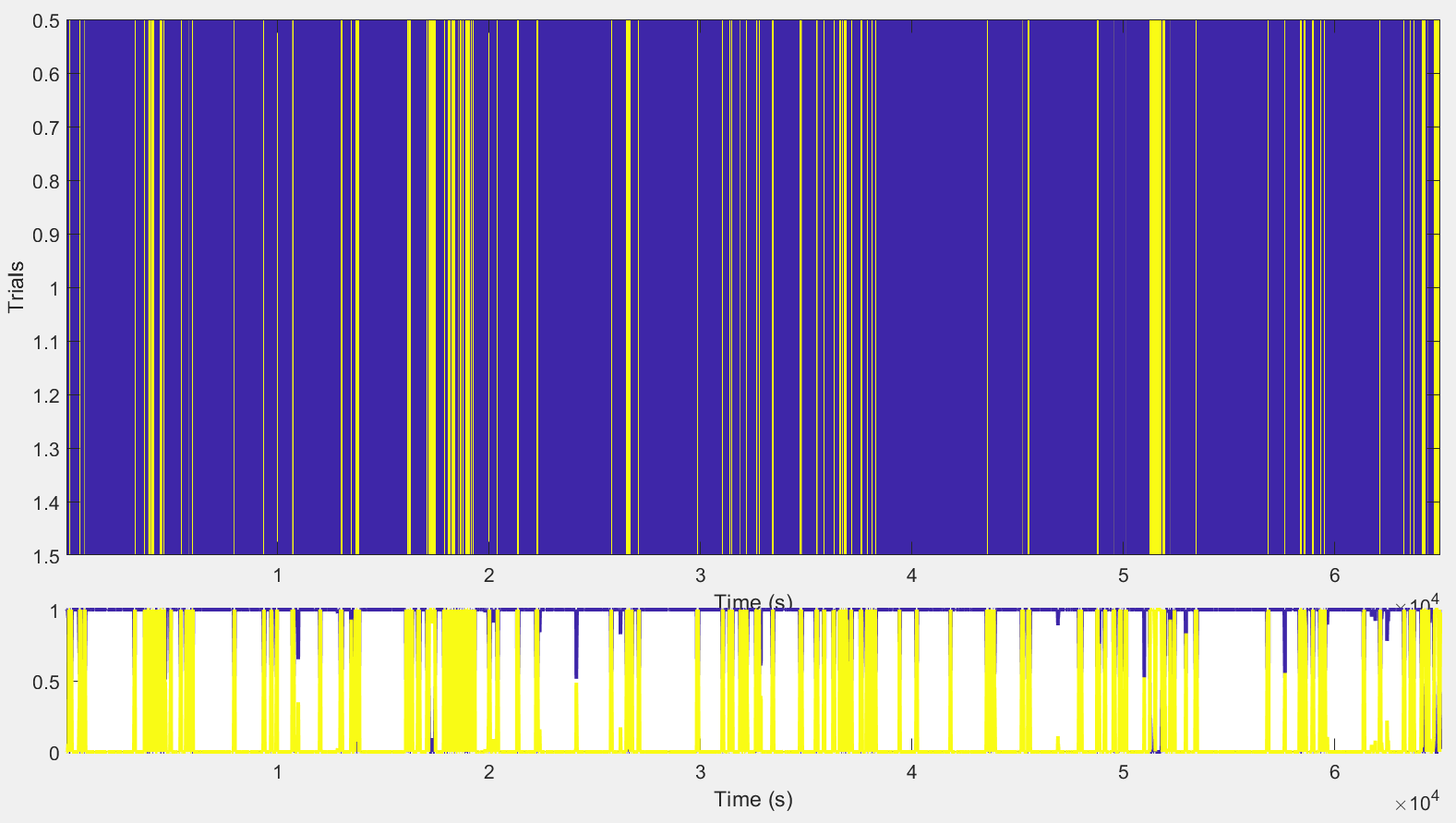}
    \caption{Estimated hidden state transition plot by HMM-VAR model.}
    \label{fig:hmm_application}
\end{figure}
It can be seen that the method estimates rapid switches between the two states for a given subject; analogous results are obtained for the other 20 subjects. One issue is that the method is most suitable to estimate frequent switches between states, which is not the case for the data at hand (only three switches).
}

\section{Discussion}
\label{section:4}
The developed TBSS algorithm is scalable to exceedingly large data sets, since an appropriate choice of the block size ($b_T\sim \sqrt{T}$) achieves \textit{sublinear} computational complexity in the number of time points. Further, it obtains very accurate estimation of both the number and the location of the break points, also supported by the theoretical analysis provided in Section \ref{sec:methods}. 
{The current presentation has focused on detection and estimation of changes in the auto-regressive parameters and ignored the impact of the contemporaneous covariance matrix $\Sigma_j$ that may be changing across time segments (see model formulation in \eqref{sec:model}). Its impact can be easily accommodated by modifying the least squares criterion in \eqref{eqn:objectivefunc} by a generalized least squares criterion (for details and algorithms, see \cite{lin2017regularized}). }

The key idea underlying TBSS is to examine blocks of time points which lead to substantial computational gains. However, the selection of the block size $b_T$ is a critical tuning parameter. On many occasions, like the EEG data application presented in Section \ref{sec:application}, the analyst has information about the spacing between consecutive break points that help select $b_T$. In absence of such information, a data-driven approach can be adopted as follows: apply TBSS with block sizes of order $T^{0.5- \frac{j}{2K}}$ for large enough $K$, starting from $j=0$, and continue increasing $j$ until the number of selected break points in Step 4 for $j$ and $j+1$ coincide (obviously, $j<K$).

{Another advantage of TBSS is that the transition matrices in the underlying VAR model is weakly sparse (as opposed to strictly sparse in previous work in the literature). In weakly sparse models, the ``small" magnitudes in the transition matrices indicate weak Granger causal effects amongst the EEG channels, while the ``large" magnitudes indicate strong effects, thus expanding the scope of the methodology and being able to accommodate more realistic settings for effective brain connectivity. Further, it is of interest to develop \textit{testing procedures} for identifying significant changes in individual entries (or groups of entries) of these networks, as well develop confidence intervals for the location of the identified break points. Both pose a number of technical challenges for developing rigorously justified procedures and constitute topics of future research.}

{
Finally, the TBSS algorithm can be easily adapted to accommodate other high dimensional models (such as graphical models) that have been used for detection of change points in brain connectivity networks.}

\newpage
\appendix
\section{Technical Proofs of Theoretical Results}
\subsection{Proof of Theorem \ref{thm:1}}
The main idea of this proof is to show that the estimated VAR model parameters in the interval $[t_{j_0-1} \vee \widehat{t}_j, \widehat{t}_{j+1} \wedge t_{j_0+1}]$ converge in the $\ell_2$ norm to both $\Phi^{(\cdot, j_0)}$ and $\Phi^{(\cdot, j_0+1)}$, which contradicts Assumption A3. 
\begin{proof}
The proof is based on arguing by contradiction. Suppose there exists an estimated break point $\widehat{t}_j$ such that $|\widehat{t}_j - t_j| > T\gamma_T$. In other words, there exists a true break point $t_{j_0}$, which is isolated from all the estimated points (there are no estimated break points close to $t_{j_0}$). 

Based on the optimization problem in \eqref{eqn:objectivefunc}, the value of the function is minimized exactly at $\widehat{\Theta}$. Therefore, for the interval $[t_{j_0-1}, t_{j_0}]$, we denote the closest block end-point $r_i$ to the right side of $t_{j_0-1}$ by $s_{j_0-1}$, and similarly, denote the closest $r_i$ to the left side of $t_{j_0}$ by $s_{j_0}$. Define a parameter sequence $\psi_k$'s, $k=1,2,\cdots, k_T$ with $\psi_k = \widehat{\theta}_k$ except for two time points $k=\widehat{t}_j$ and $k=t_{j_0}$. In this proof, we focus on the interval $[s_{j_0-1} \vee \widehat{t}_j, s_{j_0}]$. Define the following parameter sequence $\psi_k$, for $k=1,2, \cdots, k_T$ with $\psi_k = \widehat{\theta}_k$ except for two points $k=\widehat{i}_j \vee i_{j_0-1}$ and $k=i_{j_0}$. For these two points, if $\widehat{t}_j > s_{j_0-1}$, then we set $\psi_{\widehat{i}_j} = \Phi^{(\cdot, j_0)} - \widehat{\Phi}_j$ and $\psi_{{i}_{j_0}} = \widehat{\Phi}_{j+1} - \Phi^{(\cdot, j_0)}$; else if $\widehat{t}_j \leq s_{j_0-1}$, then we set $\psi_{i_{j_0-1}} = \Phi^{(\cdot, j_0)} - \widehat{\Phi}_{j+1}$ and $\psi_{i_{j_0}} = \widehat{\Phi}_{j+1} - \Phi^{(\cdot, j_0)}$. Then, we have
\begin{equation}
    \label{eq:9}
    \begin{aligned}
        \frac{1}{T}\|\mathbf{Y} - \mathbf{Z}\widehat{\Theta}\|_2^2 &+ \lambda_{1,T}\|\widehat{\Theta}\|_1 + \lambda_{2,T}\sum_{i=1}^{k_T}\left\|\sum_{j=1}^i \widehat{\theta}_j \right\|_1 \\
        &\leq \frac{1}{T}\|\mathbf{Y} - \mathbf{Z}\Psi\|_2^2 + \lambda_{1,T}\|\Psi\|_1 + \lambda_{2,T}\sum_{i=1}^{k_T}\left\| \sum_{j=1}^i \psi_j \right\|_1.
    \end{aligned}
\end{equation}

After some algebraic rearrangements for \eqref{eq:9}, we obtain that
\begin{align}
    \label{eq:10}
    0 &\leq c_0\|\Phi^{(\cdot, j_0)} - \widehat{\Phi}_{j+1}\|_2^2 \nonumber \\
    &\leq \frac{2}{s_{j_0} - s_{j_0-1} \vee \widehat{t}_{j}}\sum_{l = s_{j_0-1}\vee \widehat{t}_j}^{s_{j_0}}Y_{l-1}^\prime\left(\Phi^{(\cdot, j_0)} - \widehat{\Phi}_{j+1}\right)\epsilon_l \nonumber \\
    &+ \frac{T\lambda_{1,T}}{s_{j_0} - s_{j_0-1}\vee \widehat{t}_j}\left(\|\Phi^{(\cdot, j_0)} - \widehat{\Phi}_{j+1}\|_1 + \|\Phi^{(\cdot, j_0)} - \widehat{\Phi}_{j+1}\|_1\right) \nonumber \\
    &+ \frac{T\lambda_{2,T}}{b_T}\left(\|\Phi^{(\cdot, j_0)}\|_1 - \|\widehat{\Phi}_{j+1}\|_1\right) \\
    &\leq \left(\frac{2T\lambda_{1,T}}{s_{j_0} - s_{j_0-1} \vee \widehat{t}_j} + C\sqrt{\frac{\log p}{b_T}}\right)\|\Phi^{(\cdot, j_0)} - \widehat{\Phi}_{j+1}\|_1 \nonumber \\
    &+ \frac{T\lambda_{2,T}}{b_T}\left(\|\Phi^{(\cdot, j_0)}\|_1 - \|\widehat{\Phi}_{j+1}\|_1\right) \nonumber \\
    &\leq \frac{3}{2}\frac{T\lambda_{2,T}}{b_T}\|\Phi^{(\cdot, j_0)} - \widehat{\Phi}_{j+1}\|_{1, \mathcal{I}} - \frac{1}{2}\frac{T\lambda_{2,T}}{b_T}\|\Phi^{(\cdot, j_0)} - \widehat{\Phi}_{j+1}\|_{1,\mathcal{I}}, \nonumber
\end{align}
which implies that
\begin{equation}
    \label{eq:11}
    \|\Phi^{(\cdot, j_0)} - \widehat{\Phi}_{j+1}\|_F = \mathcal{O}_p\left( \sqrt{\frac{d^\star_{\max}\log p}{b_T}} \right),
\end{equation}
which indicates that $\|\Phi^{(\cdot, j_0)} - \widehat{\Phi}_{j+1}\|_F$ converges to zero in probability. Similarly, we can prove that in the interval $[s_{j_0}, s_{j_0+1} \wedge \widehat{t}_{j+1}]$, the quantity $\|\Phi^{(\cdot, j_0+1)} - \widehat{\Phi}_{j+1}\|_F$ converges to zero as well. This contradicts Assumption A3, which completes the proof.
\end{proof}

\subsection{Proof of Theorem \ref{thm:2}}
In this section, we illustrate the proof of Theorem \ref{thm:2}. The main idea of this proof is to show that the solution $\widetilde{t}_j$ to \eqref{eqn:final} minimizes the objective function \eqref{eq:12} defined below in the given search interval $(l_j, u_j)$. Based on the step 2 and 3, we know that there exists a true break point $t_j$ in the search interval $(l_j, u_j)$. 

Denote the objective function $\mathcal{L}(\tau)$ as follows:
\begin{equation}
    \label{eq:12}
    \mathcal{L}(\tau) = \frac{1}{u_j-l_j}\left(\sum_{t=l_j}^{\tau-1}\|y_{t+1} - \widetilde{\Phi}^{(\cdot, j)}Y_t\|_2^2
    + \sum_{t=\tau}^{u_j-1}\|y_{t+1} - \widetilde{\Phi}^{(\cdot, j+1)}Y_t\|_2^2\right).
\end{equation}
According to the definition of $\widetilde{t}_j$, we have the basic inequality $\mathcal{L}(\widetilde{t}_j) \leq \mathcal{L}(t_j)$, then by using this basic inequality and the similar argument as Proposition 4.1 in \cite{basu2015regularized}, we are able to derive that:
\begin{equation}
    \label{eq:13}
    \begin{aligned}
    &\|\widetilde{\Phi}^{(\cdot, j)} - \Phi^{(\cdot, j)}\|_1 \leq 4\sqrt{d_{\max}^\star}\|\widetilde{\Phi}^{(\cdot, j)} - \Phi^{(\cdot, j)}\|_F, \\
    &\|\widetilde{\Phi}^{(\cdot, j+1)} - \Phi^{(\cdot, j+1)}\|_1 \leq 4\sqrt{d_{\max}^\star}\|\widetilde{\Phi}^{(\cdot, j+1)} - \Phi^{(\cdot, j+1)}\|_F. 
    \end{aligned}
\end{equation}
Moreover, we obtain that
\begin{equation}
    \label{eq:14}
    \|\widetilde{\Phi}^{(\cdot, j)} - \Phi^{(\cdot, j)}\|_F = \mathcal{O}_p\left( \sqrt{\frac{d_{\max}^\star\log p}{u_j - l_j}} \right), \ 
    \|\widetilde{\Phi}^{(\cdot, j+1)} - \Phi^{(\cdot, j+1)}\|_F = \mathcal{O}_p\left( \sqrt{\frac{d_{\max}^\star\log p}{u_j - l_j}} \right).
\end{equation}

Using the results in \eqref{eq:13} and \eqref{eq:14} leads to the following result: assume that $\widetilde{t}_j > t_j$, then we denote
\begin{equation*}
    (u_j - l_j)\mathcal{L}(\widetilde{t}_j) = \sum_{t=l_j}^{t_j-1}\|y_t - \widetilde{\Phi}^{(\cdot, j)}Y_t\|_2^2 + \sum_{t=t_j}^{u_j-1}\|y_t - \widetilde{\Phi}^{(\cdot, j+1)}Y_t\|_2^2 \overset{\text{def}}{=} I_1 + I_2.
\end{equation*}
Now, we first provide the lower bounds for both $I_1$ and $I_2$ by using similar arguments in Lemma 5 case (c) in \cite{safikhani2017joint}:
\begin{equation}
    \label{eq:15}
    \begin{aligned}
    I_1 &\geq \sum_{t=l_j}^{t_j-1}\|\epsilon_t\|_2^2 + c_1v_j|\widetilde{t}_j - t_j| - c_2d_{\max}^\star\log p, \\
    I_2 &\geq \sum_{t=t_j}^{u_j-1}\|\epsilon_t\|_2^2  - c_3 d_{\max}^\star\log p. 
    \end{aligned}
\end{equation}
Hence, combining these two equations implies that
\begin{equation}
    \label{eq:17}
    \mathcal{L}(\widetilde{t}_j) \geq \sum_{t=l_j}^{u_j-1}\|\epsilon_t\|_2^2 + K_1v_j|\widetilde{t}_j - t_j| - K_2 d_{\max}^\star \log p,
\end{equation}
where $K_1$ and $K_2$ are some positive constants.

Next, we claim the upper bound for the objective function at true break point $t_j$: 
\begin{equation}
    \label{eq:18}
    \mathcal{L}(t_j) \leq \sum_{t=l_j}^{u_j-1}\|\epsilon_t\|_2^2 + Kd_{\max}^\star\log p.
\end{equation}
To show this result, we use the similar procedure as of the proof of Theorem 3 in \cite{safikhani2017joint}. Therefore, by using \eqref{eq:17} and \eqref{eq:18} together with the basic inequality, we obtain:
\begin{equation}
    \label{eq:19}
    \sum_{t=l_j}^{u_j-1}\|\epsilon_t\|_2^2 + K_1v_j|\widetilde{t}_j - t_j| - K_2 d_{\max}^\star \log p \leq \mathcal{L}(\widetilde{t}_j)
    \leq \mathcal{L}(t_j) \leq \sum_{t=l_j}^{u_j-1}\|\epsilon_t\|_2^2 + Kd_{\max}^\star\log p,
\end{equation}
which leads to the final result.

\subsection{Proof of Theorem \ref{thm:3}}
Theorem \ref{thm:3} establishes that after detecting the break points, Step 5 of the TBSS algorithms obtains consistent estimates for the VAR model parameters in each stationary segment identified. This proof is similar to that of Proposition 4.1 in \cite{basu2015regularized}. The key steps in the proof that require verification are: (1) the restricted strong convexity (RSC) condition; (2) the deviation bounds for the estimation intervals. For (1), we use analogous arguments as in the proof of Theorem 4 in \cite{safikhani2017joint} to establish the result. Further, (2) follows Lemma 1 in \cite{safikhani2017joint}. The technical details are omitted.

\section{Additional Numerical Experiments}
Next, we provide additional numerical experiments that evaluate the performance of the TBSS algorithm. Further, some extra comparisons to existing methods in the literature are also included.
\subsection{Additional Numerical Experiments}
\begin{figure}[!ht]
    \centering
    \includegraphics[scale=.275]{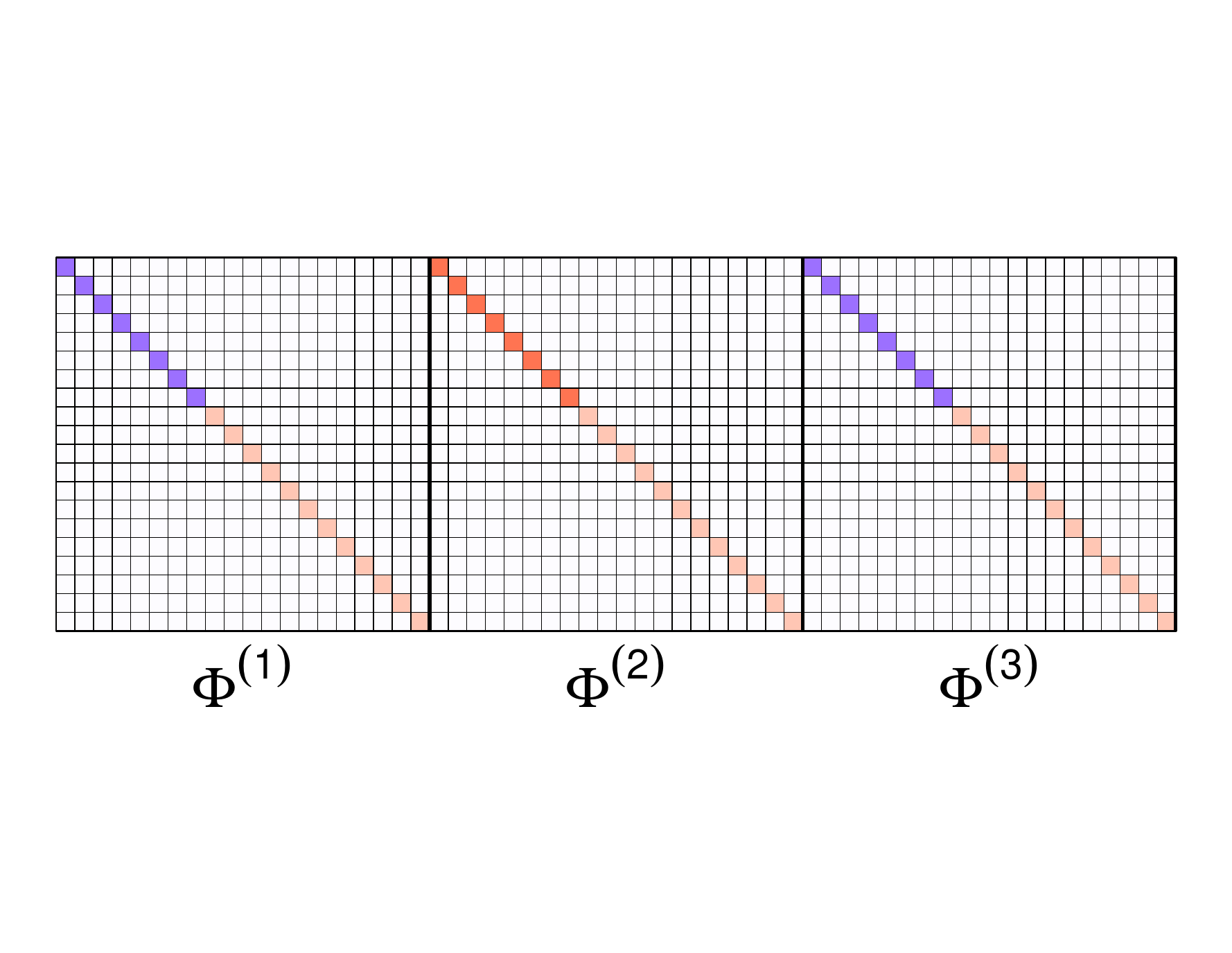}%
    \hfill
    \includegraphics[scale=.275]{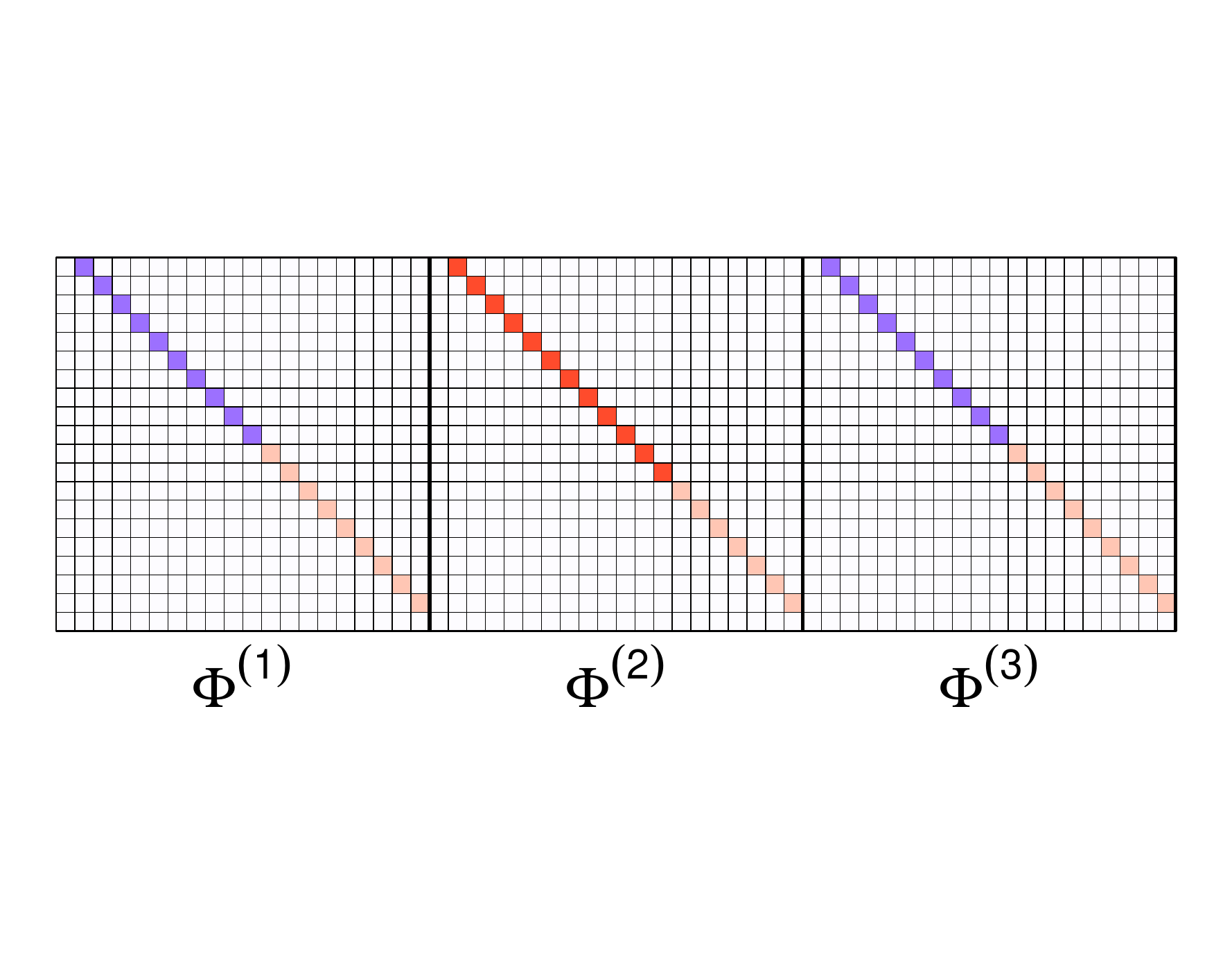}%
    \hfill
    \includegraphics[scale=.275]{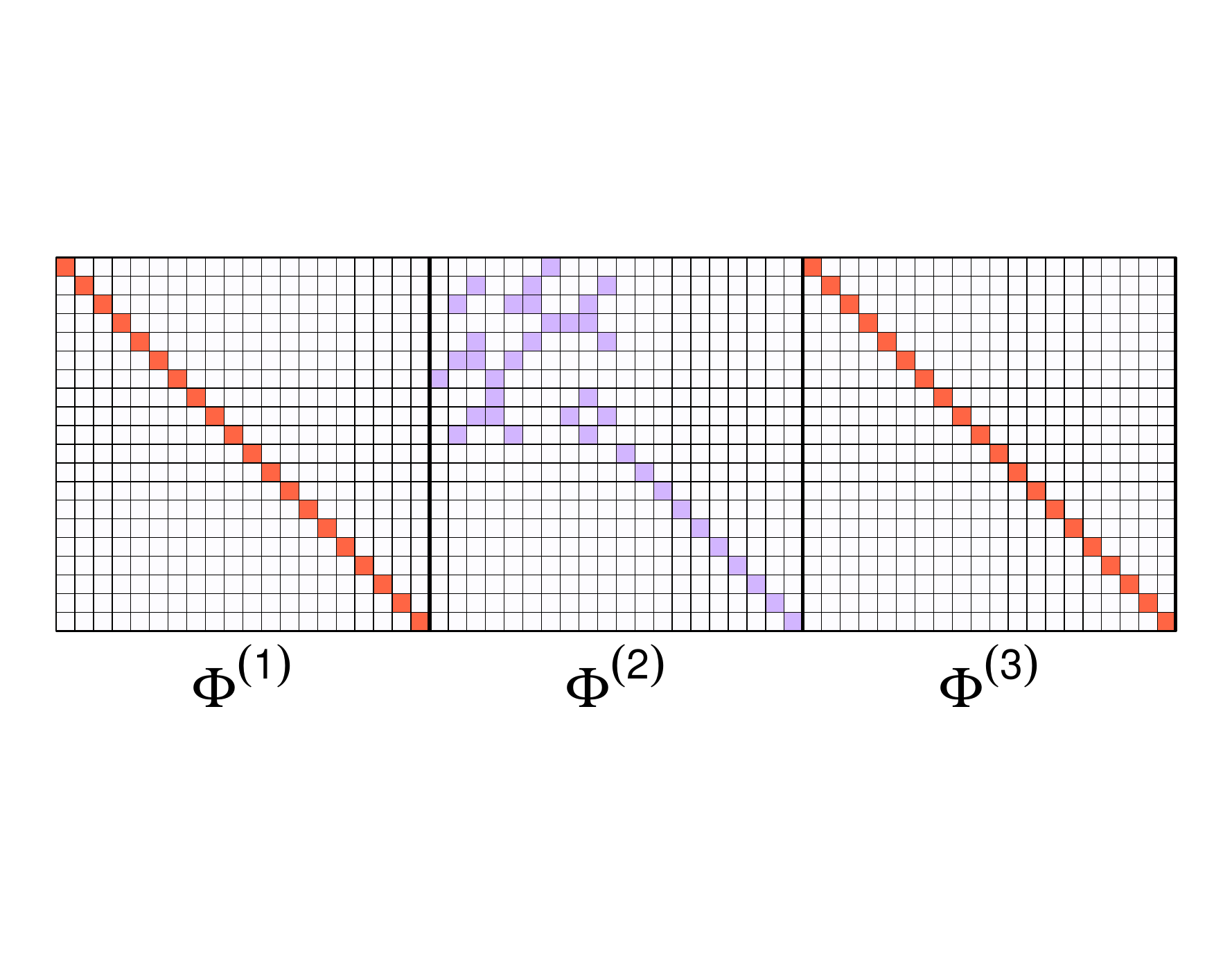}
    \vspace{-20pt}
    \caption{Transition matrices for the extra simulation settings. Left panel: extra setting 1; middle panel: extra setting 2; right panel: extra setting 3.}
    \label{fig:extra_simu}
\end{figure}

First, we consider the following settings: $p=20$, $T=600$ with two change points located at $t_1^\star=200$, $t_2^\star=400$. Then, we separately investigate three different changes in the transition matrices. In case H.1, we assume that all three transition matrices are diagonal, and only the first five entries are changing while the others remain fixed. In setting H.2, we examine the changes in 1-off diagonal transition matrices, and we assume the first 10-12 entries are time-varying and the others remain fixed. In the last case H.3, we consider randomly sparse transition matrices, as depicted in figure \ref{fig:extra_simu}.

The performance is evaluated based on the same metrics as the other scenarios in the main manuscript and presented in Table \ref{tab:simu-3.3}.
\begin{table}[!ht]
    \centering
    \caption{Performance for the extra simulation setting H.}
    \label{tab:simu-3.3}
    \resizebox{.85\textwidth}{!}{
    \begin{tabular}{c|c|c|c|c|c|c|c|c|c|c}
\hline\hline
        Case & CP & mean & sd & rate & time (sec) & SEN & SPC & ACC & MCC & RE \\
    \hline
       \multirow{2}*{H.1} & 1 & 0.330 & 0.019 & 0.94 & \multirow{2}*{2.36} & \multirow{2}*{0.91} & \multirow{2}*{0.99} & \multirow{2}*{0.98} & \multirow{2}*{0.90} & \multirow{2}*{0.31} \\
        & 2 & 0.671 & 0.018 & 0.98 & & & & & & \\
    \hline
       \multirow{2}{*}{H.2} & 1 & 0.330 & 0.014 & 0.94 & \multirow{2}*{2.45} & \multirow{2}*{0.99} & \multirow{2}*{0.98} & \multirow{2}*{0.98} & \multirow{2}*{0.96} & \multirow{2}*{0.20} \\
       & 2 & 0.668 & 0.018 & 0.96 & & & & & & \\
    \hline
       \multirow{2}*{H.3} & 1 & 0.343 & 0.036 & 0.90 & \multirow{2}*{4.71} & \multirow{2}*{0.96} & \multirow{2}*{0.97} & \multirow{2}*{0.92} & \multirow{2}*{0.88} & \multirow{2}*{0.24} \\
       & 2 & 0.654 & 0.051 & 0.94 & & & & & & \\
    \hline\hline
    \end{tabular}}
\end{table}
It can be seen that TBSS exhibits a very satisfactory performance across all metrics for all three scenarios.

Next, we consider a scenario wherein there is no change point in the given time series. The question then becomes of whether TBSS would \textit{falsely} detect change points. In scenario H, we set $p=20$, $T=600$, no change point is present, and the noise term $\{\epsilon_t\}$ is drawn according to $\epsilon_t \sim \mathcal{N}(0, 0.01\mathbf{I})$. The transition matrix $A$ is 1-off diagonal with entries whose magnitude equals 0.8. The TBSS false detection rate is listed in Table \ref{tab:simu-nocp}, based on 100 replicates and using the default block size.
\begin{table}[!ht]
    \centering
    \caption{The performance of false detection rate in the scenario M.}
    \label{tab:simu-nocp}
    \begin{tabular}{c|c|c}
    \hline\hline
        Case & \% of False CP & Running time (sec) \\
    \hline
        M & 8\% & 1.47 \\
    \hline\hline
    \end{tabular}
\end{table}
The results indicate that TBSS fails to detect a change point 92\% of the time, which is a very satisfactory outcome.

\subsection{Comparison with Other Existing Methods}
We compare the performance of TBSS to the FreSpeD method proposed in \cite{schroder2019fresped} using synthetic data. In this additional experiment, we set $p=20$, $T=6000$ with four change points located at $t_1^\star=1200$, $t_2^\star=2400$, $t_3^\star=3600$, and $t_4^\star=4800$, respectively, and the transition matrices are 1-off diagonal and the magnitudes are the same as simulation case A.1. Here we use the default block size $b_T = \lfloor \sqrt{T} \rfloor = 77$. The following Table \ref{tab:compare_table} summarizes the comparison results by using the Hausdorff distance to measure the difference between the set of estimated change points $\widehat{\mathcal{A}}$ and the true change points $\mathcal{A}^\star$.
\begin{table}[!ht]
    \centering
    \caption{Comparison between FreSpeD and TBSS algorithm.}
    \label{tab:compare_table}
    \begin{tabular}{c|c|c|c}
    \hline\hline
       Model & Running time & mean $d_H(\widehat{\mathcal{A}}, \mathcal{A}^\star)$ & sd $d_H(\widehat{\mathcal{A}}, \mathcal{A}^\star)$ \\
    \hline
       FreSpeD & 6.54 & 1151.50 & 738.46 \\
    \hline
       TBSS & \textbf{2.56} & \textbf{195.80} & \textbf{677.68} \\
    \hline\hline
    \end{tabular}
\end{table}

It can be seen that TBSS clearly outperforms FreSpeD both in terms of accuracy, as well as computational speed.

\section{Additional Details for the EEG Data}
\subsection{EEG data pre-processing}
\begin{figure}[!ht]
    \centering
    \includegraphics[scale=.22]{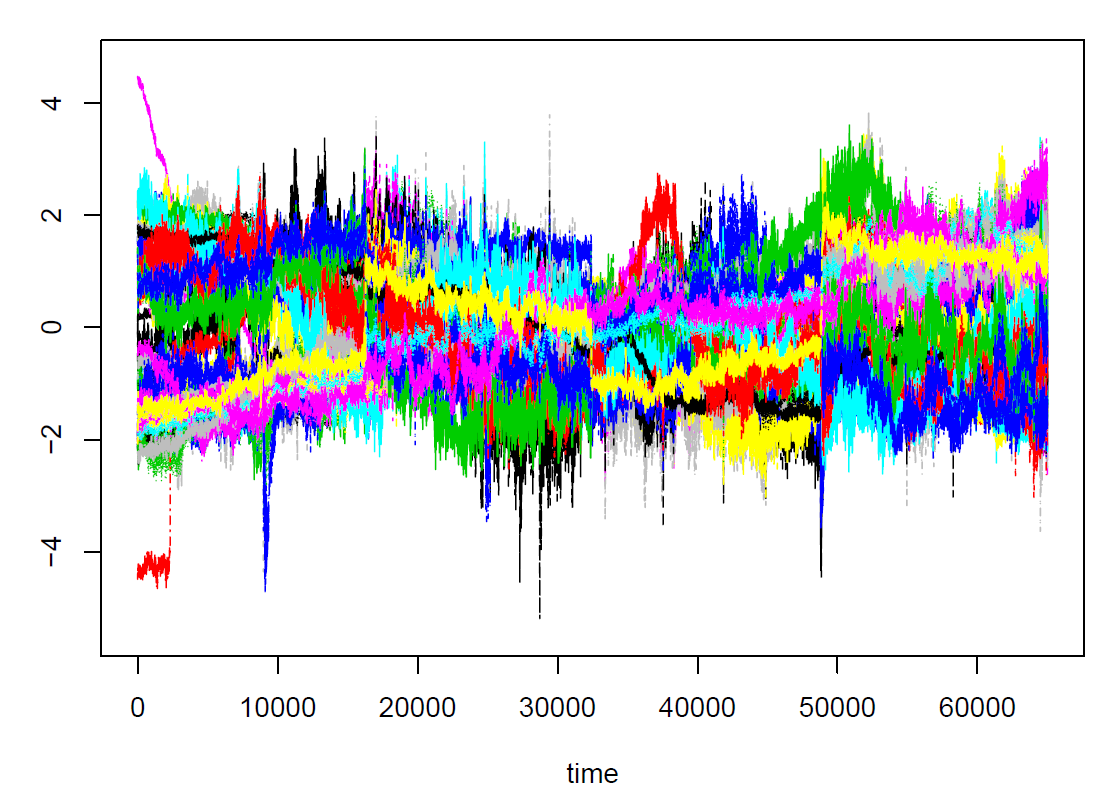}%
    \hfill
    \includegraphics[scale=.22]{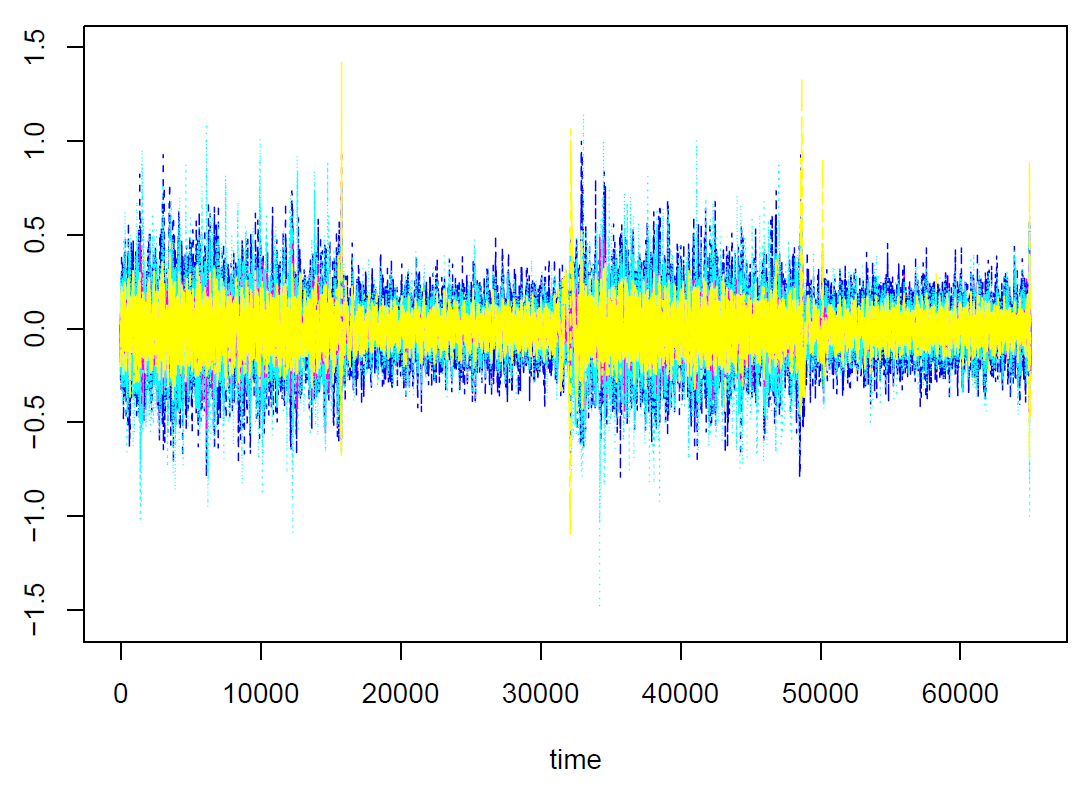}
    \caption{Left panel: raw EEG data for the selected 71 channels; Right panel: all data after robust detrending.}
    \label{fig:rawdata}
\end{figure}

The raw EEG signals data are depicted in Figure~\ref{fig:rawdata} and exhibit obvious trend patterns due to known recording artifacts \cite{de2018robust}. The trend patterns were removed by employing the robust detrending method \cite{de2018robust} based on fitting local polynomials to the data.

\begin{figure}[!ht]
    \centering
    \includegraphics[scale=.22]{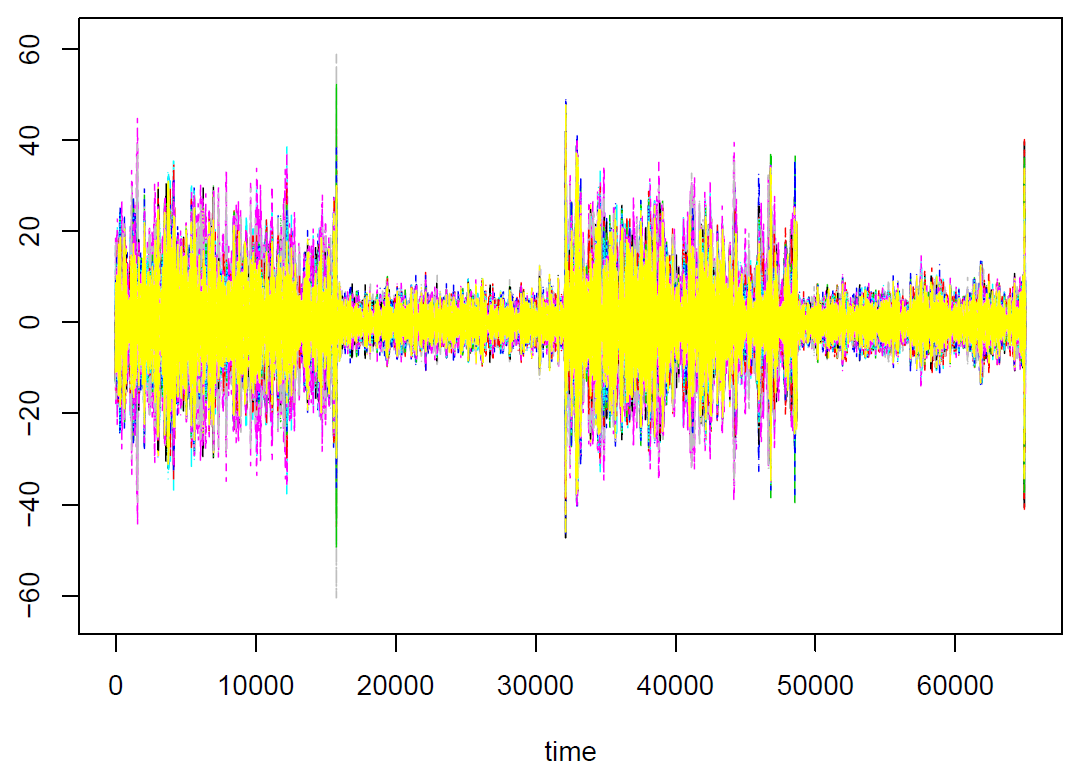}%
    \hfill
    \includegraphics[scale=.22]{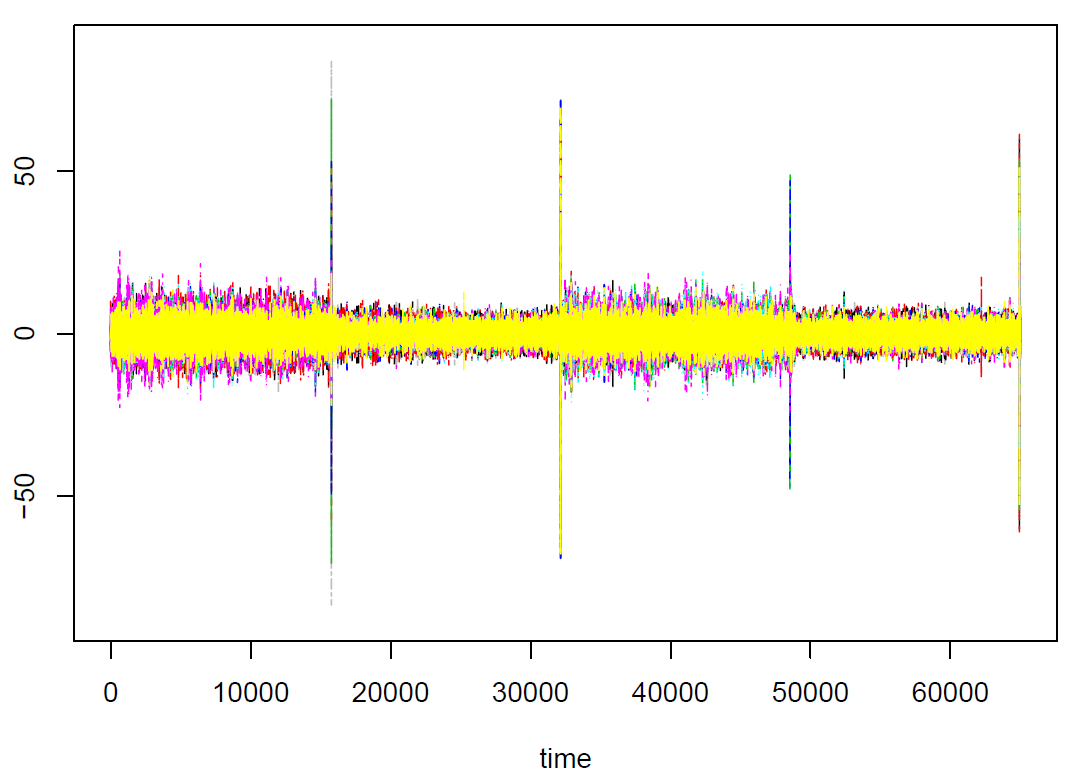}
    \caption{Filtered data in specific frequency band. Left panel: Alpha-band data; Right panel: Beta-band data.}
    \label{fig:freqdata}
\end{figure}

We also extract data corresponding to specific frequency bands: (i) the alpha-band (8-13 Hz) and (ii) the beta-band (14-30Hz); the resulting time series are depicted in Figure~\ref{fig:freqdata}

These plots show that both the robust detrended and the alpha-band data exhibit strong changes in their patterns across different open and closed eyes segments. These patterns are less pronounced in the beta-band data.

\subsection{Evidence of VAR Models}
\subsubsection{Linearity of the Data}
\begin{figure}[!ht]
    \centering
    \includegraphics[scale=.24]{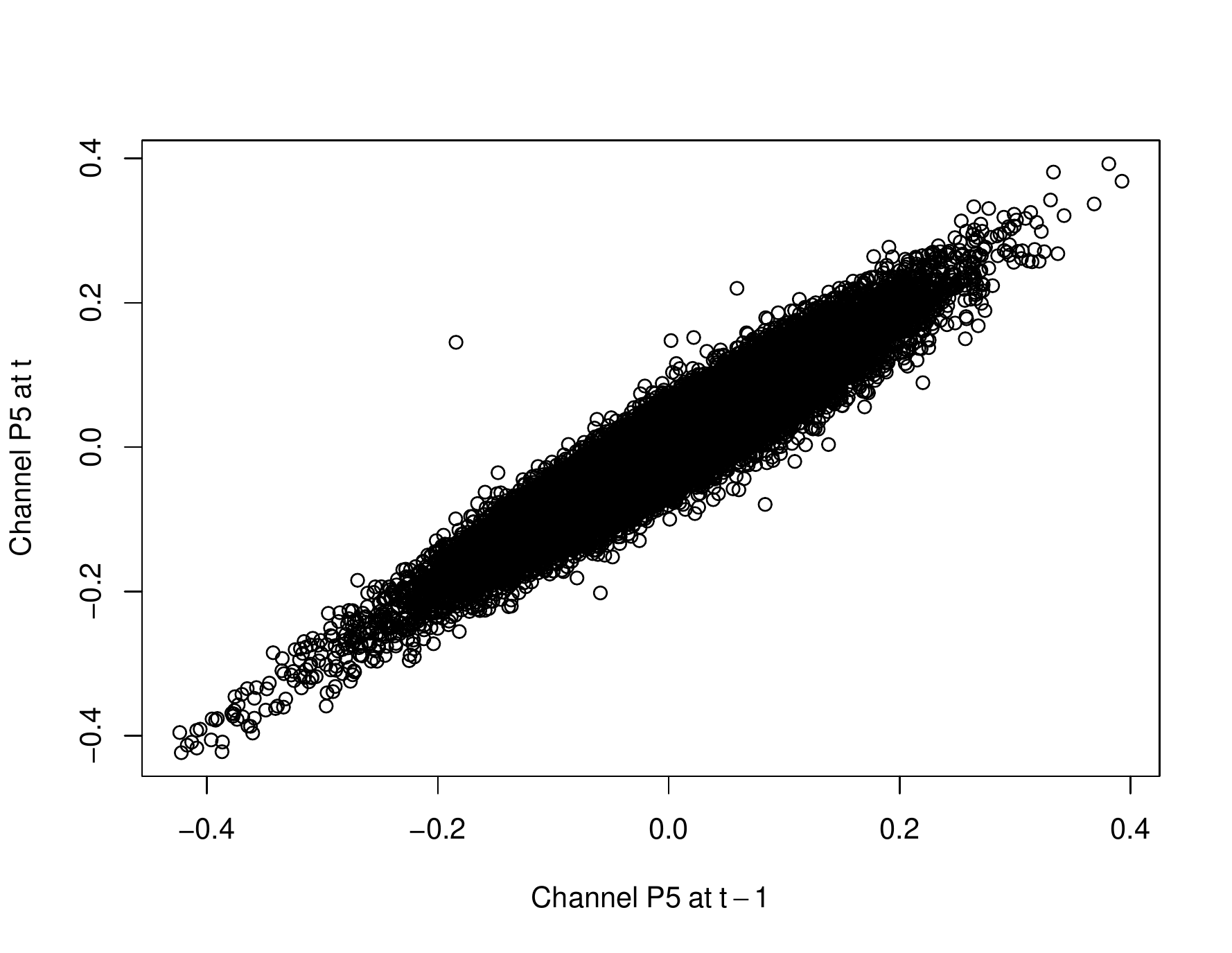}%
    \includegraphics[scale=.24]{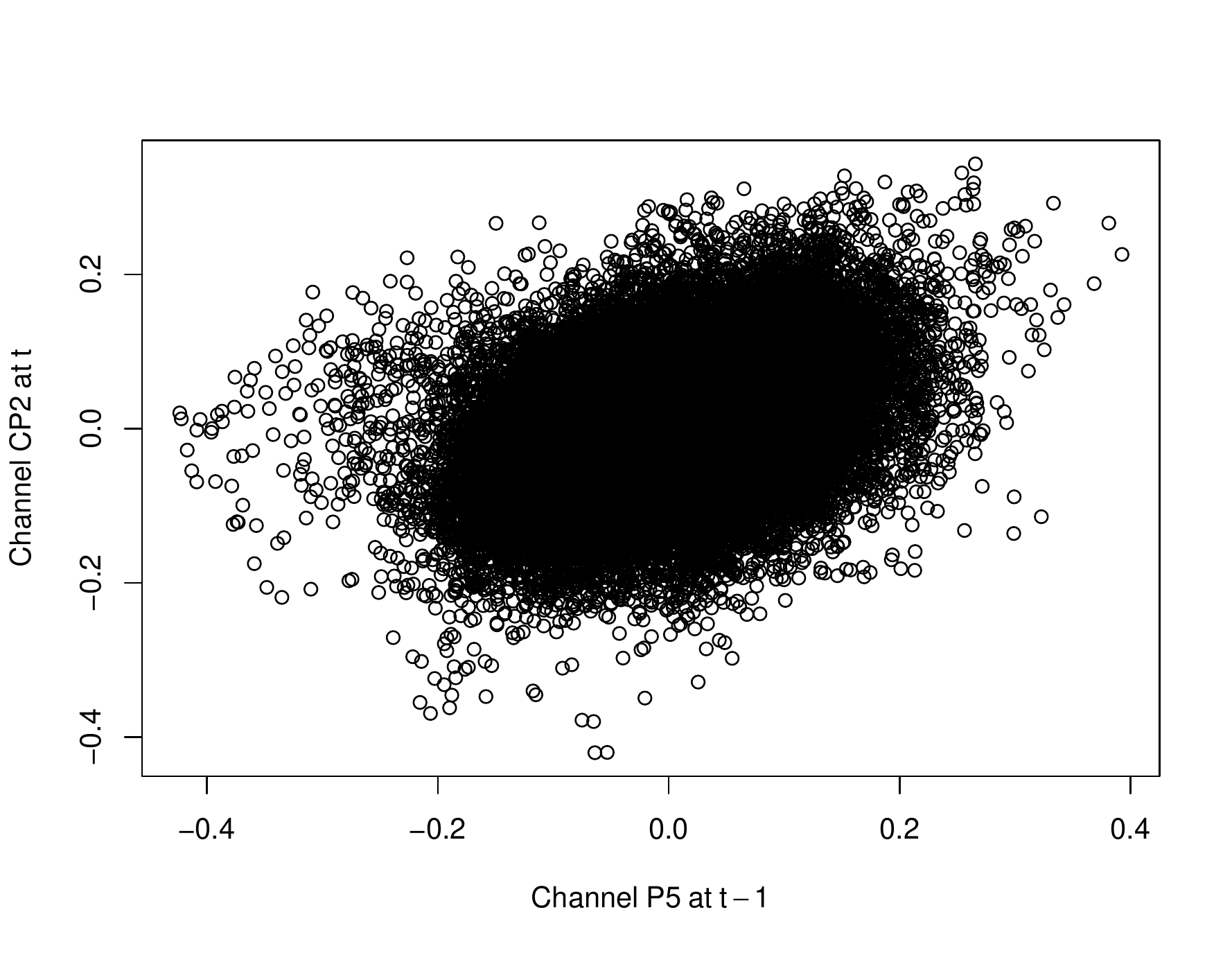}%
    \vfill
    \includegraphics[scale=.24]{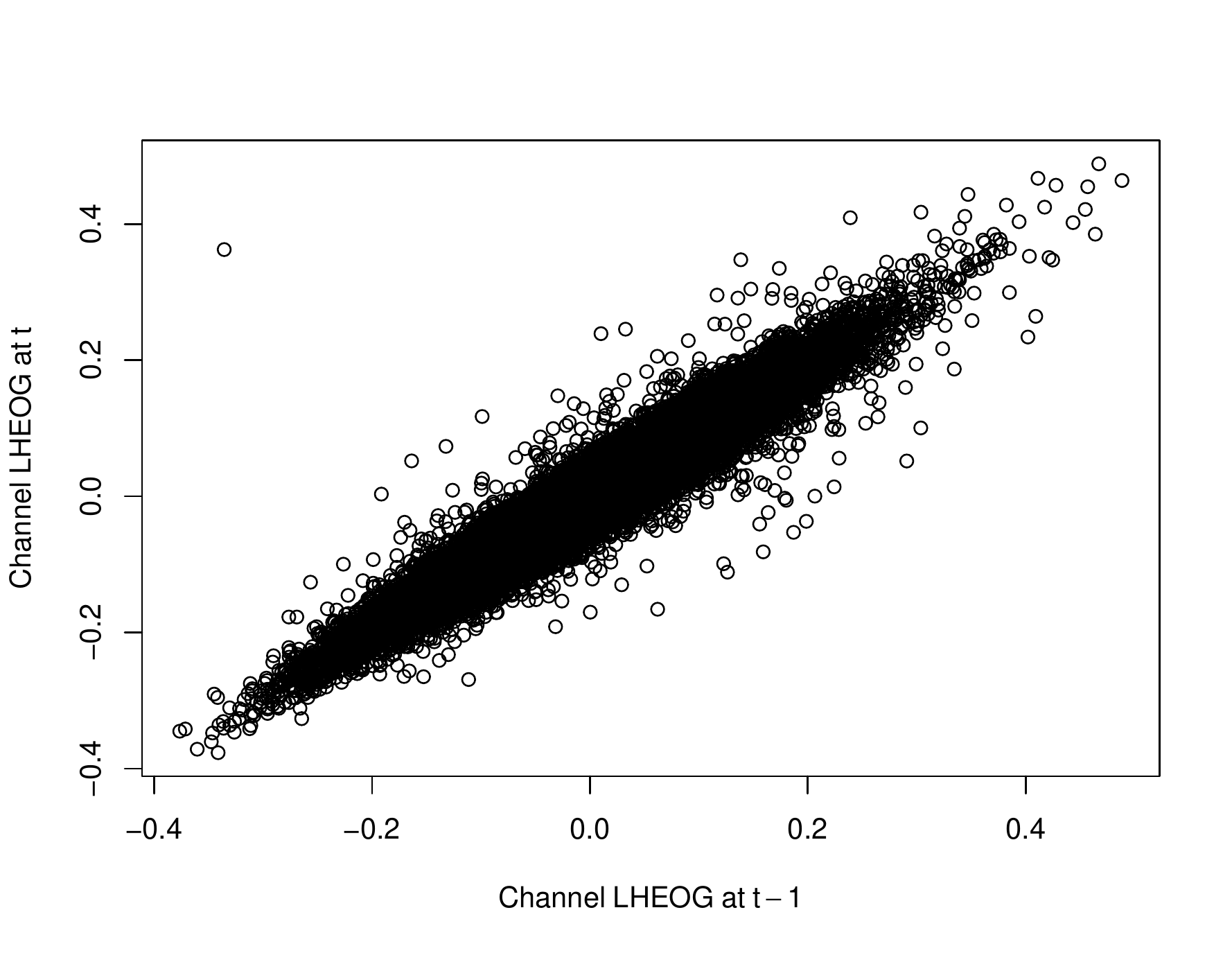}%
    \includegraphics[scale=.24]{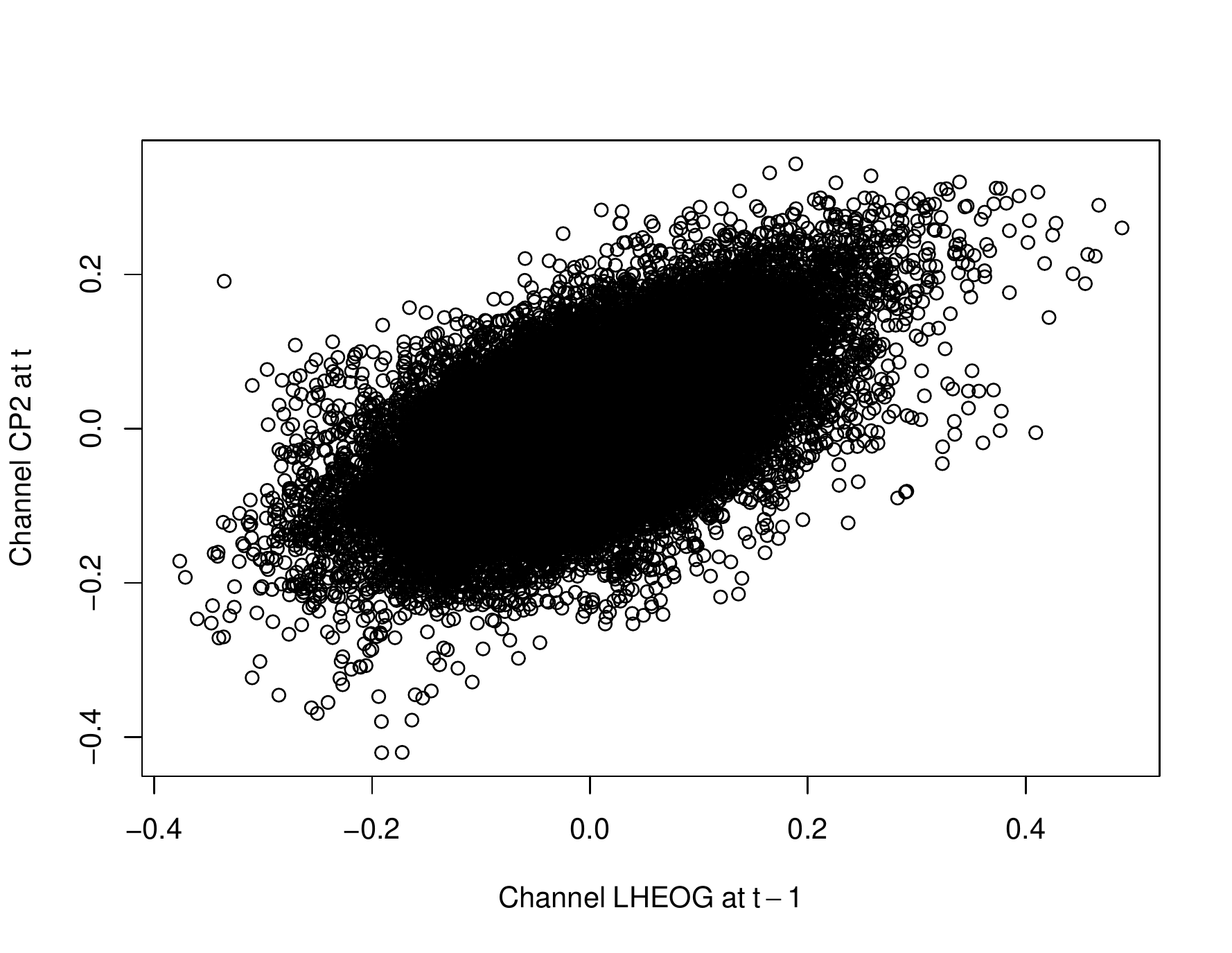}
    \caption{Scatter plots among selected channels in the EEG signals data. }
    \label{fig:scatters}
\end{figure}

In the main text, we have provided strong evidence to justify the sufficiency of a linear VAR model. Some additional evidence is included next. First, scatter plots of other channels for one subject are shown in Figure \ref{fig:scatters} based on the robust detrended data. The top and bottom left panels show scatter plots between the P5 and LHEOG channels and their 1-lag data, respectively. The very strong linear pattern observed confirms the adequacy of using a linear VAR model, since it indicates a strong autoregressive effect (diagonal entries of the transition matrix). The top and bottom right panels show scatter plots of the CP2 channel vs the P5 and LHEOG ones, respectively, indicate strong cross autoregressive effects (off-diagonal entries of the transition matrix), thus, also strongly supporting the adequacy of a linear VAR model.

\subsubsection{Selection Procedure for Time Lags of the VAR Models}
We consider the time lag selection procedure for the VAR models. The key steps for using the BIC criterion are:
\begin{itemize}
    \item[1.] Employ VAR($d$) models based on for different lags $d=1,2,\dots$ to detect change points in the given time series data;
    \item[2.] Based on the detected change points, we fit a VAR($d$) model for the segment $[\widehat{t}_j, \widehat{t}_{j+1})$, for $j=1,2,\dots, \widehat{m}$, and calculate the BIC value as follows:
    \begin{equation*}
        \text{BIC}_j = \log\det(\widehat{\Sigma}_j) + \log (\widehat{t}_{j+1} - \widehat{t}_j)d_j,
    \end{equation*}
    where $\widehat{\Sigma}_j$ is the estimated covariance matrix for the $j$th segment $[\widehat{t}_j, \widehat{t}_{j+1})$, $d_j$ is the number of non-zero elements in the transition matrices. Then, we define the total BIC is:
    \begin{equation*}
        \text{BIC} = \sum_{j=1}^{\widehat{m}}\text{BIC}_j.
    \end{equation*}
\end{itemize}
Based on the developed BIC selection procedure, the results for the selection of lags are given in Table \ref{tab:bic-table}.
\begin{table}[!ht]
    \centering
    \caption{BIC values for different time lag VAR models. }
    \label{tab:bic-table}
    \begin{tabular}{c|c|c|c|c|c}
    \hline\hline
        Lags & 1 & 2 & 3 & 4 & 5 \\
    \hline
        Total BIC value $(\times 10^6)$ & 1.566 & 5.777 & 6.544 & 9.097 & 9.203 \\
    \hline\hline
    \end{tabular}
\end{table}
It can be seen that a VAR(1) is selected by the BIC criterion.



\subsection{Granger causal networks obtained from Stability Selection (Step 6 of TBSS)}
\begin{figure}[!ht]
    \centering
    \includegraphics[scale=.24]{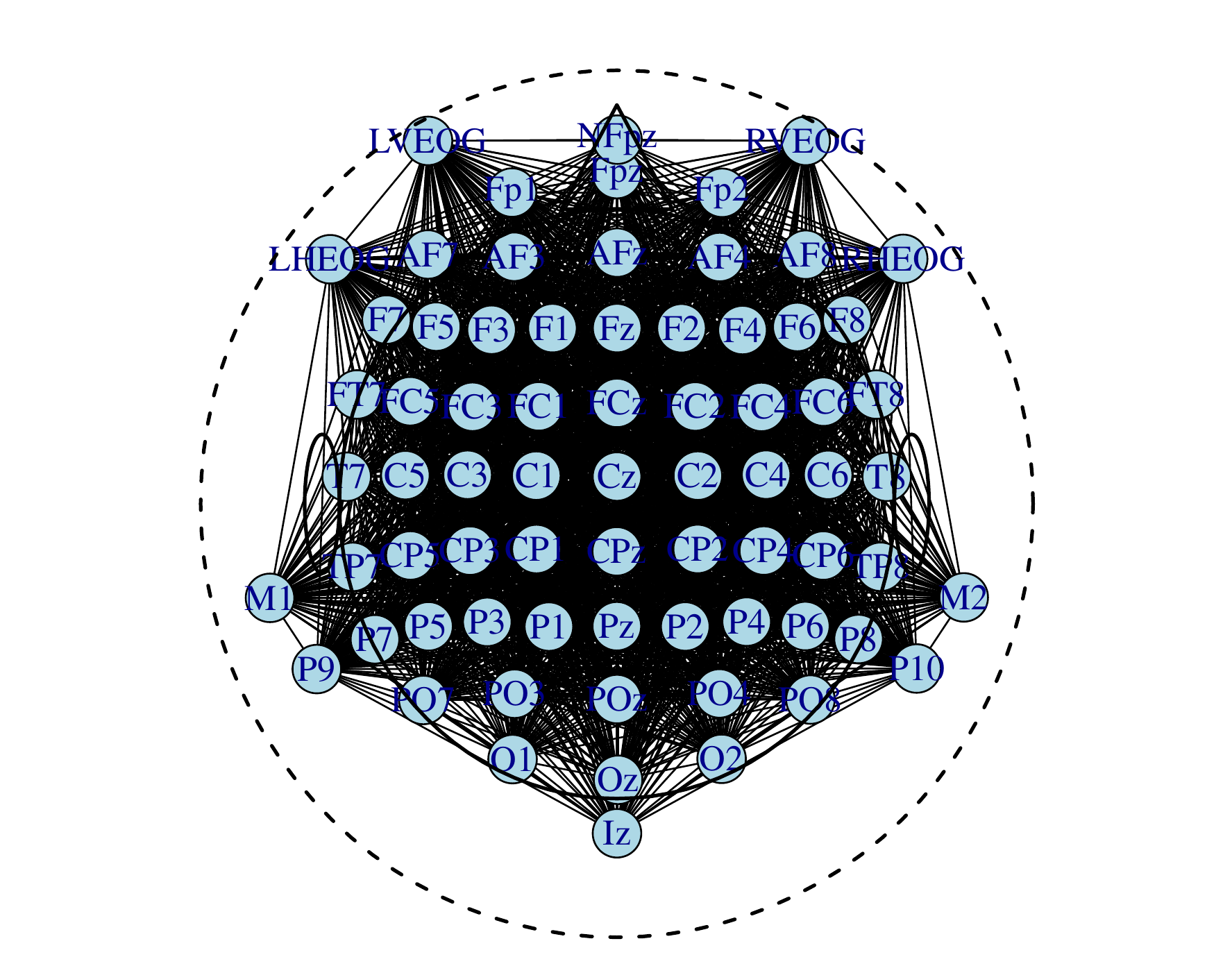}%
    \includegraphics[scale=.24]{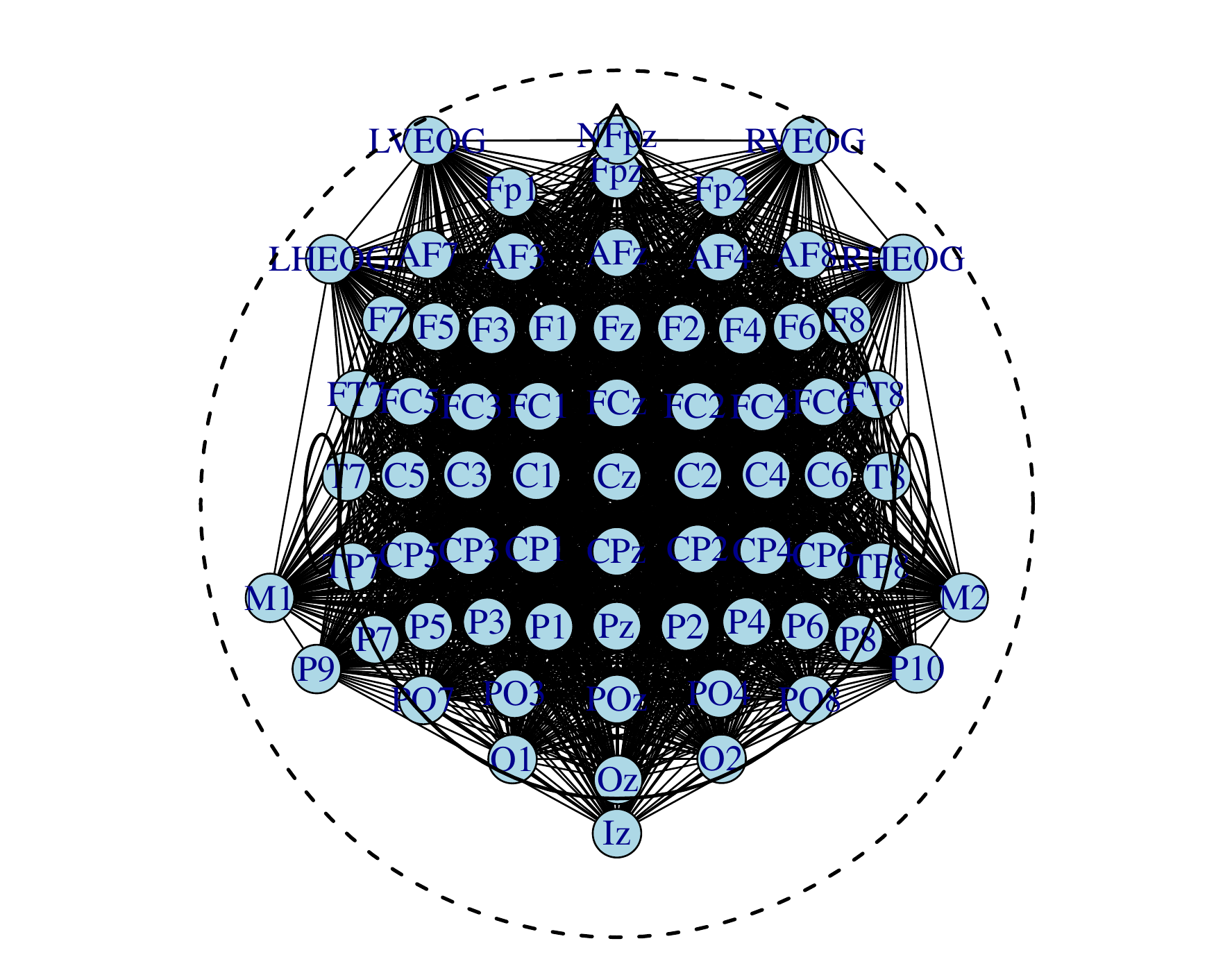}%
    \vfill
    \includegraphics[scale=.24]{images/stabs_open_network.pdf}%
    \includegraphics[scale=.24]{images/stabs_close_network.pdf}
    \caption{Estimated Granger causal networks for open (left column) / close (right column) segments from Steps 5 (top panel) and 6 (stability selection - bottom panel) of TBSS.}
    \label{fig:stabs_select}
\end{figure}

Figure \ref{fig:stabs_select} depicts the average across all 21 subjects of the estimated Granger causal networks for the open and closed eyes states, obtained without (top panel) and with (bottom panel) Step 6 of TBSS, based on 50 subsamples and a cutoff threshold of 0.8. It can be seen that the resulting stable networks are more sparse and hence easier to identify key connectivity patterns.

\subsection{Additional Comparison with Other Existing Methods}
In this section, we compared our TBSS algorithm with DCR algorithm proposed in \cite{cribben2013detecting}. The following figure \ref{fig:compare_dcr} illustrates the estimated change points via their method DCR.
\begin{figure}[!ht]
    \centering
    \includegraphics[scale=.42]{images/revised_cps_histogram.pdf}%
    \includegraphics[scale=.42]{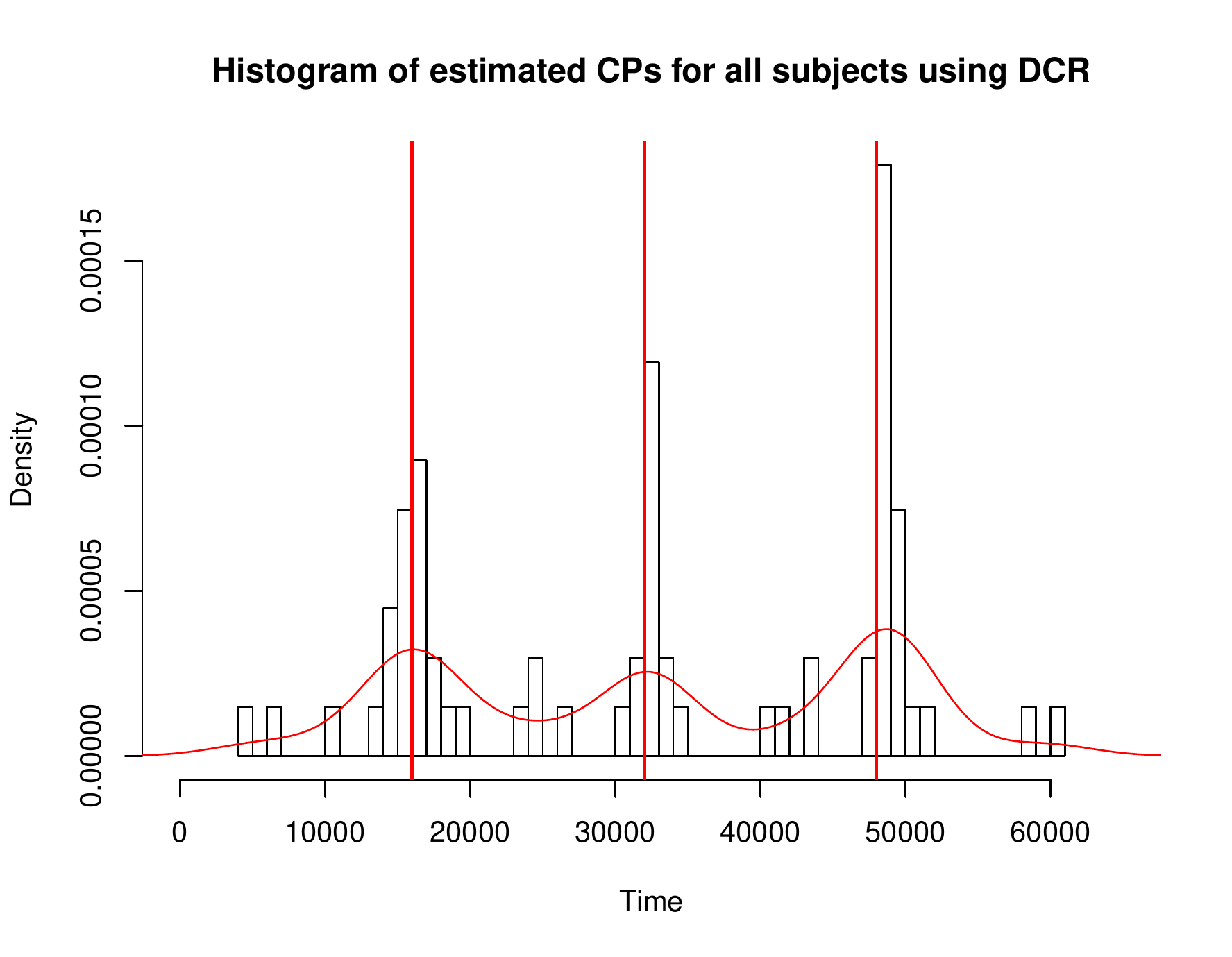}
    \caption{Histogram of selected change points for all 21 subjects by using TBSS and DCR algorithms. Left panel: histogram for TBSS algorithm; Right panel: histogram for DCR algorithm.}
    \label{fig:compare_dcr}
\end{figure}
As one can see, both TBSS algorithm and DCR algorithm provide satisfactory estimated change points. However, for the given EEG signals dataset, DCR algorithm requires nearly $>24$ hours running time per subject, TBSS only needs $\approx 25$ minutes to get final result for each subject. It's worth to note that in \cite{cribben2013detecting}, the authors only examined data sets with $p=15$ and $T=500$ model dimensions, which is far smaller than our application scenario. Hence, for examining all 21 subjects, we applied High Performance Computing (HPC) platform to allocate 21 CPU cores, which is expensive for normal users. 

\subsection{Additional Results for the EEG data}
\begin{figure}[!ht]
    \centering
    \includegraphics[scale=.42]{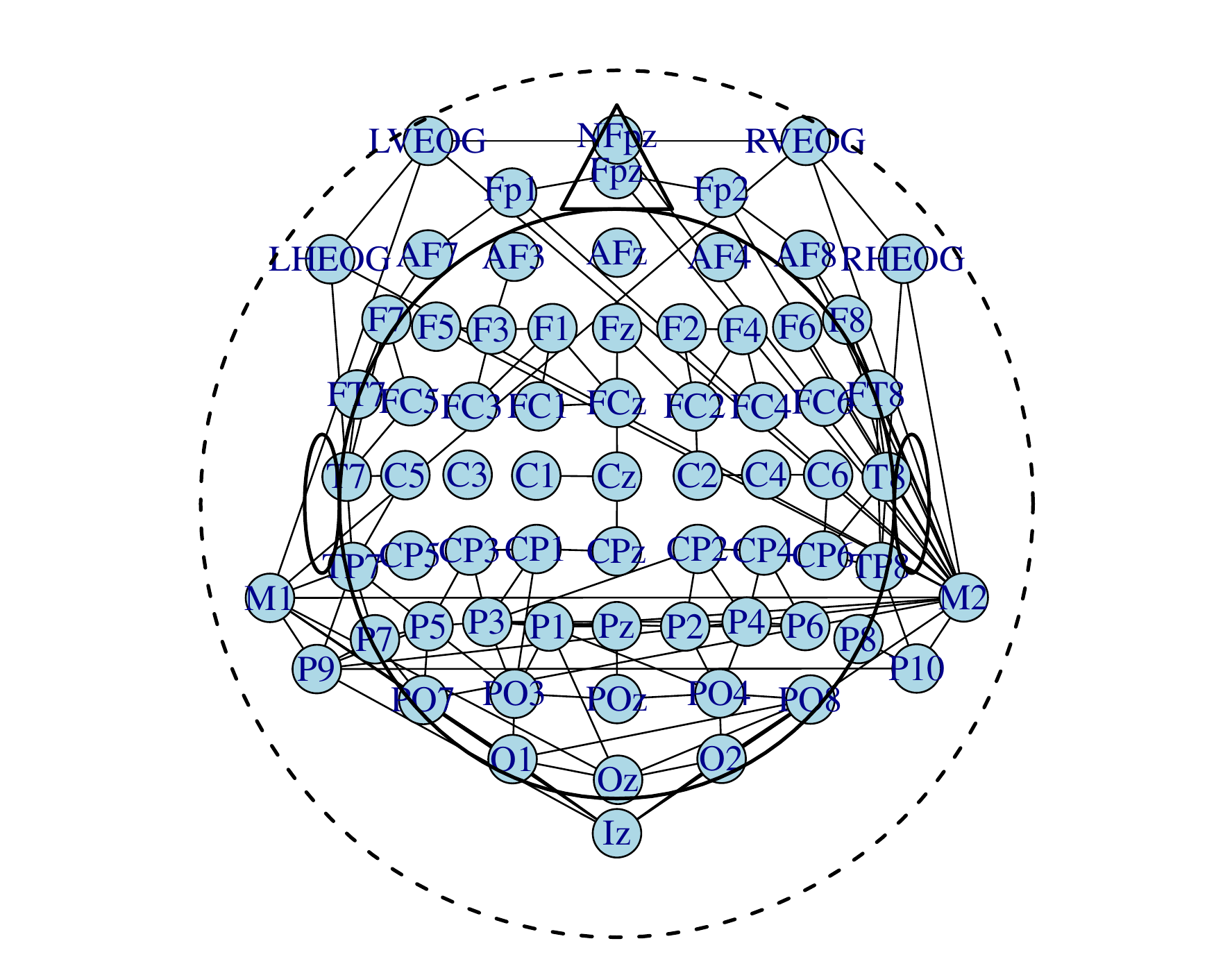}%
    \includegraphics[scale=.42]{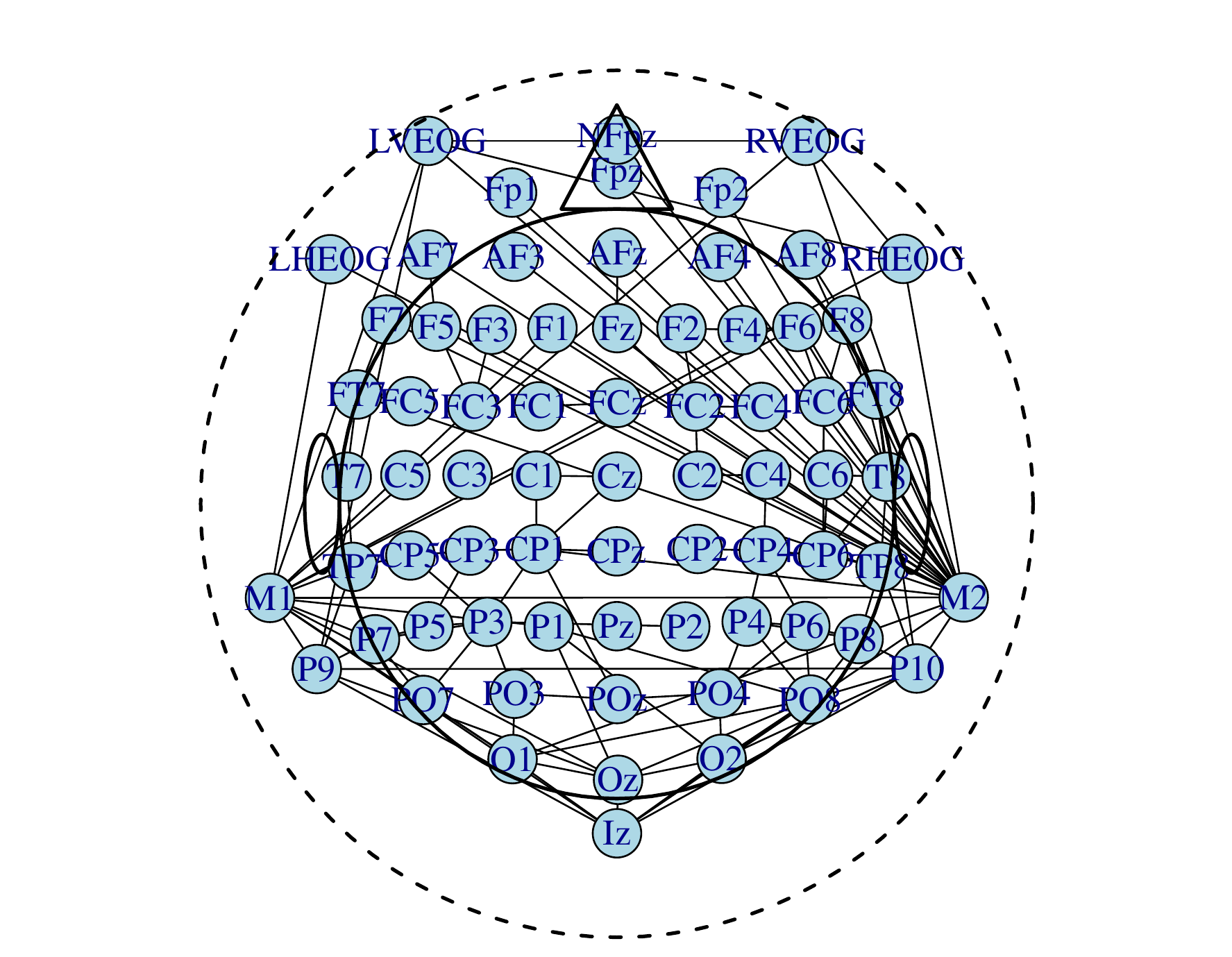}
    \caption{Estimated Granger causal networks for open (left panel) and closed (right panel) states by using filtered beta-band data.}
    \label{fig:beta-band-network}
\end{figure}

In the main text, the Granger causal networks obtained from the full data and those extracted from the alpha band were shown. Figure \ref{fig:beta-band-network} shows their counterparts obtained from the beta-band data. 

Further, Table \ref{tab:beta-result} provides some additional details. It can be seen that the total connectivity is not different across the two states, unlike the case for the full and the alpha-band data that show marked differences. On the other hand, we observed differences between the most connected channels. Further, the Hamming distance (HD) between these two networks is $\text{HD}_{beta}(open, close)=0.032$, which is much smaller compared to the distance calculated using the robust detrended and alpha-band data given in the main text.
\begin{table}[!ht]
    \centering
    \caption{Additional results for the estimated networks for eye-open/eye-closed states for all 21 subjects.}
    \label{tab:beta-result}
    \resizebox{0.85\textwidth}{!}{
    \begin{tabular}{c|c|c|c}
    \hline\hline
    Data & State & No. edges & Top 5 most connected channels (degrees) \\
    \hline
    \multirow{2}*{Beta-band} & Open & 203 & M2(20), P3(10), M1(10), TP7(8), P5(8)  \\
         & Closed & 205 & M2(27), M1(15), PO8(10), Iz(9), P10(9) \\
    \hline\hline
    \end{tabular}}
\end{table}

Finally, we present coherence and partial coherence plots for selected highly connected EEG channels. The original data are subsampled, by selecting one observation every eight time points. Therefore, the true break points are located at $t_1=2040, t_2=4080$ and $t_3=6120$, respectively. In the following figure \ref{fig:coherence}, we examine two channel pairs: (PO7, Iz) and (Oz, PO8), which are the most connected ones listed in Table III in the main text. 
\begin{figure}[!ht]
    \centering
    \includegraphics[scale=.42]{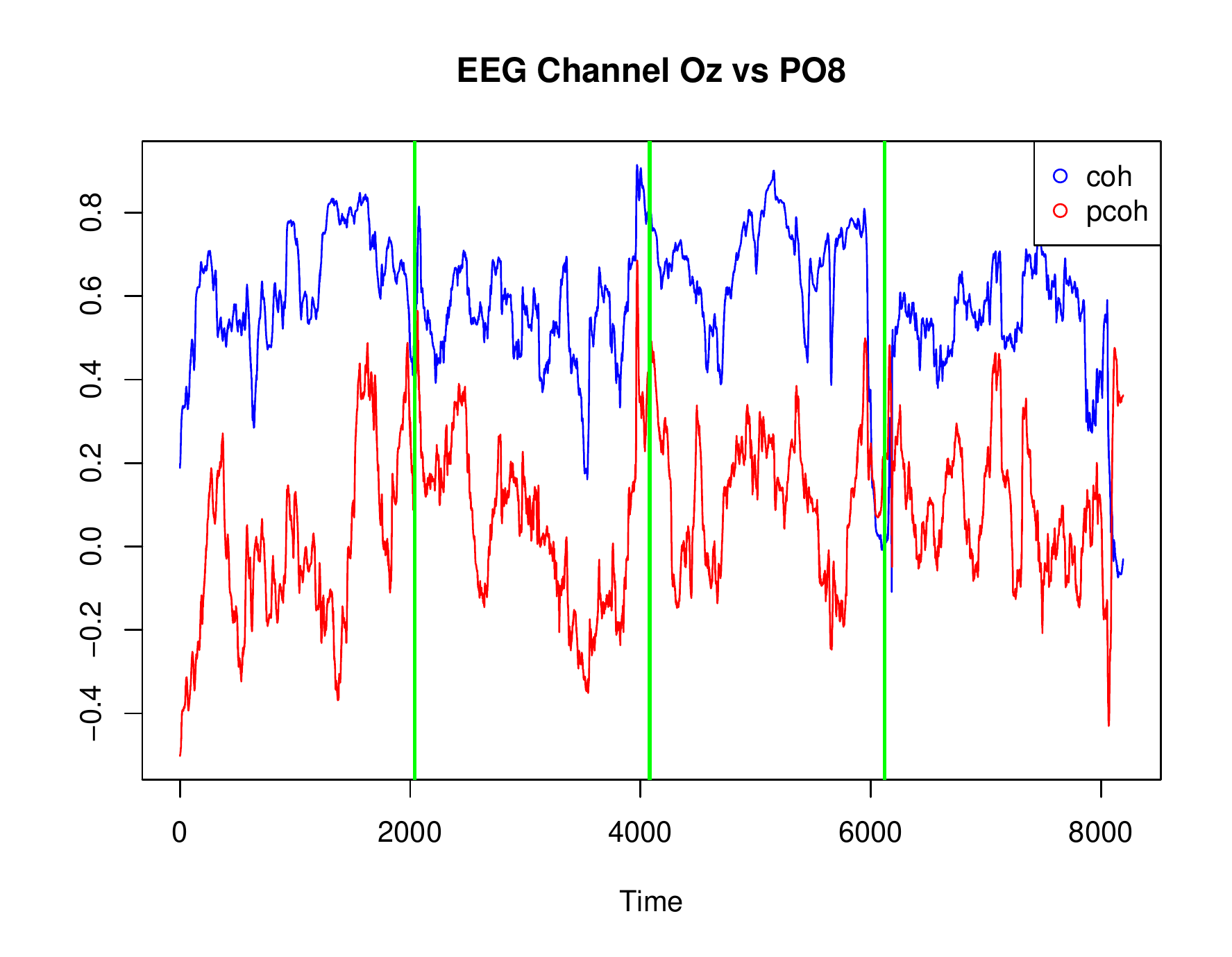}%
    \includegraphics[scale=.42]{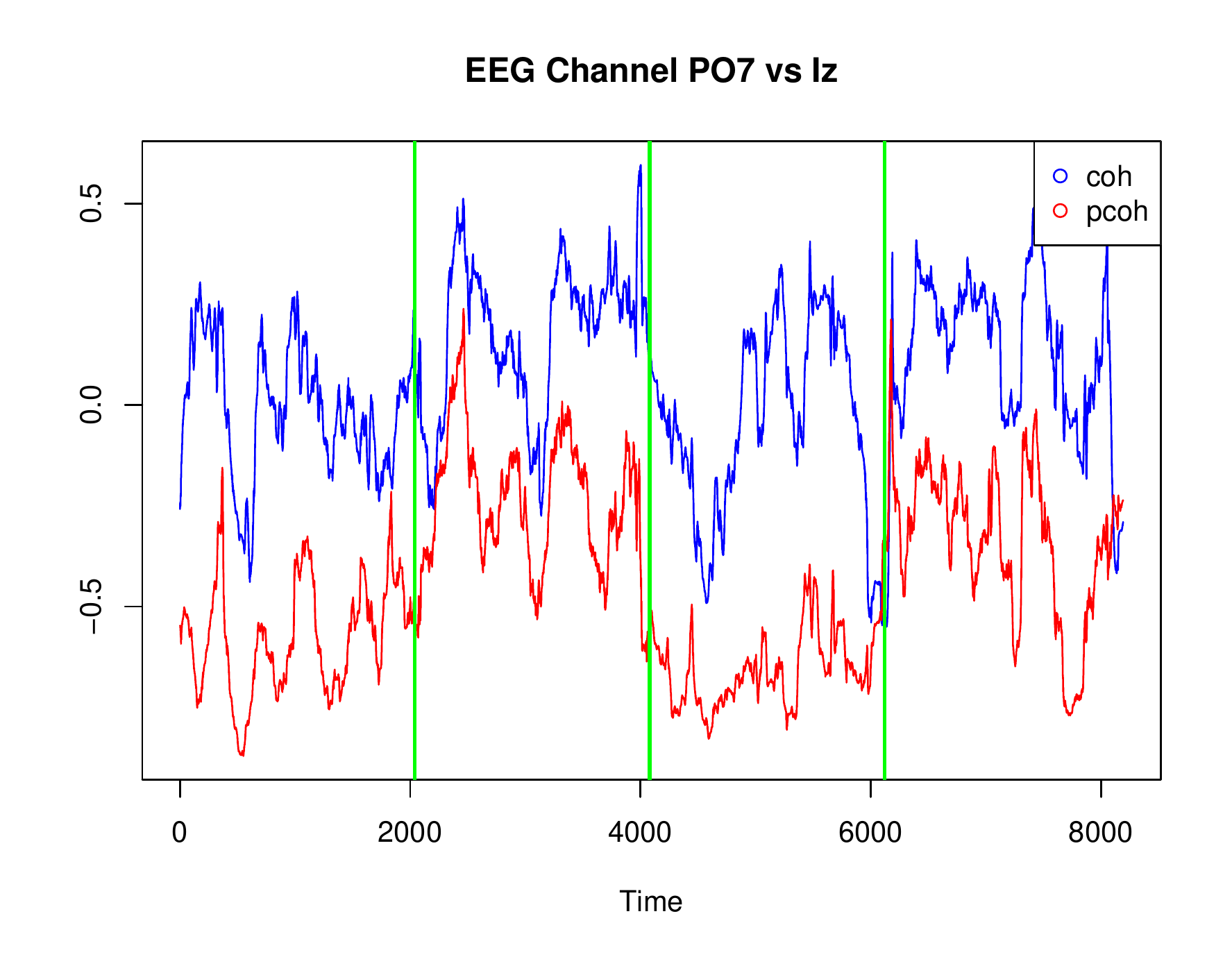}
    \vspace{-20pt}
    \caption{Coherence plots for selected EEG channel pairs. Blue curve: coherence plot; Red curve: partial coherence plot; green lines: locations of true break points.}
    \label{fig:coherence}
\end{figure}

From Figure \ref{fig:coherence}, one can see that for pair Oz and PO8, both the coherence and partial coherence functions increase to around 0.8 when they get close to the break points, and decrease afterwards to less than 0.7. As described in details in \cite{ombao2006coherence}, absolute coherence of above 0.7 is considered ``strong" in the EEG literature while absolute coherence between 0.3 and 0.7 is considered as ``moderate". Using this terminology, we can state that the coherence between channels Oz and PO8 becomes strong near true location of break points while it remains weak ($<0.3$) or moderate at other time points.  For pair PO7 and Iz, a clear pattern is observed within the eyes-closed segments (0-2000 and 4000-6000 time periods); namely, coherence is around 0 and partial coherence is around -0.5. Contrary in the eyes-open segments (2000-4000 and 6000-8000 time periods), the coherence increases to almost 0.5, and the partial coherence changes to close to 0. For the last pair PO8 and Iz, it is obvious that both coherence and partial coherence drops at the break points. 

\section{Additional Details for the Weakly Sparse VAR Model}
We assume that the transition matrices of the linear VAR model are weakly sparse. A $p \times p$ matrix $A$ is weakly sparse \cite{negahban2012unified}, if it satisfies
\begin{equation}
    \label{eq:lq-ball}
    \mathbb{B}_q(R_q) := \left\{ A \in \mathbb{R}^{p\times p}: \sum_{i=1}^p\sum_{j=1}^p |a_{ij}|^q \leq R_q \right\},
\end{equation}
for some $q \in (0,1)$; i.e., its entries are restricted in an $\ell_q$ ball of radius $R_q$. Note that when $q \to 0^+$, this set converges to an exact sparse model, if and only if $A$ has at most $R_0$ nonzero entries. Otherwise, the resulting matrix is relatively dense, with most of its entries being of small magnitude. 

For a weakly sparse model, the following optimization problem is solved in Step 5 of TBSS. Suppose we obtain $\widetilde{m}$ break points after Steps 1-4 of TBSS. After removing data in each $R_T$-radius neighborhood for each break point $\widetilde{t}_j$, the resulting penalized regression problem \eqref{eq:estimation_third} is modified as:
\begin{equation}
    \label{eq:weakly-sparse-estimate}
    \widetilde{\textbf{B}} = \argmin_{\textbf{B}}\frac{1}{N}\left\| \textbf{Y}_{\textbf{s}} - {\textbf{Z}_{\textbf{s}}} \textbf{B} \right\|_2^2,\quad \text{s.t.}\quad \textbf{B} \in \mathbb{B}_q(R_q),
\end{equation}
where all notation is the same as defined in Step 5. Problem \eqref{eq:weakly-sparse-estimate} can be solved using the same lasso algorithm, as for the strictly sparse case \cite{negahban2012unified}, and replacing the resulting zero entries with small magnitudes that satisfy the constraint in
\eqref{eq:weakly-sparse-estimate}.

\section{The TBSS Algorithm}
The main detection steps 1-4 of TBSS are summarized in the following Algorithm \ref{algo:1}. They are also illustrated in Figure~\ref{fig:steps} in the main text.

\begin{center}
\scalebox{0.75}{
\begin{minipage}{\textwidth}
\begin{algorithm}[H]
    \DontPrintSemicolon
    \KwInput{Time series data $\{X_t\}$, $t=1,2,\dots, T$; regularization parameters $\lambda_{1,T}$ and $\lambda_{2,T}$; block size $b_T$.}
    
    \ {\bf Block Fused Lasso:} Partition time series into blocks of size $b_T$ and fix the coefficient parameters within each block. Then we estimate model parameters for all blocks $\widehat{\Theta}$ by solving:
    \begin{equation*}
        \widehat{\mathbf{\Theta}} 
        = \argmin_{\mathbf{\Theta}}\frac{1}{T-q+1} \| \textbf{Y} - \textbf{Z} \mathbf{\Theta} \|_2^2 + \lambda_{1,T}  \| \mathbf{\Theta} \|_1 + \lambda_{2,T} \sum_{i=1}^{k_T} \left \| \sum_{j=1}^{i} \theta_j \right \|_1.
    \end{equation*}
    Then, obtain candidate break points
    \begin{equation}
        \label{eq:candidate_set}
        \widehat{\mathcal{A}}_T = \left\{i \geq 2: \widehat{\theta}_i \neq 0 \right\}.
    \end{equation}
    
    \ {\bf Hard-thresholding:} We firstly denote the set of selected blocks with large jumps as $J$ and initialize it as $J = \emptyset$, and initialize the value of the BIC function by $\text{BIC}^{\text{old}}=\infty$.
    
    \While{$\text{BIC}^{\text{diff}} < 0$}{
    \ Apply $k$-means clustering to the obtained jumps set $V$ with $k=2$ clusters. Use $V_S$ to denote the set with small jump values, and $V_L$ to denote the set with large jump values. 
    
    \ Define $\eta = (\min V_L + \max V_S)/2$ as the threshold. 
    
    \ Add the corresponding blocks in $V_L$ into $J$ and compute the BIC by using the estimated parameter $\widehat{\Theta}$ with $\widehat{\theta}_i = 0$ for blocks $i \notin J$ and denoted as $\text{BIC}^{\text{new}}$.
    
    \ Compute the difference between the BIC values: $\text{BIC}^{\text{diff}} = \text{BIC}^{\text{new}} - \text{BIC}^{\text{old}}$. Set $\text{BIC}^{\text{old}} = \text{BIC}^{\text{new}}$ and $V = V\backslash J$.
    }
    
    \ {\bf Block Clustering:} In this step, we use the Gap statistics to determine the number of clusters of the candidate break points screened by Step 2. The output is $C = \{C_1, C_2, \dots, C_{\widetilde{m}}\}$. 
    
    \ {\bf Exhaustive Search:} For each selected cluster of break points $C_i$, for $i=1,2,\dots, \widetilde{m}$ from Step 3. We define the search interval $(l_i, u_i)$, and the lower and upper bound $l_i$ and $u_i$ are defined in the main context. Denote the subset of corresponding block indices in the interval $(l_i, u_i)$ by $J_i$ with $J_0 = \{1\}$ and $J_{\widetilde{m}+1} = \{k_T\}$. Then denote the closest block end to $(\max J_{i-1} + \min J_{i})/2$ as $w_i$. 
    
    \For{$i=1,2,\dots, \widetilde{m}+1$}{
    \ Define \textit{local} parameter estimators are:
    \begin{equation*}
        \widetilde{\Phi}^{(\cdot, i)} = \sum_{k=1}^{w_i}\widehat{\theta}_k,
    \end{equation*}
    where $\widehat{\theta}_k$ is derived by fused lasso program in Step 1. 
    
    \ The final selected break points is defined as:
    \begin{equation}
        \label{eq:final_cps}
        \widetilde{t}_i 
        = \argmin_{s\in (l_i, u_i)}\left\{ \sum_{t=l_i}^{s-1}\|y_{t+1} - \widetilde{\Phi}^{(.,i)}Y_t \|_2^2 + \sum_{t=s}^{u_i-1}\|y_{t+1} - \widetilde{\Phi}^{(.,i+1)}Y_t \|_2^2 \right\},
    \end{equation}
    }
    where $\widetilde{\Phi}^{(\cdot, i)}$ are the local parameter estimates obtained previously. The final set of estimated break points obtained by solving \eqref{eq:final_cps} are denoted as $\widetilde{\mathcal{A}}_T = \{\widehat{t}_1, \dots, \widetilde{t}_{\widetilde{m}}\}$.
    
    \ {\bf Model Parameters Estimation:} Based on the final estimated break points $\widetilde{\mathcal{A}}_T$, we partition input time series into $\widetilde{m}$ \textit{strictly stationary} segments, and estimate model parameters on all estimated segments via solving \eqref{eq:estimation_third}.
    
    \KwOutput{The final estimated break points $\widetilde{\mathcal{A}}_T = \{\widetilde{t}_1, \dots, \widetilde{t}_{\widetilde{m}}\}$, and corresponding model parameters $\widehat{\textbf{B}}$.}
    \caption{Threshold Block Segmentation Scheme (TBSS) Algorithm}
    \label{algo:1}
\end{algorithm}
\end{minipage}%
}
\end{center}

\printbibliography

\end{document}